\documentclass[hidelinks,3p,times]{elsarticle}
\journal{Journal of Computational Physics}

\usepackage{color}
\usepackage{natbib}
\usepackage{amsmath}
\usepackage{amssymb}
\usepackage{graphicx}
\usepackage{amsthm}
\usepackage{empheq}
\usepackage{multirow}
\usepackage{subfig}
\usepackage{moreverb}
\usepackage{nicefrac} 
\usepackage{dsfont}
\usepackage{bm}
\graphicspath{{}} 

\usepackage{dsfont}
\usepackage{float}
\usepackage{multicol}
\usepackage{hyperref}
\biboptions{longnamesfirst}
\newcommand{\A}{\mathbf{A}} 
\newcommand{\AAA}{\mathbf{A}} 
\newcommand{\J}{\mathbf{J}}
\newcommand{\G}{\mathbf{G}}

\usepackage{color} 
\usepackage{psfrag}

\newcommand{\halb}{{\frac{1}{2}}}
\newcommand{\quarter}{{\frac{1}{4}}}

\newcommand{\Q}{{\mathbf{Q}}}
\newcommand{\F}{{\mathbf{F}}}
\renewcommand{\G}{{\mathbf{G}}}

\newcommand{\B}{{\mathbf{B}}}
\renewcommand{\A}{{\mathbf{A}}}
\renewcommand{\S}{{\mathbf{S}}}
\newcommand{\pd}{\partial}

\begin{document}
	
	\begin{frontmatter}
		
		\title{A structure-preserving staggered semi-implicit finite volume scheme for continuum mechanics}

		\author[UniFE]{W. Boscheri}
		\ead{walter.boscheri@unife.it}
		\author[UniTN]{M. Dumbser \corref{mycorrespondingauthor}}
		\cortext[mycorrespondingauthor]{Corresponding author}
		\ead{michael.dumbser@unitn.it}
		\author[UniTN]{M. Ioriatti}
		\ead{matteo.ioriatti@unitn.it}
		\author[UniTN]{I. Peshkov}
		\ead{ilya.peshkov@unitn.it}		
		\author[UniTN,Sob]{E. Romenski}
		\ead{evrom@math.nsc.ru}
		
		\address[UniFE]{Department of Mathematics and Computer Science, University of Ferrara, via Machiavelli 30, I-44121 Ferrara, Italy}
		\address[UniTN]{Department of Civil, Environmental and Mechanical Engineering, University of Trento, Via Mesiano 77, I-38123 Trento, Italy}
		\address[Sob]{Sobolev Institute of Mathematics, 4 Acad. Koptyug Avenue, 630090 Novosibirsk, Russia} 
		
		\begin{abstract}
			
We propose a new \textit{pressure-based} structure-preserving (SP) and quasi asymptotic preserving (AP) staggered semi-implicit finite volume scheme for the unified first order hyperbolic formulation of continuum mechanics \cite{GPRmodel}, which goes back to the pioneering work of Godunov \cite{God1961} and further work of Godunov and Romenski \cite{GodunovRomenski72} and Peshkov \& Romenski \cite{PeshRom2014}. 
The unified model is based on the theory of symmetric-hyperbolic and thermodynamically compatible (SHTC) systems \cite{God1961,Rom1998} and includes the description of elastic and elasto-plastic solids in the  nonlinear large-strain regime as well as viscous and inviscid heat-conducting fluids, which correspond 
to the stiff relaxation limit of the model. In the absence of relaxation source terms, the homogeneous 
PDE system is endowed with two stationary linear differential constraints  (\textit{involutions}), 
which require the curl of distortion field and the curl of the thermal impulse to be zero for 
all times. In the stiff relaxation limit, the unified model tends asymptotically to the compressible
Navier-Stokes equations. 

The new structure-preserving scheme presented in this paper can be proven to be \textit{exactly curl-free} 
for the homogeneous part of the PDE system, i.e. in the absence of relaxation source terms. 
We furthermore prove that the scheme is quasi asymptotic preserving  
in the stiff relaxation limit, in the sense that the numerical scheme reduces to a consistent second 
order accurate discretization of the compressible Navier-Stokes equations when the relaxation times tend 
to zero. Last but not least, the proposed scheme is suitable for the simulation of all Mach number flows 
thanks to its conservative formulation and the implicit discretization of the pressure terms. 
			
		\end{abstract}

		
	\end{frontmatter}

\setlength\parindent{10pt} 
\setlength{\parskip}{5pt} 

\section{Introduction}
\label{sec.intro}

The need of structure-preserving schemes for hyperbolic conservation laws with involution constraints is very well known in the context of numerical methods for the solution of the Maxwell and MHD equations in the time domain. There, the involution consists in the divergence-free condition of the magnetic field, which is a stationary extra conservation law that is automatically satisfied by the governing PDE system for all times 
if the initial magnetic field was divergence-free. Exactly divergence-free schemes (so-called constrained transport schemes) usually employ 
a \textit{staggered mesh}, see the pioneering work of Yee \cite{Yee66}, where in two space-dimensions the edge-normal components of the magnetic field are directly stored and evolved on the edges of the primary control volumes at the aid of an electric field that is defined in the vertices of each edge. For further developments in the context of constrained transport schemes, see e.g. the following list of references, 
which does not pretend to be complete, \cite{DeVore,BalsaraSpicer1999,Balsara2004,GardinerStone,balsarahlle2d,ADERdivB}. 
An alternative to the use of exactly divergence-free schemes is the use of 
\textit{divergence cleaning} techniques, which add extra terms to the governing 
PDE system. This can be either achieved by the so-called Powell 
terms \cite{PowellMHD1}, which are actually based on the symmetric form of the MHD equations found by 
Godunov in 1972, see \cite{God1972MHD}, or the hyperbolic generalized Lagrangian multiplier (GLM) approach 
of Munz \textit{et al.} \cite{MunzCleaning,Dedneretal}. 

The governing PDE system discussed in the present paper goes back to \cite{God1961,GodunovRomenski72,PeshRom2014,GPRmodel} and in the absence of relaxation source terms it is also endowed with two involution constraints, but here the  \textit{curl} of some quantities is required to be zero for all times rather than the \textit{divergence}.  Much less is known about exactly or approximately curl preserving finite volume schemes, probably because this type of involution is not yet as frequent as the well-known divergence constraints on the magnetic and electric field in computational electromagnetics. It definitely  arises in nonlinear hyperelasticity, see e.g. the discussions in \cite{Rom1998,Favrie2014,Haider2017}. A  rather general framework for the construction of structure-preserving schemes (including curl-preserving methods) was developed by Hyman and Shashkov \cite{HymanShashkov1997} and Jeltsch and Torrilhon  \cite{JeltschTorrilhon2006,Torrilhon2004}. Further work on mimetic and structure-preserving finite difference schemes can be found e.g. in \cite{Margolin2000,Lipnikov2014,Carney2013}. For curl-free wavelets
the reader is referred to \cite{Deriaz2009}, while compatible finite elements are discussed, for example, in 
\cite{Nedelec1,Nedelec2,Cantarella,Hiptmair,Monk,Arnold,Alonso2015}.  
The GLM approach of Munz \textit{et al.} has been very recently also generalized to PDE with curl involutions in 
\cite{FOCCZ4GLM,SHTCSurfaceTension}, while a comparison of different approaches to treat curl-free PDE has 
been provided in \cite{Godunov90}. 

Common to almost all the previously-mentioned exactly structure-preserving schemes is the fact that they require the use of 
a \textit{staggered grid} in order to provide natural and compatible definitions of the discrete curl, gradient
and divergence operators. But staggered grids are not only used in the context of structure-preserving schemes.
They are also widely used in the context of semi-implicit schemes for the solution of the incompressible Navier-Stokes equations since the pioneering work of Harlow and Welch \cite{markerandcell}. 
For a non-exhaustive overview of some of the most important contributions concerning pressure-based staggered semi-implicit finite difference schemes for the Navier-Stokes and shallow water equations the reader is referred to  
\cite{chorin1,chorin2,patankar,patankarspalding,BellColellaGlaz,vanKan,HirtNichols,Casulli1990,CasulliCheng1992,Casulli1999,CasulliWalters2000,Casulli2009,CasulliVOF}. For a new family of staggered hybrid finite volume / finite element schemes for incompressible and weakly compressible flows, see e.g. \cite{BFTVC2018,Hybrid1} 
and references therein.  
It is therefore a very natural choice to employ staggered meshes when constructing a new semi-implicit  structure-preserving scheme, which is the declared objective of this paper. 

As already stated before, staggered semi-implicit schemes are typically used in the context of incompressible or low Mach number flows. The first semi-implicit scheme for the compressible Euler equations was the method of Casulli and Greenspan \cite{CasulliCompressible}, but this scheme was not conservative and thus not suitable for the simulation of shock waves. For the compressible high Mach number flows, usually explicit density-based Godunov-type finite volume schemes are employed, see 
\cite{lax,godunov,roe,osherandsolomon,hll,munz91,munz94,Toro:1994,LeVeque:2002a,toro-book}, because of their intrinsic conservation property that allows the correct computation of shock waves. Up to now, semi-implicit  methods are only rarely used for the simulation of compressible flows with shock waves, but some recent 
developments can be found in \cite{MunzPark,CordierDegond,FedkiwSI,IterUpwind,DumbserCasulli2016,RussoAllMach,SIMHD}, where new families of \textit{conservative} 
pressure-based semi-implicit schemes were introduced and which are therefore also suitable for shock waves and compressible flows at all Mach numbers. However, to the best knowledge of the authors currently there exists no numerical scheme
for the model \cite{GPRmodel} (called GPR model in the following) that satisfies all involution constraints exactly on the discrete level, which is furthermore asymptotic preserving (AP) for vanishing relaxation times and which is suitable for all Mach number 
flows. A very recent all speed scheme for nonlinear hyperelasticity can be found
in \cite{AbateIolloPuppo}. For a review about Lagrangian and Eulerian schemes for nonlinear 
hyperelasticity, see  
\cite{GodunovRomenski72,Godunov:1995a,Rom1998,Godunov:2003a,Bartonetal,favrie2009solid,Kluth2010,
	FavrGavr2012,PeshRom2014,Ndanou2014,GPRmodel,BoscheriDumbser2016,SHTC-GENERIC-CMAT,Iollo2017,
	HypoHyper2,Jackson2019,Jackson2019a,Barton2010,Barton2011,Barton2019,FrontierADERGPR,Gil2015,Gil2017}.

The declared aim of this paper is therefore to develop a new, conservative, pressure-based semi-implicit finite volume method on staggered meshes for the solution of the model \eqref{eqn.GPR}, which is at the same time structure-preserving (SP) for all involution constraints, (quasi) asymptotic-preserving (AP) in the stiff relaxation limit and suitable for all Mach number flows. 
It is well-known that explicit density-based solvers become inefficient and  
inaccurate in the low Mach number regime and for these reasons an implicit time discretization is needed. However, discretizing all terms implicitly would in general lead to a \textit{highly nonlinear} 
non-symmetric system with a large number of unknowns (density, velocity, pressure, distortion field and thermal impulse), for which convergence is very difficult to control. Therefore, the new structure-preserving 
semi-implicit finite volume (SPSIFV) method proposed in this paper uses instead (i) an explicit 
discretization for all nonlinear 
convective terms, (ii) a compatible and structure-preserving explicit discretization for the 
distortion field and the thermal impulse on a \textit{vertex-based} staggered mesh, (iii) while an 
implicit discretization is only employed for the pressure terms and for the stiff algebraic 
relaxation source terms. This judicious combination leads in the end to only one  
\textit{mildly-nonlinear} and \textit{symmetric positive definite} system for the fluid pressure as 
the only unknown.  
The properties of the resulting pressure system allow the use of the Newton-type techniques of Casulli \textit{et al.} 
\cite{BrugnanoCasulli,BrugnanoCasulli2,CasulliZanolli2010,CasulliZanolli2012}, for which convergence has been \textit{rigorously proven}. 
Due to the implicit pressure terms, the time step of our new scheme is only restricted by the fluid velocity and characteristic wave speeds of shear and heat wave propagation, and not by the adiabatic sound speed. 
For this reason, the method proposed in this paper is a true structure-preserving \textit{all Mach number} flow solver. 

We underline that a new family of \textit{explicit} high order curl-preserving Godunov-type finite volume schemes, which makes use of an \textit{edge-based} staggering in combination with a high order curl-preserving  WENO reconstruction and multi-dimensional Riemann solvers, has been very recently introduced by Balsara \textit{et al.} in \cite{BalsaraCurlFree}, while the method proposed in the present paper makes use of a \textit{vertex-based} staggering and a \textit{semi-implicit} pressure-based formulation. 

The rest of the paper is organized as follows: in Section \ref{sec.model} we briefly recall the unified first order hyperbolic GPR model of continuum mechanics. In Section \ref{sec.method} we present the new staggered semi-implicit structure-preserving finite volume scheme. The mathematical properties of the numerical method are analyzed in Section \ref{sec.analysis} and computational results for a large set of test problems are shown in Section \ref{sec.results}.  
The paper closes with Section \ref{sec.concl}, in which we give some concluding remarks and an outlook to future work.

\section{Unified first order hyperbolic model of continuum mechanics} 
\label{sec.model} 

\subsection{Governing PDE system}\label{sec.govPDEs}

The unified first order hyperbolic GPR model of continuum mechanics including heat conduction reads 
as follows, see 
also \cite{Rom1998,PeshRom2014,GPRmodel} 

\begin{subequations}\label{eqn.GPR}
	\begin{align}
	& \frac{\partial \rho}{\partial t}+\frac{\partial (\rho v_k)}{\partial 
	x_k}=0,\label{eqn.conti}\\[2mm]
	&\frac{\partial \rho v_i}{\partial t}+\frac{\partial \left(\rho v_i v_k + p \, \delta_{ik} - 
	\sigma_{ik} \right)}{\partial x_k}=0, \label{eqn.momentum}\\[2mm]
	&\frac{\partial A_{i k}}{\partial t}+\frac{\partial (A_{im} v_m)}{\partial x_k} + 
	v_m \left(\frac{\partial A_{ik}}{\partial x_m}-\frac{\partial A_{im}}{\partial x_k}\right)
	=-\dfrac{ \psi_{ik} }{\theta_1(\tau_1)},\label{eqn.deformation}\\[2mm]
	&\frac{\partial J_k}{\partial t}+\frac{\partial \left( J_m v_m + T \right)}{\partial x_k} + 
	v_m \left(\frac{\partial J_{k}}{\partial x_m}-\frac{\partial J_{m}}{\partial x_k}\right)  =
	-\dfrac{H_k}{\theta_2(\tau_2)}, \label{eqn.heatflux}\\[2mm]
	& \frac{\partial \rho  E}{\partial t}+\frac{\partial \left( v_k \rho E + v_i (p \, \delta_{ik} 
	- \sigma_{ik}) + q_k \right)}{\partial x_k}=0. \label{eqn.energy} 
	\end{align}
\end{subequations}
Here, $\rho$ is the mass density, $v_i$ is the velocity field, $A_{ik}$ is 
the distortion field 
(which is a basis triad and thus transforms as a set of three vectors and not as a tensor under coordinate transforms), $J_k$ 
is the thermal impulse density (it has the SI units 
$\text{K}\cdot\frac{\text{s}}{\text{m}}$), $\rho E$ is the 
total energy 
density, 
$p$ 
is the fluid 
pressure, 
$\sigma_{ik}$
is the stress tensor that contains shear stress as well as thermal stresses, and $q_k$ is the heat 
flux. 

Furthermore, the energy potential $ E $ plays the role of the closure for system~\eqref{eqn.GPR}. 
Thus, thermodynamic consistency of \eqref{eqn.GPR} requires that the pressure be $ p = \rho^2 
E_{\rho} $, the temperature $ T=E_S $, the stress 
tensor 
\begin{equation}\label{eqn.stress}
\sigma_{ik} = -\rho A_{ji}E_{A_{jk}} - \rho J_i E_{J_k} = -\rho A_{ji}\psi_{jk} - \rho J_i H_k,
\end{equation}
$ \psi_{ik} := E_{A_{ik}} $, $ H_k := E_{J_k} $, and 
the heat flux \begin{equation}\label{eqn.heat.flux}
q_k = \rho E_S E_{J_{k}} = \rho\, T H_k,
\end{equation}
where $ E_\rho $, $ E_S $, $ E_{A_{ik}} $, etc. denote partial derivative of $ E $ with respect to 
the state 
variables $ E_\rho = \pd E/\pd \rho$, $ E_S = \pd E/\pd S $, $ E_{A_{ik}} = \pd E/\pd A_{ik} $, 
etc. Therefore, to close 
system~\eqref{eqn.GPR}, one has to provide the specific expression for the energy $ E $. 

We remark that the thermal impulse equation \eqref{eqn.heatflux} is different from the one used in 
our previous 
papers~\cite{GPRmodel,GPRmodelMHD}. Both equations, although they look slightly different, are compatible with the Fourier law of heat conduction in the equilibrium limit (small relaxation times $ \tau_2 \to 0$). The 
thermal impulse equation \cite{GPRmodel,GPRmodelMHD} has a convenient divergence form (i.e. it can 
be written as four-divergence of a vector field) and free of involution constraints which is good 
for numerical purposes, while the equation  
\eqref{eqn.heatflux} can not be represented in a fully conservative flux divergence form. Nevertheless, 
the heat conduction equation \eqref{eqn.heatflux} proposed by Romenski in~\cite{Rom1998} should be 
considered more preferable from the  theoretical 
standpoint. First of all, it admits a variational formulation~\cite{SHTC-GENERIC-CMAT}, while the 
thermal impulse
equation from \cite{GPRmodel,GPRmodelMHD} does not. Second, \eqref{eqn.heatflux} is consistent 
with the Hamiltonian formulation for non-equilibrium thermodynamics known as  
GENERIC~\cite{Ottinger2005,PKG-Book2018}, see \cite{SHTC-GENERIC-CMAT}. This implies that 
\eqref{eqn.heatflux} should be more 
advantageous for describing heat transfer in non-equilibrium settings. Last, the equations with 
exactly the same structure of differential terms as \eqref{eqn.heatflux} and thus, with the curl 
involution appear in many other 
physical settings, e.g. multi-phase 
flows~\cite{Romenski2016}, continuum modeling of surface tension~\cite{HypSurfTension}, and 
hyperbolic reformulation of the nonlinear Schr\"{o}dinger's equation
\cite{Dhaouadi2018}.

\subsection{Consistency with thermodynamics}\label{sec.thermodynamics}
 
The model is also endowed with the following evolution equation for the entropy density $ 
\rho S $
\begin{equation}\label{eqn.entropy}
	\frac{\partial \rho S}{\partial t}+\frac{\partial \left( \rho S v_k + \rho H_k 
	\right)}{\partial x_k}=\dfrac{\rho}{\theta_1(\tau_1) T} \psi_{ik} \psi_{ik} + 
	\dfrac{\rho}{\theta_2(\tau_2) T} H_i H_i \geq 0  
\end{equation} 
 which establishes the \textit{second law of thermodynamics} for system~\eqref{eqn.GPR}.

System \eqref{eqn.GPR} belongs to the class of so-called Symmetric Hyperbolic and Thermodynamically 
Compatible (SHTC) systems~\cite{SHTC-GENERIC-CMAT} proposed by Godunov and Romenski in a series of  
papers~\cite{God1961,God1972MHD,Godunov1996,Rom1998,Rom2001}. In particular, this means that the 
over-determined\footnote{This system is over-determined because we have more equations than 
unknowns.} system \eqref{eqn.GPR} is compatible that is, the energy equation \eqref{eqn.energy}, 
in fact, can be obtained as the linear combination of the remaining equations multiplied by certain 
factors 
(e.g. see~\cite{SHTC-GENERIC-CMAT})
\begin{equation}\label{eqn.summation}
\eqref{eqn.energy} = (\rho E)_{\rho}\cdot\eqref{eqn.conti} + (\rho E)_{\rho 
v_i}\cdot\eqref{eqn.momentum} +
(\rho E)_{A_{ik}}\cdot\eqref{eqn.deformation} + (\rho E)_{J_{k}} \cdot\eqref{eqn.heatflux} + (\rho 
E)_{\rho 
S}\cdot\eqref{eqn.entropy}
\end{equation}
In particular, the source terms in \eqref{eqn.deformation} and \eqref{eqn.heatflux} are designed in 
such a way that, on one hand energy is conserved (there is no source term in 
\eqref{eqn.energy}) and, on the other hand the physical 
entropy is non-decreasing (see the \textit{entropy inequality} in~\eqref{eqn.entropy}).

\subsection{Involution constraints}\label{sec.inv.constr}

An important feature of the distortion and thermal impulse equations is that, e.g. see 
\cite{SHTC-GENERIC-CMAT,PRD-Torsion2018}, 
\begin{equation}\label{eqn.involutions.full}
\frac{\pd}{\pd t}(\nabla\times\A - \B) = 0, \qquad \frac{\pd}{\pd t}(\nabla\times\J - \bm{\Omega}) 
= 0,
\end{equation}
where $ \B = [B_{ij}]$ and $ \bm{\Omega} = [\Omega_i] $ are the solutions to
\begin{subequations}\label{eqn.PDEs.curl}
	\begin{align}
		&\frac{\pd B_{ij}}{\pd t} + \frac{\pd}{\pd x_k} \left(
		B_{ij} v_k - v_j B_{ik} + \varepsilon _{jkm} \theta_1^{-1} \psi_{im}
		\right) + v_j \frac{\pd B_{ik}}{\pd x_k} = 0,\label{eqn.Burg}\\[2mm]
		&\frac{\pd \Omega_{j}}{\pd t} + \frac{\pd}{\pd x_k} \left(
		\Omega_j v_k - v_j \Omega_k + \varepsilon_{jkm} \theta_2^{-1} H_m
		\right) + v_j \frac{\pd \Omega_k}{\pd x_k} = 0,\label{eqn.therm.curl}
	\end{align}
\end{subequations}
which can be derived from \eqref{eqn.deformation} and \eqref{eqn.heatflux} by applying the curl 
operator ``$ 
\nabla\times $'' to them.
It follows from these equations that, in the absence of source terms ($\theta_1\sim\tau_1 \to 
\infty$ and $\theta_2\sim\tau_2 \to 
\infty$), the solution to \eqref{eqn.GPR} satisfies the following two stationary linear constraints 
(involutions)  
\begin{equation}
  \frac{\partial A_{ik}}{\partial x_m}-\frac{\partial A_{im}}{\partial x_k} = 0, \qquad 
  \textnormal{ and } \qquad 
  \frac{\partial J_{k}}{\partial x_m}-\frac{\partial J_{m}}{\partial x_k} = 0,
\label{eqn.involutions} 
\end{equation}
if these constraints are satisfied by the initial data at $ t=0 $. For finite values of $ \tau_1 $ 
and $ \tau_2 $ these curls are not zero in general, but satisfy the time evolution equations 
\eqref{eqn.PDEs.curl}. 

It is important to emphasize that even if the source terms in \eqref{eqn.deformation} and 
\eqref{eqn.heatflux} are absent and thus, $ \pd_m A_{ik} - \pd_k A_{im} = 0 $ and $ \pd_m J_k - 
\pd_k J_m = 0 $, the PDEs \eqref{eqn.deformation} and \eqref{eqn.heatflux} should not be 
replaced with equations in a conservative form such as
\begin{equation}\label{eqn.conservative.AJ}
\frac{\partial A_{i k}}{\partial t}+\frac{\partial (A_{im} v_m)}{\partial x_k} = 0,\qquad 
\frac{\partial J_k}{\partial t}+\frac{\partial \left( J_m v_m + T \right)}{\partial x_k} = 0,
\end{equation}
because this would immediately destroy both the thermodynamic consistency, as well as the Galilean 
invariance. For example, equations \eqref{eqn.conservative.AJ} are \textit{not} compatible with the 
energy conservation law \eqref{eqn.energy} in the sense of 
\eqref{eqn.summation}. Also, the conservative equations \eqref{eqn.conservative.AJ} have a 
characteristic structure that is not compatible with Galilean invariance of the PDE system, similar 
to the equations of magnetohydrodynamics (MHD) \cite{PowellMHD2}, which also belong to the SHTC 
class of equations \cite{God1972}.

Finally, we note that, generally speaking, an involution preserving scheme should in principle 
guarantee that the numerical solution to the full GPR model (i.e. with the dissipative source terms 
in 
\eqref{eqn.deformation} and \eqref{eqn.heatflux}) should also satisfy equations 
\eqref{eqn.involutions.full} and \eqref{eqn.PDEs.curl}. So far, we are not there yet, hence we only 
propose a numerical scheme which guarantees the fulfillment of the \textit{stationary} involution 
constraints \eqref{eqn.involutions} in the absence of dissipative source terms. 



\subsection{Structural aspects of the governing PDEs}

The SHTC theory starts from the question on the admissible structure of mechanically and 
thermodynamically consistent 
equations in continuum mechanics, e.g. see \cite{SHTC-GENERIC-CMAT}. Therefore, there are several 
\textit{structural aspects} of the governing equations that, ideally, have to be respected also at 
the discrete level. 
In what follows, we summarize three main structural features (SF) of the SHTC equations, the 
fulfillment of which at the discrete level may potentially be beneficial for the 
quality of the numerical solution.  
\begin{itemize}
	\item[SF1.] \textbf{Overdetermination}: Equations \eqref{eqn.GPR}, \eqref{eqn.entropy} form an 
	overdetermined system of PDEs, that is there is one more equation than unknowns. But in fact, 
	the 
	total energy $ E $ is not an unknown  but a potential $ E = E(\rho,S,v_i,A_{ik},J_k) $ while 
	the 
	entropy is a true unknown. However, usually it is not the entropy PDE \eqref{eqn.entropy} but 
	the 
	energy conservation 
	law \eqref{eqn.energy} which is considered within the equations to be discretized in order 
	to guaranty the energy conservation at the discrete level. Nevertheless, one may think of a new 
	class 
	of numerical schemes that explicitly takes into account the summation property  
	\eqref{eqn.summation} also at the discrete level. Such a scheme would discretize the entropy 
	inequality  \eqref{eqn.entropy} instead 
	of the energy conservation law \eqref{eqn.energy}, but due to the fulfillment of the summation 
	property 
	\eqref{eqn.summation} at the discrete level it would automatically also guarantee the discrete 
	energy  
	conservation. We note that 
	a general purpose scheme usually cannot automatically guarantee \eqref{eqn.summation} at the 
	discrete level.
	\\[-2mm]
	\item[SF2.] \textbf{Involution constraints}: Homogeneous SHTC equations are usually endowed 
	with stationary 
	involution constraints of the type \eqref{eqn.involutions} which are usually just a part of a
	more general \textit{integrability condition} like \eqref{eqn.PDEs.curl}, e.g. see 
	\cite{SHTC-GENERIC-CMAT}. In this paper, we deal directly with 
	stationary involution constraints of SHTC equations, while the general case of an 
	integrability condition compatible scheme will be covered in future publications.
	\\[-2mm]
	\item[SF3.] \textbf{Equilibrium subsystem}: SHTC equations posses two limiting PDE structures 
	formally 
	corresponding to $ \tau_1 \to \infty $, $ \tau_2 \to \infty  $ (short wave-length limit) and $ 
	\tau_1 \to 0 $, $ \tau_2 \to 0 
	$ (long wave-length limit). The former has the full SHTC structure and corresponds to the most 
	non-equilibrium state of the 
	system 
	and is described by the homogeneous part of \eqref{eqn.GPR} with involution constraints 
	\eqref{eqn.involutions}, while the later has the reduced structure of five Euler equations of 
	ideal fluid\footnote{Or eight equations of ideal magnetohydrodynamics 
	\cite{GPRmodelMHD} if coupled with 
	the electromagnetic fields.} ($ 
	\tau_1=\tau_2=0 $, global thermodynamic equilibrium) or five 
	Navier-Stokes-Fourier equations ($ \tau_1\ll1, 
	\tau_2\ll1 $, local thermodynamic equilibrium). Thus, a proper structure preserving scheme 
	should be 
	able to reproduce both limits of SHTC equations as is the case with the proposed novel SPSIFV 
	scheme.
	\\[-2mm]
	\item[SF4.] \textbf{Hamiltonian structure and symplectic integrators}: as it was shown recently 
	\cite{SHTC-GENERIC-CMAT}, the SHTC equations have an underlying Hamiltonian and thus 
	geometrical  
	structure. This means that the 
	reversible part of the time evolution (i.e. all the differential terms in the left hand-side of 
	\eqref{eqn.GPR}, \eqref{eqn.entropy}) is actually generated by corresponding Poisson brackets. 
	In 
	fact, in Hamiltonian mechanics, there is a class of methods, called \textit{symplectic 
	integrators} 
	\cite{Morrison2017},  
	which aims in retaining the underlying geometrical structure of the governing equations. 
	Adoption 
	of a similar strategy for the numerical solution of the SHTC equations may, at least in 
	principle, 
	improve the overall quality and physical consistency of the numerical solution. 
\end{itemize}

\subsection{Equation of state}\label{sec.EOS}

Throughout this paper we assume that the specific total energy can be written as a sum of three contributions as 
\begin{equation}
\label{eqn.energy.sum} 
 E(\rho,v_i,p,A_{ik},J_k) = E_1(\rho,p) + E_2(A_{ik}, J_k) + E_3(v_i),   
\end{equation}
with the specific internal energy given by the ideal gas equation of state
\begin{equation}
   E_1(\rho,p) = \frac{p}{\rho (\gamma-1)}, \quad \text{or} \quad 
   E_1(\rho,s) = \frac{c_0^2}{\gamma(\gamma-1)}, \ c_0^2 = \gamma\rho^{\gamma-1} e^{s/c_v} 
\end{equation}
in the case of gases, and given by the so-called stiffened gas equation of state 
\begin{equation}
E_1(\rho,p) = \frac{c_0^2}{\gamma(\gamma-1)}\left (\frac{\rho}{\rho_0}\right )^{\gamma-1} e^{s/c_v} 
+ \frac{\rho_0 c_0^2 -\gamma p_0}{\gamma\rho}
\end{equation}
in the case of solids ($ c_v $ is the specific heat at constant volume, $ \gamma $ is the ration of 
specific heats, and $ p_0 $ and $ \rho_0 $ are the reference pressure and mass density, 
respectively).
The specific energy stored in material deformations and in the thermal impulse is
\begin{equation}
E_2(A_{ik},J_k) = \frac{1}{4} c_s^2 \mathring{G}_{ij} \mathring{G}_{ij} + \frac{1}{2} \alpha^2 J_k J_k, 
\label{eqn.E2} 
\end{equation}
where $ c_s $ is the characteristic velocity (assumed to be constant in this 
paper) of propagation of shear perturbations, while $ \alpha $ (also constant) relates to the 
characteristic velocity of thermal perturbations $ c_h $ as $ c_h^2 \sim \alpha^2 T/c_v $ (the 
SI units of $ \alpha $ are $ \frac{\text{m}^2}{\text{s}^2} \cdot \text{K}^{-1} \sim [c_v]$). 
Furthermore, $G_{ij} 
= A_{ki} A_{kj}$ is the Riemannian metric tensor induced by the 
mapping from
Eulerian coordinates to the current stress-free reference configuration and $\mathring{G}_{ij}$ 
is its trace-free part, defined as usual by 
\begin{equation}
  \mathring{G}_{ij} = G_{ij} - \frac{1}{3} G_{kk} \, \delta_{ij}. 
\end{equation} 
For an alternative equation of state in nonlinear hyperelasticity, see \cite{Ndanou2014}. 
The specific kinetic energy is contained in the third
contribution to the total energy and reads 
\begin{equation}
E_3(v_k) = \frac{1}{2} v_i v_i. 
\end{equation}
With the equation of state chosen above, we get the following expressions for the stress tensor, 
the heat flux and the functions $\psi_{ik}=E_{A_{ik}}$ and $H_k=E_{J_k}$ present in the relaxation 
source terms: 
\begin{equation}
 \sigma_{ik} = \rho c_s^2 G_{ij} \mathring{G}_{jk} + \rho \alpha^2 J_i J_k, \qquad  
 q_k = \rho T \alpha^2 J_k, 
\end{equation}
\begin{equation}
\psi_{ik} = c_s^2 A_{ij} \mathring{G}_{jk}, \qquad 
H_k = \alpha^2 J_k. 
\end{equation} 
The functions $\theta_1$ and $\theta_2$ are chosen in such a way that a \textit{constant} shear 
viscosity $ \mu $ and 
thermal conductivity $ \kappa $ are obtained in the stiff relaxation limit, see \cite{GPRmodel} for 
a  
formal asymptotic analysis: 
\begin{equation}
  \theta_1(\tau_1) = \frac{1}{3} \tau_1 c_s^2 |\mathbf{A}|^{\frac{5}{3}}, \qquad \theta_2(\tau_2) = 
  \tau_2 \frac{\alpha^2}{\rho T},  
\end{equation}
Following the procedure detailed in \cite{GPRmodel}, one can show via formal asymptotic expansion that 
in the stiff relaxation limit $\tau_1 \to 0$, $\tau_2 \to 0$ (i.e. small but fixed relaxation times 
$ \tau_1 \ll 1 $, $ \tau_2 \ll 1 $), the stress tensor and the heat flux reduce to 
\begin{equation}
\boldsymbol{\sigma} = -\frac{1}{6} \rho_0 c_s^2 \tau_1 \left( \nabla \mathbf{v} + \nabla \mathbf{v}^T
- \frac{2}{3} \left( \nabla \cdot \mathbf{v}\right) \mathbf{I} \right) - \frac{\alpha^2}{\rho T^2} 
\tau^2_2 \nabla T \otimes \nabla T,
\label{eqn.stress.limit}   
\end{equation} 
and 
\begin{equation}
\boldsymbol{q} = -  \alpha^2 \tau_2 \nabla T,
\label{eqn.heatflux.limit}    
\end{equation} 
so that
\begin{equation}\label{eqn.mu.kappa}
	\mu = \frac{1}{6}\rho_0 \tau_1 c_s^2, 
	\qquad
	\kappa = \tau_2 \alpha^2.
\end{equation}

One can see that the leading terms in the asymptotic expansions of the heat flux and stress tensor 
correspond to the Fourier law and Navier-Stokes viscous stress respectively. The second order with 
respect to $\tau_2$ term in \eqref{eqn.stress.limit} is negligible for small $\tau_2 \ll 1$.

\section{Numerical method} 
\label{sec.method}

System \eqref{eqn.GPR} can be written more compactly in the following matrix-vector notation   
\begin{equation}
\label{eqn.pde} 
\partial_t \Q + \nabla \cdot \mathbf{F}(\Q) + \B(\Q) \cdot \nabla \Q = \S(\Q), 
\end{equation}
with the state vector $\Q = (\rho, \rho v_i, A_{ik}, J_k, \rho E)^T$, the flux tensor $\F(\Q) $, 
the 
non-conservative product $\B(\Q) \cdot \nabla \Q$ containing the curl terms and the vector of potentially
stiff algebraic relaxation source terms $\S(\Q)$. As proposed in \cite{DumbserCasulli2016,SIMHD} we now
\textit{split} the flux tensor into a convective part and a pressure part. However, the equations for
the new objects $A_{ik}$, $J_k$ as well as their respective contributions to the momentum equation 
and to the total energy conservation law need a special \textit{compatible} and 
structure-preserving discretization using a vertex-based grid staggering. Hence, 
eqn.~\eqref{eqn.pde} is rewritten as  
\begin{equation}
\label{eqn.pde.split} 
\partial_t \Q + \nabla \cdot \left( \mathbf{F}_c(\Q) + \mathbf{F}_p(\Q) + \mathbf{F}_v(\Q) \right) + 
\nabla \G_v(\Q) + \B_v(\Q) \cdot \nabla \Q = \S(\Q), 
\end{equation}
where $\mathbf{F}_c(\Q) $ refers to purely convective fluxes that will be discretized explicitly 
and  
$\mathbf{F}_p(\Q) $ are the pressure fluxes that will be discretized implicitly using an edge-based 
staggered grid. The resulting splitting into pressure and convective fluxes is identical to the 
flux-vector splitting scheme of Toro and V\'azquez-Cend\'on recently forwarded in 
\cite{ToroVazquez}. The  remaining terms $\mathbf{F}_v(\Q) $,  $\nabla \G_v(\Q)$ and $\B_v(\Q) 
\cdot \nabla \Q$ 
will be carefully discretized in a structure-preserving manner using an explicit scheme on an 
appropriate vertex-based staggered grid. The relaxation source terms $\S(\Q)$  
can become stiff and thus require an implicit discretization on the vertex-based staggered mesh. 

The split fluxes read 
\begin{equation} 
\label{eqn.split.def} 
  \mathbf{F}_c = \left( \begin{array}{c} \rho v_k \\ \rho v_i v_k \\ 0 \\ 0 \\ \rho v_k ( E_2 + E_3 ) \end{array} \right),  \qquad 
  \mathbf{F}_p = \left( \begin{array}{c} 0 \\ p \delta_{ik} \\ 0 \\ 0 \\ h \rho v_k \end{array} \right), \qquad 
  \mathbf{F}_v = \left( \begin{array}{c} 0 \\ -\sigma_{ik}  \\ 0 \\ 0 \\ -v_i \sigma_{ik} + q_k \end{array} \right), 
\end{equation} 
with the specific enthalpy $h=E_1 + p/\rho$. The terms involving the gradient operator $\nabla \G_v(\Q)$ and the non-conservative 
product containing the curl terms read 
\begin{equation} 
   \G_v(\Q) = \left( \begin{array}{c} 0 \\ 0 \\ A_{im} v_m \\ J_m v_m + T \\ 0 \end{array} \right),  
 \qquad 
  \B_v(\Q) \cdot \nabla \Q = \left( \begin{array}{c} 0 \\ 0 \\ 
  v_m \left(\frac{\partial A_{ik}}{\partial x_m}-\frac{\partial A_{im}}{\partial x_k}\right) \\ 
  v_m \left(\frac{\partial J_{k}}{\partial x_m}-\frac{\partial J_{m}}{\partial x_k}\right)   \\ 
  0 \end{array} \right). 
\end{equation} 

The following subsystem  
\begin{equation}
\label{eqn.pde.ex} 
\partial_t \Q + \nabla \cdot \left( \mathbf{F}_c(\Q) + \mathbf{F}_v(\Q) \right) + 
\nabla \G_v(\Q) + \B_v(\Q) \cdot \nabla \Q = \S(\Q), 
\end{equation}
will be discretized explicitly, apart from the potentially stiff algebraic source terms in $\S$, 
which are discretized implicitly with a simple backward Euler scheme. The discretization method 
presented in the next section will consist in a combination of a classical second order 
MUSCL-Hancock type \cite{toro-book} TVD finite volume scheme for the convective fluxes $\F_c$, a 
curl-free discretization for the terms $\G_v$ and $\B_v \cdot \nabla \Q$ using compatible gradient 
and curl operators as well as a vertex-based discretization of the terms $\F_v$. The eigenvalues of 
subsystem \eqref{eqn.pde.ex} in $x$ direction are 
\begin{equation}
   \lambda^{c,v}_{1,2} = \halb u \pm \halb \sqrt{ \frac{4 T}{c_v} \alpha^2 + u^2 }, \qquad 
   \lambda^{c,v}_{3,4} = u \pm \frac{2}{3} \sqrt{3} c_s, \qquad 
   \lambda^{c,v}_{5,6,7,8} = u \pm c_s, \qquad   
   \lambda^{c,v}_{9,10,\cdots,17} = u. 
   \label{eqn.eval.c}  
\end{equation}
The remaining pressure subsystem, which will be discretized implicitly, reads as follows: 
\begin{equation}
\label{eqn.pde.im} 
\partial_t \Q + \nabla \cdot \mathbf{F}_p(\Q) =0.  
\end{equation}
As already mentioned before, the resulting pressure subsystem is formally identical to the  Toro-V\'azquez pressure system \cite{ToroVazquez}, hence its eigenvalues in $x$ direction are 
\begin{equation}
       \lambda^p_{1,2} = \frac{1}{2} \left( u \pm \sqrt{u^2 + 4 c_0^2 } \right), 
       \qquad \lambda^p_{3,4,5,\cdots,17} = 0, \qquad 
   \label{eqn.eval.p}  
\end{equation}
with the adiabatic sound speed $c_0$, e.g. for the ideal gas EOS we have, as usual, $c_0^2 = \gamma 
p / \rho$.

\begin{figure}[!htbp]
	\begin{center}
		\includegraphics[width=0.5\textwidth]{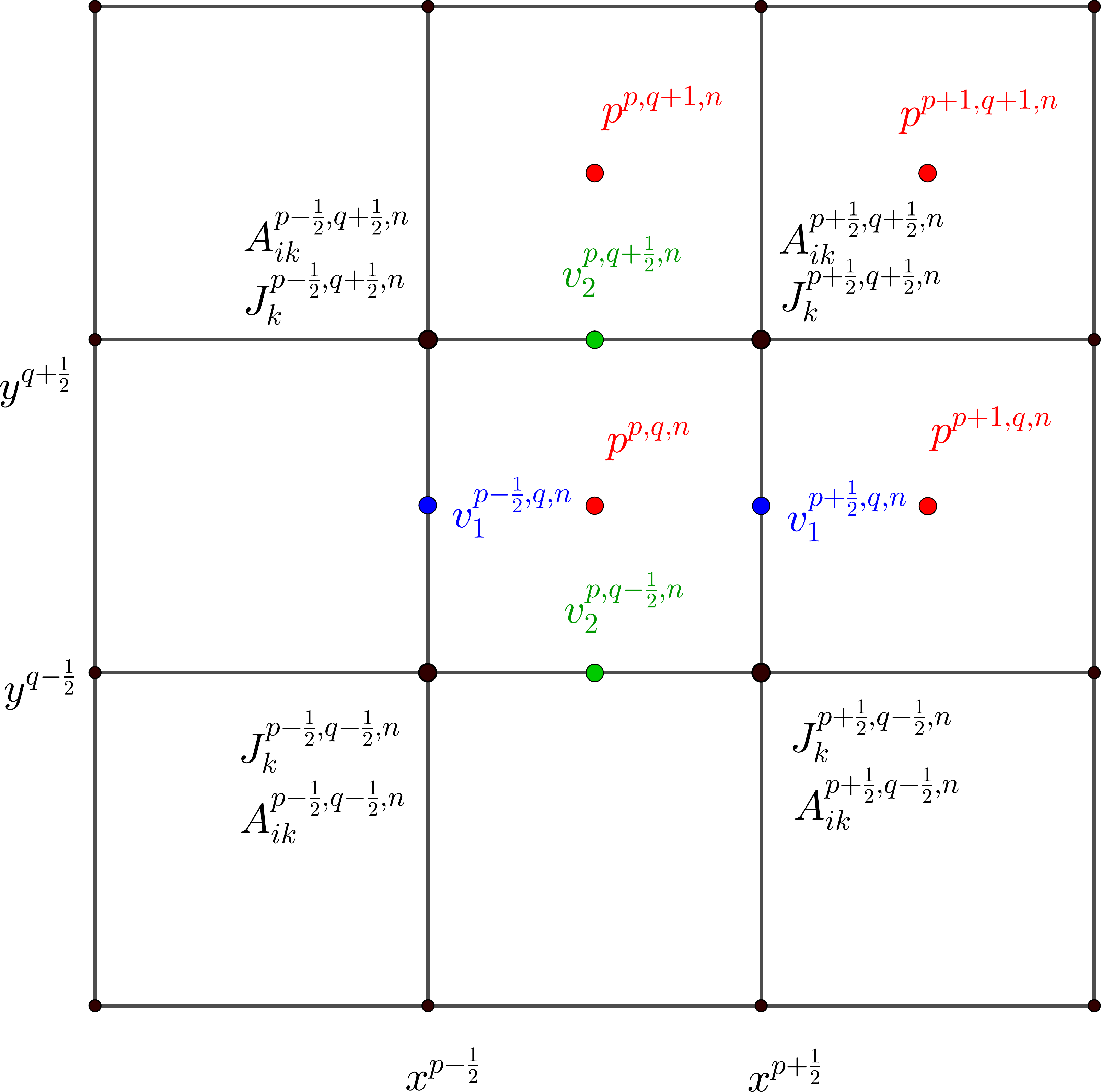}  
		\caption{Staggered mesh configuration with the pressure field $p^{p,q,n}$ defined in the cell barycenters, the velocity field components $v_1^{p+\halb,q,n}$ and $v_2^{p,q+\halb,n}$ defined on the edge-based staggered dual grids, respectively, and the distortion field $A_{ik}^{p+\halb,q+\halb,n}$ as well as the specific thermal impulse $J_k^{p+\halb,q+\halb,n}$ defined on the vertices of the main grid.}  
		\label{fig.staggered}
	\end{center}
\end{figure}

\subsection{Staggered mesh and discrete divergence, curl and gradient operators}

To simplify the description of the numerical scheme, we restrict the discussion to two-dimensional 
motion, i.e. we assume that $ \frac{\pd}{\pd x_3} $ vanishes for all fields and thus, we assume a 
two-dimensional physical domain $ \Omega $ spread in $ x_1 = x $ and $ x_2 = y $ and which is 
covered by a set of equidistant and non-overlapping 
Cartesian control volumes $\Omega^{p,q} = [x^{p-\halb},x^{p+\halb}] \times 
[y^{q-\halb},y^{q+\halb}]$ with 
uniform mesh spacings $\Delta x = x^{p+\halb} - x^{p-\halb}$ and $\Delta y = y^{q+\halb} - 
y^{q-\halb}$ 
in $x$ and $y$ direction, respectively, and with $x^{p \pm \halb}=x^p \pm \Delta x / 2$ and 
$y^{q \pm \halb} = y^q \pm \Delta y /2$. Nevertheless, we keep all 3-\textit{rd} components of 
vectors and tensors in the discussion. The 3D extension of the scheme is 
straightforward. We will furthermore use the notation $\mathbf{e}_x = 
(1,0,0)$, 
$\mathbf{e}_y = (0,1,0)$ and $\mathbf{e}_z = (0,0,1)$ for the unit vectors pointing into the 
directions 
of the Cartesian coordinate axes.

To avoid confusion between tensor indices and discretization indices, throughout this paper we will 
use the \textit{subscripts}  $i,j,k,l,m$ for \textit{tensor indices} and the \textit{superscripts} 
$n,p,q,r,s$ for the \textit{discretization indices} in time 
and space, respectively. 
The discrete spatial coordinates will be denoted by $x^p$ and $y^q$, while the set of discrete 
times will be denoted by $t^n$. For a sketch of the employed staggered grid arrangement of the main 
quantities, see Fig. \ref{fig.staggered}.

The main ingredients of the new structure-preserving staggered semi-implicit scheme proposed in this paper are the definitions of
appropriate discrete divergence, gradient and curl operators acting on quantities that are arranged in different and judiciously 
chosen locations on the staggered mesh. The discrete pressure field at time $t^n$ is denoted by $p^{h,n}$ and its degrees of freedom are located in the center of each control volume as $p^{p,q,n}=p(x^p,y^q,t^n)$. Throughout this paper we denote 
with the superscript $h$ the set of all degrees of freedom of the discrete solution and all degrees of freedom generated by a 
discrete operator, in order to ease notation. The discrete  velocities $v_1^{h,n}$ and $v_2^{h,n}$ are arranged in an edge-based staggered fashion, i.e.  
$u^{p+\halb,q,n}:=v_1^{p+\halb,q,n}=v_1(x^{p+\halb},y^q,t^n)$ and 
$v^{p,q+\halb,n}:=v_2^{p,q+\halb,n}=v_2(x^p,y^{q+\halb},t^n)$. 
The discrete distortion field $\AAA^{h,n}=A_{ik}^{h,n}$ and the discrete thermal impulse $\J^{h,n} = J_k^{h,n}$ are defined on 
the \textit{vertices} of each spatial control volume as  
$A_{ik}^{p+\halb,q+\halb,n} = A_{ik}(x^{p+\halb},y^{q+\halb},t^n)$ and 
$J_{k}^{p+\halb,q+\halb,n} = J_{k}(x^{p+\halb},y^{q+\halb},t^n)$, respectively. For clarity, 
see again Fig. \ref{fig.staggered}.  

\begin{figure}[!htbp]
		\begin{minipage}[c]{0.5\textwidth} 
			\includegraphics[width=0.8\textwidth]{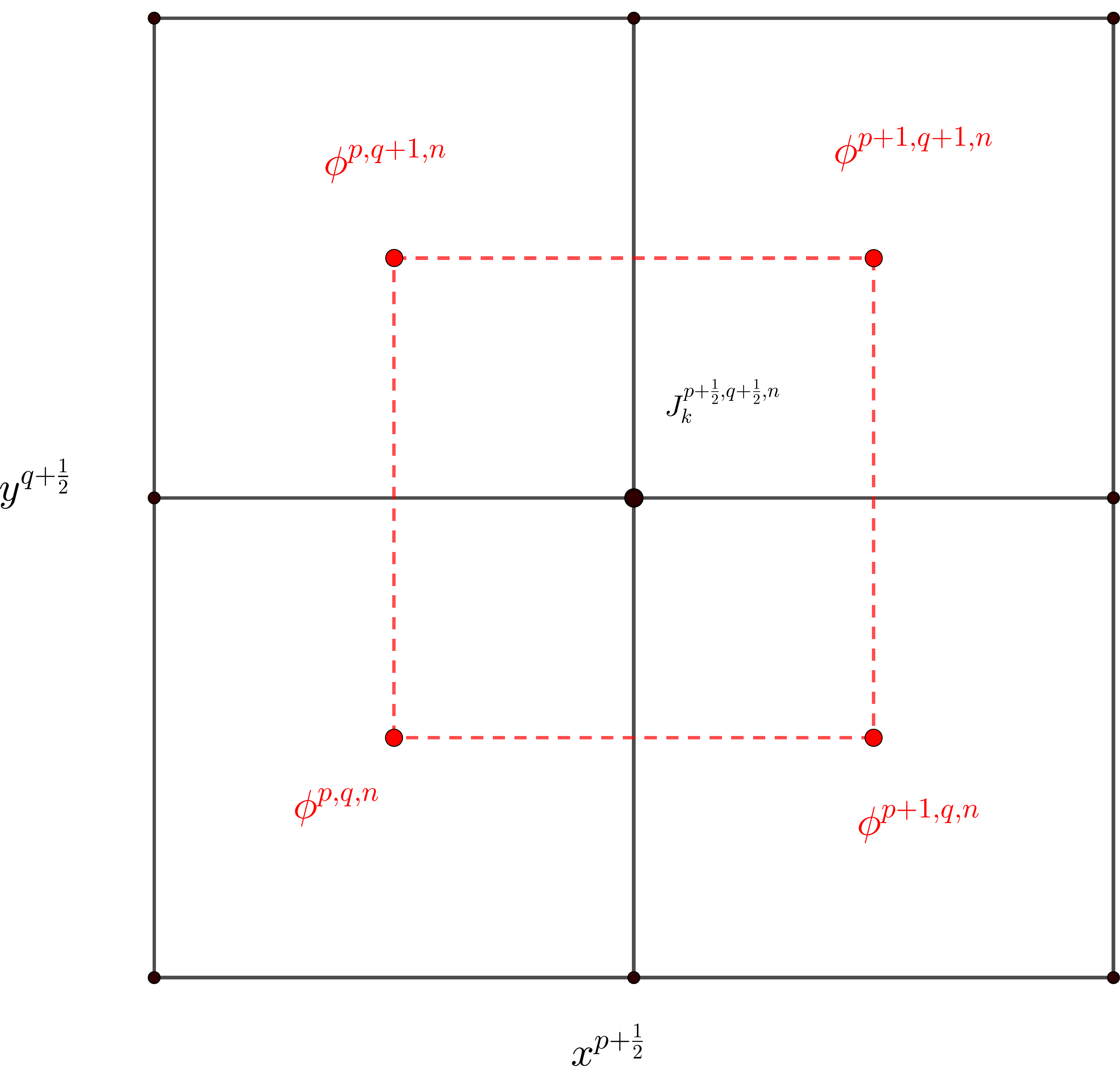} 
		\end{minipage} 
		\begin{minipage}[c]{0.5\textwidth} 
			\includegraphics[width=0.75\textwidth]{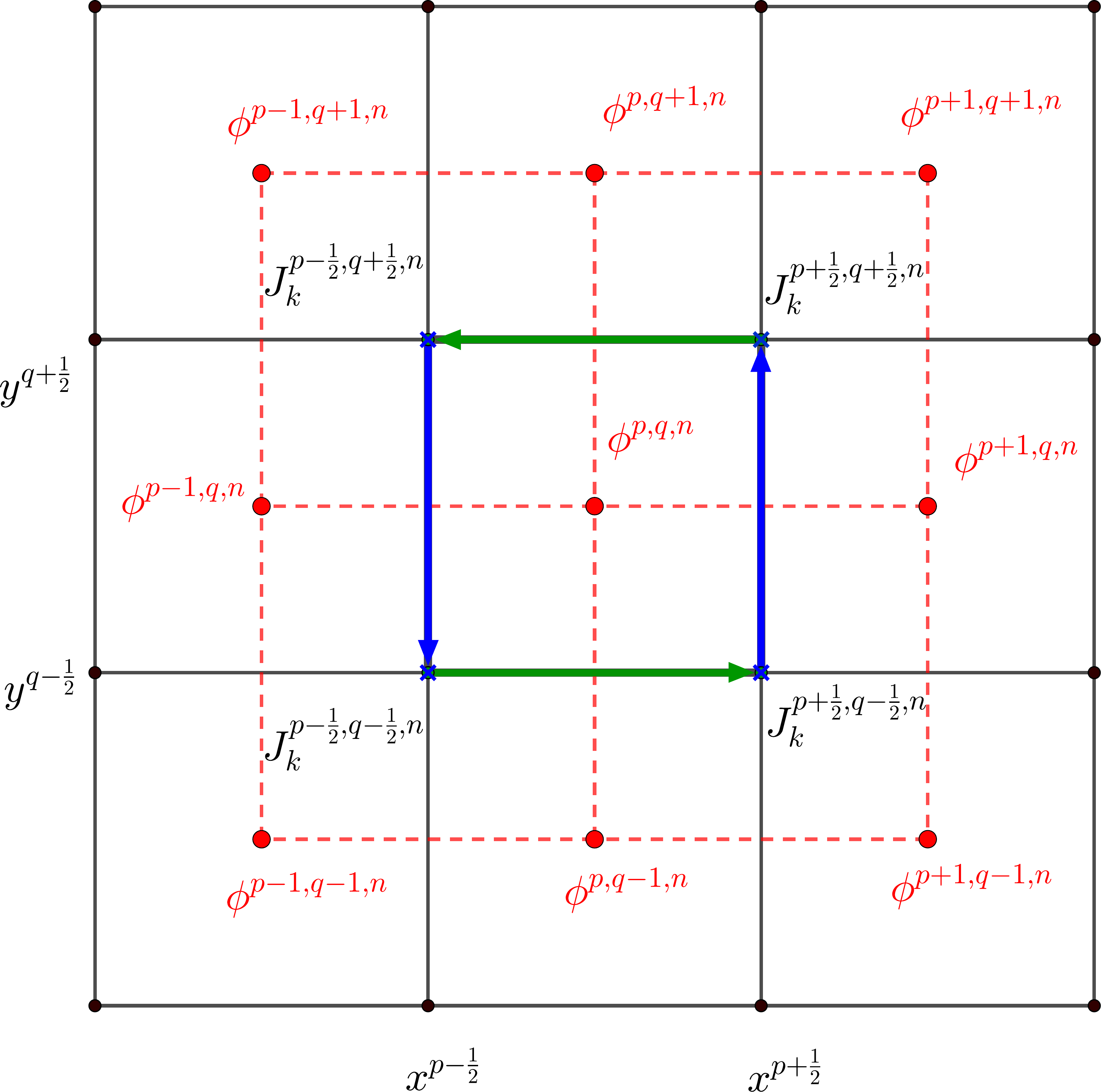} 
		\end{minipage}    
		\caption{Left: stencil of the discrete gradient operator, which computes the corner gradient of a scalar field defined in the cell barycenter. Right: stencil of the discrete curl operator, which defines the curl inside the cell barycenter using the vector field components defined in the corners of the primary control volume. In the
		right panel we also show the total 9-point stencil that is needed for the discrete identity $\nabla^h \times \nabla^h \phi^{h,n} = 0$.}  
		\label{fig.grad.curl}
\end{figure}

The \textit{discrete divergence operator}, $\nabla^h \cdot$, acting on a discrete vector field 
$\mathbf{v}^{h,n}$ is abbreviated 
by $\nabla^h \cdot \mathbf{v}^{h,n}$ and its degrees of freedom are given by  
\begin{equation}
 \nabla^{p,q} \cdot \mathbf{v}^{h,n}  =  \partial_k^{p,q} v_k^{h,n} =  
 \frac{v_1^{p + \halb,q,n} - v_1^{p - \halb,q,n}}{\Delta x} + 
 \frac{v_2^{p,q + \halb,n} - v_2^{p,q - \halb,n}}{\Delta y}, 
 \label{eqn.div} 
\end{equation}
i.e. it is based on the \textit{edge-based} staggered values of the field $\mathbf{v}^{h,n}$. It defines  
a discrete divergence on the control volume $\Omega^{p,q}$ via the Gauss theorem, 
\begin{equation}
   \nabla^{p,q} \cdot \mathbf{v}^{h,n} = \frac{1}{\Delta x \Delta y} \int \limits_{\Omega^{p,q}} \nabla \cdot \mathbf{v} d\mathbf{x} = \frac{1}{\Delta x \Delta y} \int \limits_{\partial \Omega^{p,q}} \mathbf{v} \cdot \mathbf{n} \, dS,  
   \label{eqn.gauss}
\end{equation}
based on the mid-point rule for the computation of the integrals along each edge of $\Omega^{p,q}$. In \eqref{eqn.gauss} the
outward pointing unit normal vector to the boundary $\partial \Omega^{p,q}$ of $\Omega^{p,q}$ is denoted by $\mathbf{n}$. 
In a similar manner, the $z$ component of the \textit{discrete curl}, $\nabla^h \times $, of a 
discrete vector field 
$\J^{h,n}$
is denoted by $\left( \nabla^h \times \J^{h,n} \right) \cdot \mathbf{e}_z$ and its degrees of freedom are naturally defined as
\begin{eqnarray}
\left( \nabla^{p,q} \times \mathbf{J}^{h,n} \right) \cdot \mathbf{e}_z &=& \epsilon_{3jk} 
\partial_j^{p,q} J_k^{h,n} \nonumber \\ 
 &=&  
 \halb \left (\frac{J_2^{p + \halb,q +\halb,n} - J_2^{p - \halb,q + \halb,n}}{\Delta x} + 
 \frac{J_2^{p + \halb,q -\halb,n}  - J_2^{p - \halb, q -\halb,n} }{\Delta x} \right ) - 
 \nonumber \\ &&  
 \halb \left (\frac{J_1^{p + \halb,q +\halb,n} - J_1^{p + \halb,q - \halb,n}}{\Delta y} 
 + \frac{J_1^{p - \halb,q +\halb,n} - J_1^{p - \halb, q -\halb,n} }{\Delta y} \right ), 
 \label{eqn.rot} 
\end{eqnarray}
making use of the \textit{vertex-based} staggered values of the field $\J^{h,n}$, see the right panel in Fig. \ref{fig.grad.curl}. In Eqn. \eqref{eqn.rot} the symbol $\epsilon_{ijk}$ is the usual Levi-Civita tensor. Eqn. \eqref{eqn.rot} defines a discrete
curl on the control volume $\Omega^{p,q}$ via the Stokes theorem
\begin{equation}
\left( \nabla^h \times \J^{h,n} \right) \cdot \mathbf{e}_z = \frac{1}{\Delta x \Delta y} \int \limits_{\Omega^{p,q}} 
\left( \nabla \times  \mathbf{J} \right) \cdot \mathbf{e}_z \, d\mathbf{x} = \frac{1}{\Delta x \Delta y} \int \limits_{\partial \Omega^{p,q}} \mathbf{J} \cdot \mathbf{t} \, dS,  
\label{eqn.stokes}
\end{equation}
based on the trapezoidal 
rule for the computation of the integrals along each edge of $\Omega^{p,q}$. Since the distortion 
field $\AAA$ transforms as a vector and not as a rank 2 tensor ($\AAA$ is a triad and thus a set
of three vectors), the degrees of freedom of the $z$ component of the discrete curl of $\AAA^{\!h,n}$ simply read  
\begin{eqnarray}
\left( \nabla^{p,q} \times \mathbf{A}^{\!h,n} \right) \cdot \mathbf{e}_z &=&  \epsilon_{3jk} \partial_j^{p,q} A_{ik}^{h,n}  \nonumber \\ 
 &=&  
\halb \left (\frac{A_{i2}^{p + \halb,q +\halb,n} - A_{i2}^{p - \halb,q + \halb,n}}{\Delta x} + 
\frac{A_{i2}^{p + \halb,q -\halb,n} - A_{i2}^{p - \halb, q -\halb,n} }{\Delta x} \right )- 
\nonumber \\ &&  
\halb \left (\frac{A_{i1}^{p + \halb,q +\halb,n} - A_{i1}^{p + \halb,q - \halb,n}}{\Delta y} + 
\frac{A_{i1}^{p - \halb,q +\halb,n} - A_{i1}^{p - \halb, q -\halb,n} }{\Delta y} \right ). 
\label{eqn.rotA} 
\end{eqnarray}
Last but not least, we need to define a discrete gradient operator that is compatible with the discrete curl,
so that the continuous identity
\begin{equation}
   \nabla \times \nabla \phi = 0
   \label{eqn.rotgrad} 
\end{equation}
also holds on the discrete level. If we define a scalar field in the barycenters of the control volumes $\Omega^{p,q}$ as
$\phi^{p,q,n}=\phi(x^p,y^q,t^n)$ then the corner gradient generates a natural discrete gradient operator $\nabla^{h}$ 
of the discrete scalar field $\phi^{h,n}$  that defines a discrete gradient in all vertices of the mesh. 
The corresponding degrees of freedom generated by $\nabla^{h} \phi^{h,n}$ read (see the left panel 
of Fig. \ref{fig.grad.curl})
\begin{equation}
\label{discr.grad}
	\nabla^{p+\halb,q+\halb}  \phi^{h,n} = \partial_k^{p+\halb,q+\halb} \phi^{h,n} = \left( 
	\begin{array}{c}  
	\halb \left( \frac{\phi^{p + 1, q + 1,n} - \phi^{p, q + 1,n}}{\Delta x} 
		+ \frac{\phi^{p + 1, q,n} - \phi^{p, q,n} }{\Delta x} \right) \\
	\halb \left( \frac{\phi^{p + 1, q + 1,n} - \phi^{p + 1, q,n}}{\Delta y} + \frac{\phi^{p, q + 
	1,n} 
	- \phi^{p,q,n} }{\Delta 
	y} \right)  \\ 
	\displaystyle 0
	\end{array} \right).
\end{equation}

It is then straightforward to verify that an immediate consequence of \eqref{eqn.rot} and 
\eqref{discr.grad} is 
\begin{equation}
  \nabla^h \times \nabla^h \phi^{h,n} = 0, 
  \label{eqn.discrete.curl} 
\end{equation}
i.e. one obtains a discrete analogue of \eqref{eqn.rotgrad}. This can be easily seen by computing 
\begin{eqnarray}
\left( \nabla^{p,q} \times \nabla^{p+\halb,q+\halb}  \phi^{h,n} \right) \cdot \mathbf{e}_z  & = &   \nonumber \\ 
\quarter \frac{ \left( \phi^{p + 1, q + 1,n} - \phi^{p + 1, q,n} + \phi^{p, q + 1,n} - \phi^{p,q,n} \right) + \left( \phi^{p + 1, q ,n} - \phi^{p + 1, q-1,n} + \phi^{p, q ,n} - \phi^{p,q-1,n} \right)  }{\Delta x \Delta y}  & -& \nonumber \\  
\quarter \frac{\left( \phi^{p , q + 1,n} - \phi^{p , q,n} + \phi^{p-1, q + 1,n} - \phi^{p-1,q,n} \right) + \left( \phi^{p , q ,n} - \phi^{p , q-1,n} + \phi^{p-1, q ,n} - \phi^{p-1,q-1,n} \right) }{\Delta x \Delta y}  &-& \nonumber \\ 
\quarter \frac{\left( \phi^{p + 1, q + 1,n} - \phi^{p, q + 1,n} + \phi^{p + 1, q,n} - \phi^{p, q,n} \right)  + \left( \phi^{p , q + 1,n} - \phi^{p-1, q + 1,n} + \phi^{p , q,n} - \phi^{p-1, q,n} \right)    }{\Delta y \Delta x}  &+& \nonumber \\  
\quarter \frac{\left( \phi^{p + 1, q ,n} - \phi^{p, q ,n} + \phi^{p + 1, q-1,n} - \phi^{p, q-1,n} \right)  + \left( \phi^{p , q ,n} - \phi^{p-1, q ,n} + \phi^{p , q-1,n} - \phi^{p-1, q-1,n} \right)    }{\Delta y \Delta x} &=& 0.
\label{eqn.rot.grad} 
\end{eqnarray}

We furthermore define the following averaging operators from the three different staggered meshes 
to the cell barycenter $(x^p,y^q)$: 
\begin{eqnarray}
  J_k^{p,q,n} &=& \frac{1}{4} \left( J_k^{p-\halb,q-\halb,n} + J_k^{p+\halb,q-\halb,n} +  
                                   J_k^{p-\halb,q+\halb,n} + J_k^{p+\halb,q+\halb,n} \right), 
                                   \nonumber \\ 
  v_1^{p,q,n} &=& \frac{1}{2} \left( v_1^{p-\halb,q,n} + v_1^{p+\halb,q,n} \right),     
								   \nonumber \\                                 								   
  v_2^{p,q,n} &=& \frac{1}{2} \left( v_2^{p,q-\halb,n} + v_2^{p,q+\halb,n} \right).                                   								   
\end{eqnarray}


\subsection{Explicit, compatible discretization of the distortion field and of the thermal impulse} 

The key ingredient of the numerical method proposed in this paper is the proper discretization of the terms
$\nabla \mathbf{G}_v(\mathbf{Q})$ and $\mathbf{B}_v(\mathbf{Q}) \nabla \mathbf{Q}$ present in \eqref{eqn.pde.split}. 
We propose the following compatible discretization for the thermal impulse equation: 
\begin{eqnarray}
J_{k}^{p+\halb,q+\halb,n+1} &=&   J_{k}^{p+\halb,q+\halb,n} - \Delta t \, \partial^{p+\halb,q+\halb}_k \left( J_{m}^{h,n} v_m^{h,n} + T^{h,n} \right)  \nonumber \\ 
&& - \Delta t \frac{1}{4} \sum \limits_{r=0}^{1} \sum \limits_{s=0}^{1} v_m^{p+r,q+s,n} \left( \partial_m^{p+r,q+s} J_{k}^{h,n} - \partial_k^{p+r,q+s} J_{m}^{h,n} \right) \nonumber \\
&& - \Delta t \frac{\rho^{p+\halb,q+\halb,n} \, T^{p+\halb,q+\halb,n}}{\tau_2} J_{k}^{p+\halb,q+\halb,n+1}. 
\label{eqn.Jh} 
\end{eqnarray}
It is easy to check that in the homogeneous case (when $\tau_2 \to \infty$ and therefore the algebraic source term vanishes)  
for an initially curl-free vector field $\mathbf{J}^{h,n}$ that satisfies 
$\nabla^{h} \times \mathbf{J}^{h,n} = 0$ also $\nabla^{h} \times \mathbf{J}^{h,n+1} = 0$ holds. To see this, one needs  
to apply the discrete curl operator $\nabla^h \times$ to Eqn. \eqref{eqn.Jh}.  One realizes 
that the second row of \eqref{eqn.Jh}, which contains the discrete curl of $\mathbf{J}^{h,n}$ vanishes immediately, 
due to $\nabla^{h} \times \mathbf{J}^{h,n} = 0$. The third row vanishes because $\tau_2 \to \infty$. 
The curl of the first term on the right hand side in the first row of Eqn. \eqref{eqn.Jh} is zero because of 
$\nabla^{h} \times \mathbf{J}^{h,n} = 0$ and the curl of the second term is zero because of 
$\nabla^h \times \nabla^h \phi^{h,n} = 0$, with the auxiliary scalar field $\phi^{h,n} = J_m^{h,n} v_m^{h,n} + T^{h,n} $, whose 
degrees of freedom are computed as $\phi^{p,q,n} = J_m^{p,q,n} v_m^{p,q,n} + T^{p,q,n}$ after averaging of the velocity vector and 
the thermal impulse vector into the barycenters of the control volumes $\Omega^{p,q}$. The key ingredient of our compatible  discretization for the $\mathbf{J}$ equation is indeed the use of a discrete gradient operator that is compatible with the discrete curl operator, see Eq. \eqref{eqn.rot.grad}.

The discrete form of the evolution equation of the distortion field is very similar to \eqref{eqn.Jh} and reads 
\begin{eqnarray}
A_{ik}^{p+\halb,q+\halb,n+1} &=&   A_{ik}^{p+\halb,q+\halb,n} - \Delta t \, \partial^{p+\halb,q+\halb}_k \left( A_{im}^{h,n} v_m^{h,n} \right)  \nonumber \\ 
&& - \Delta t \frac{1}{4} \sum \limits_{r=0}^{1} \sum \limits_{s=0}^{1} v_m^{p+r,q+s,n} \left( \partial_m^{p+r,q+s} A_{ik}^{h,n} - \partial_k^{p+r,q+s} A_{im}^{h,n} \right) \nonumber \\
&& - \Delta t \frac{\left| A_{ij}^{p+\halb,q+\halb,n+1} \right|^{\frac{5}{3}}}{\tau_1}  
A_{ij}^{p+\halb,q+\halb,n+1} \mathring{G}_{jk}^{p+\halb,q+\halb,n+1}. 
\label{eqn.Ah}
\end{eqnarray}
Following exactly the same reasoning as for the discrete thermal impulse equation, it is easy to check that in the 
homogeneous case (when $\tau_1 \to \infty$) for an initially curl-free distortion field $\mathbf{A}^{h,n}$ that satisfies 
$\nabla^{h} \times \mathbf{A}^{h,n} = 0$ also $\nabla^{h} \times \mathbf{A}^{h,n+1} = 0$ holds.

\subsection{Compatible numerical viscosity} 

The previous discretizations were all \textit{central} and thus without artificial numerical viscosity. In order to add a \textit{compatible numerical viscosity} operator, we need to recall the definition of the vector Laplacian at the continuous level, which reads: 
\begin{equation}
 \nabla^2 \mathbf{J} = \nabla \left( \nabla \cdot \mathbf{J} \right) - \nabla \times \nabla \times \mathbf{J} 
  \label{vector.laplace} 
\end{equation}
In order to define a discrete analogue of \eqref{vector.laplace} we define another discrete divergence operator
as follows: 
\begin{eqnarray}
\nabla^{p+\halb,q+\halb} \cdot \mathbf{J}^{h,n} &=& \partial_k^{p+\halb,q+\halb} J_k^{h,n}   = \nonumber \\ 
&&  
\halb \left (\frac{J_1^{p+1,q+1,n} - J_1^{p,q+1,n}}{\Delta x} + \frac{J_1^{p+1,q,n} - 
J_1^{p,q,n}}{\Delta x} \right )+ 
\nonumber \\ 
&& 
\halb \left (\frac{J_2^{p+1,q+1,n} - J_2^{p+1,q,n}}{\Delta y}  + \frac{J_2^{p,q+1,n} - 
J_2^{p,q,n}}{\Delta y} \right ). 
\label{vector.divc} 
\end{eqnarray}
The discrete vector Laplacian then simply reads 
\begin{equation}
\nabla^{p+\halb,q+\halb} \cdot \nabla^h \mathbf{J}^{h,n} = \nabla^{p+\halb,q+\halb} \left( \nabla^h \cdot \mathbf{J}^{h,n} \right) - \nabla^{p+\halb,q+\halb} \times \nabla^h \times \mathbf{J}^{h,n},   
\label{disc.vector.laplace} 
\end{equation}
i.e. it is composed of a grad-div contribution minus a curl-curl term. 
Taking \eqref{disc.vector.laplace} into account, a compatible discretization of $\mathbf{J}$ \textit{with} numerical viscosity then reads 
\begin{eqnarray}
J_{k}^{p+\halb,q+\halb,n+1} &=&   J_{k}^{p+\halb,q+\halb,n} - \Delta t \, 
\partial^{p+\halb,q+\halb}_k \left( J_{m}^{h,n} v_m^{h,n} + T^{h,n} 
\textcolor{red}{- h\, c_a  \, \partial^h_k J_k^{h,n}} \right)  
\textcolor{red}{-\Delta t\, h \, c_a \, \epsilon_{kj3} \, 
\partial_j^{p+\halb,q+\halb} \epsilon_{3lm} \, \partial_l^{h} J_m^{h,n}  } 
\nonumber \\ 
&& - \Delta t \frac{1}{4} \sum \limits_{r=0}^{1} \sum \limits_{s=0}^{1} v_m^{p+r,q+s,n} \left( \partial_m^{p+r,q+s} J_{k}^{h,n} - \partial_k^{p+r,q+s} J_{m}^{h,n} \right) \nonumber \\
&& 
- \Delta t \frac{\rho^{p+\halb,q+\halb,n} \, T^{p+\halb,q+\halb,n}}{\tau_2} J_{k}^{p+\halb,q+\halb,n+1}, 
\label{eqn.Jh.visc} 
\end{eqnarray}
where $h = \max( \Delta x, \Delta y)$ is a characteristic mesh spacing and $c_a$ is a characteristic velocity
related to the artificial viscosity that one would like to add to the scheme, e.g. $c_a = c_s$. For the sake of clarity, the additional numerical viscosity terms have been highlighted in red. It is obvious that also \eqref{eqn.Jh.visc} satisfies the curl-free property $\nabla^{h} \times \mathbf{J}^{h,n+1} = 0$ under the assumptions that $\tau_2 \to \infty$ and $\nabla^{h} \times \mathbf{J}^{h,n} = 0$.   
In analogy, the final evolution equation for $\mathbf{A}$ including the compatible numerical viscosity reads 
\begin{eqnarray}
A_{ik}^{p+\halb,q+\halb,n+1} &=&   A_{ik}^{p+\halb,q+\halb,n} - \Delta t \, \partial^{p+\halb,q+\halb}_k \left( A_{im}^{h,n} v_m^{h,n} \textcolor{red}{- h\, c_a  \, \partial^h_k A_{ik}^{h,n}} \right)  
\textcolor{red}{-\Delta t\, h \, c_a \, \epsilon_{kj3} \, 
	\partial_j^{p+\halb,q+\halb} \epsilon_{3lm} \, \partial_l^{h} A_{im}^{h,n}  } 
\nonumber \\ 
&& - \Delta t \frac{1}{4} \sum \limits_{r=0}^{1} \sum \limits_{s=0}^{1} v_m^{p+r,q+s,n} \left( \partial_m^{p+r,q+s} A_{ik}^{h,n} - \partial_k^{p+r,q+s} A_{im}^{h,n} \right) 
\nonumber \\
&& - \Delta t \frac{\left| A_{ij}^{p+\halb,q+\halb,n+1} \right|^{\frac{5}{3}}}{\tau_1}  
A_{ij}^{p+\halb,q+\halb,n+1} \mathring{G}_{jk}^{p+\halb,q+\halb,n+1}. 
\label{eqn.Ah.visc}
\end{eqnarray}
It is easy to check that one has $\nabla^{h} \times \mathbf{A}^{h,n+1} = 0$ as a consequence of \eqref{eqn.Ah.visc} for $\tau_1 \to \infty$ and  $\nabla^{h} \times \mathbf{A}^{h,n} = 0$. In order
to reduce the numerical dissipation, it is possible to employ a piecewise linear reconstruction 
and insert the barycenter extrapolated values into the discrete divergence operator under the 
discrete gradient.

\subsection{Explicit discretization of the nonlinear convective terms and of the corner fluxes}

The semi-implicit scheme proposed in this paper applies an explicit discretization 
of the nonlinear convective terms contained in $\mathbf{F}_c = (\mathbf{f}_c(\Q),\mathbf{g}_c(\Q))$ 
and of the corner (vertex) fluxes $\mathbf{F}_v = (\mathbf{f}_v(\Q),\mathbf{g}_v(\Q))$, 
starting from the known solution $\mathbf{Q}^{p,q,n}$ at time $t^n$. 
The result is a new intermediate state vector $\mathbf{Q}^{p,q,*}$ that is computed via a conservative finite volume formulation 
\begin{equation}
\mathbf{Q}^{p,q,*} = \mathbf{Q}^{p,q,n} 
- \frac{\Delta t}{\Delta x} \left( \mathbf{f}_{c,v}^{p+\halb,q} - \mathbf{f}_{c,v}^{p-\halb,q} \right) 
- \frac{\Delta t}{\Delta y} \left( \mathbf{g}_{c,v}^{p,q+\halb} - \mathbf{g}_{c,v}^{p,q-\halb} \right), 
\label{eqn.Qstar} 
\end{equation} 
with the numerical fluxes defined as 
\begin{eqnarray}
\mathbf{f}_{c,v}^{p+\halb,q} &=& \phantom{+} \halb \left( \mathbf{f}_{c}\left( \mathbf{Q}^{p+\halb,q,n+\halb}_{-} \right)  + \mathbf{f}_{c}\left( \mathbf{Q}^{p+\halb,q,n+\halb}_{+} \right) \right) 
- \halb s^x_{\max} \left( \mathbf{Q}^{p+\halb,q,n+\halb}_{+} - \mathbf{Q}^{p+\halb,q,n+\halb}_{-} \right) \nonumber \\ 
&& 
+ \halb \left( \mathbf{f}_{v}\left( \mathbf{Q}^{p+\halb,q+\halb,n} \right)  + \mathbf{f}_{v}\left( \mathbf{Q}^{p+\halb,q-\halb,n} \right) \right), 
\label{eqn.numflux.f} 
\end{eqnarray} 
and 
\begin{eqnarray}
\mathbf{g}_{c,v}^{p,q+\halb} &=& \phantom{+} \halb \left( \mathbf{g}_{c}\left( \mathbf{Q}^{p,q+\halb,n+\halb}_{-} \right)  + \mathbf{g}_{c}\left( \mathbf{Q}^{p,q+\halb,n+\halb}_{+} \right) \right) 
- \halb s^y_{\max} \left( \mathbf{Q}^{p,q+\halb,n+\halb}_{+} - \mathbf{Q}^{p,q+\halb,n+\halb}_{-} \right) \nonumber \\ 
&& 
+ \halb \left( \mathbf{g}_{v}\left( \mathbf{Q}^{p+\halb,q+\halb,n} \right)  + \mathbf{g}_{v}\left( \mathbf{Q}^{p-\halb,q+\halb,n} \right) \right). 
\label{eqn.numflux.g} 
\end{eqnarray} 
Note that the fluxes above contain the nonlinear convective terms as well as the vertex fluxes $\mathbf{f}_v$ and $\mathbf{g}_v$, which contain the stress tensor $\boldsymbol{\sigma}$ and the heat flux $\mathbf{q}$. In Eqns. \eqref{eqn.numflux.f} and \eqref{eqn.numflux.g} the maximum signal speeds are computed as 
\begin{eqnarray} 
s_{\max}^x &=& \max \left( |\boldsymbol{\Lambda}^{c,v}_x(\mathbf{Q}^{p+\halb,q,n+\halb}_{-})|, |\boldsymbol{\Lambda}^{c,v}_x(\mathbf{Q}^{p+\halb,q,n+\halb}_{+})|\right), \nonumber \\ 
s_{\max}^y &=& \max \left( |\boldsymbol{\Lambda}^{c,v}_y(\mathbf{Q}^{p,q+\halb,n+\halb}_{-})|, |\boldsymbol{\Lambda}^{c,v}_y(\mathbf{Q}^{p,q+\halb,n+\halb}_{+})|\right), 
\end{eqnarray} 
with $\boldsymbol{\Lambda}^{c,v}_{k}$ the diagonal matrix of eigenvalues of the explicit subsystem 
\eqref{eqn.pde.ex} in direction $x$ and $y$, respectively.   

The boundary-extrapolated values are simply computed via a standard MUSCL-Hancock scheme (see \cite{toro-book}) as follows: 
\begin{eqnarray}
\mathbf{Q}^{p+\halb,q,n+\halb}_{-} &=& \mathbf{Q}^{p,q,n}   + \halb \Delta x \, \partial^h_x \mathbf{Q}^{p,q,n}   + \halb \Delta t \partial^h_t \mathbf{Q}^{p,q,n}, \nonumber \\
\mathbf{Q}^{p+\halb,q,n+\halb}_{+} &=& \mathbf{Q}^{p+1,q,n} - \halb \Delta x \, \partial^h_x \mathbf{Q}^{p+1,q,n} + \halb \Delta t \partial^h_t \mathbf{Q}^{p+1,q,n}, \nonumber  
\end{eqnarray} 
and 
\begin{eqnarray}
\mathbf{Q}^{p,q+\halb,n+\halb}_{-} &=& \mathbf{Q}^{p,q,n}   + \halb \Delta y \, \partial^h_y \mathbf{Q}^{p,q,n}   + \halb \partial_t \mathbf{Q}^{p,q,n}, \nonumber \\
\mathbf{Q}^{p,q+\halb,n+\halb}_{+} &=& \mathbf{Q}^{p,q+1,n} - \halb \Delta y \, \partial^h_y \mathbf{Q}^{p,q+1,n} + \halb \partial_t \mathbf{Q}^{p,q+1,n}, \nonumber   
\end{eqnarray} 
with the discrete gradients in space and time computed via  
\begin{eqnarray}
\partial^h_x \mathbf{Q}^{p,q,n} &=& \textnormal{minmod} \left( \frac{\mathbf{Q}^{p+1,q,n} - \mathbf{Q}^{p,q,n}}{\Delta x}, \frac{\mathbf{Q}^{p,q,n} - \mathbf{Q}^{p-1,q,n}}{\Delta x}  \right), \nonumber \\
\partial^h_y \mathbf{Q}^{p,q,n} &=& \textnormal{minmod} \left( \frac{\mathbf{Q}^{p+1,q,n} - 
\mathbf{Q}^{p,q,n}}{\Delta y}, \frac{\mathbf{Q}^{p,q,n} - \mathbf{Q}^{p-1,q,n}}{\Delta y}  \right), 
\nonumber 
\end{eqnarray} 
and
\begin{eqnarray}
\partial^h_t \mathbf{Q}^{p,q,n} &=& - \frac{\mathbf{f}_{c}\left( \mathbf{Q}^{p,q,n} + \halb \Delta x \, \partial^h_x \mathbf{Q}^{p,q,n}\right) - \mathbf{f}_{c}\left( \mathbf{Q}^{p,q,n} - \halb \Delta x \, \partial^h_x \mathbf{Q}^{p,q,n}\right)  }{\Delta x} \nonumber \\    
&&  - \frac{\mathbf{g}_{c}\left( \mathbf{Q}^{p,q,n} + \halb \Delta y \, \partial^h_y \mathbf{Q}^{p,q,n}\right) - \mathbf{g}_{c}\left( \mathbf{Q}^{p,q,n} - \halb \Delta y \, \partial^h_y \mathbf{Q}^{p,q,n}\right)  }{\Delta y}. \nonumber  
\end{eqnarray}

\subsection{Implicit solution of the pressure equation} 

Up to now, the contribution of the pressure to the momentum and to the total energy conservation laws 
has been excluded, i.e. the terms contained in the pressure fluxes $\mathbf{F}_p$. The discrete momentum 
equations including the pressure terms read 
\begin{eqnarray}
\label{eqn.rhou2d} 
(\rho v)_{1}^{p+\halb,q,n+1} &=& (\rho v)_{1}^{p+\halb,q,*} - \frac{\Delta t}{\Delta x} \left( p^{p+1,q,n+1} - p^{p,q,n+1} \right),  \nonumber \\ 
(\rho v)_{2}^{p,q+\halb,n+1} &=& (\rho v)_{2}^{p,q+\halb,*} - \frac{\Delta t}{\Delta y} \left( p^{p,q+1,n+1} - p^{p,q,n+1} \right),  
\end{eqnarray} 
where pressure is taken \textit{implicitly}, while all nonlinear convective terms and the vertex fluxes have already been discretized \textit{explicitly} via the operators $(\rho v)_{1}^{p+\halb,q,*} $ and 
$(\rho v)_{2}^{p,q+\halb,*}$ given in \eqref{eqn.Qstar} and after averaging of the obtained 
quantities back to the 
edge-based staggered dual grid. A preliminary form of the discrete total energy equation reads 
\begin{eqnarray}
\label{eqn.rhoE2d.prelim} 
\rho E_1\left( p^{p,q,n+1} \right) + \rho E_2^{p,q,n+1} + \rho \tilde{E}_{3}^{p,q,n+1}  &=& \rho E^{p,q,*}  
\nonumber \\      
- \frac{\Delta t}{\Delta x} \left(  \tilde{h}^{p+\halb,q,n+1} (\rho v)_1^{p+\halb,q,n+1} - \tilde{h}^{p-\halb,q,n+1} (\rho v)_{1}^{p-\halb,q,n+1} \right)\phantom{.}    && \nonumber \\
- \frac{\Delta t}{\Delta y} \left(  \tilde{h}^{p,q+\halb,n+1} (\rho v)_{2}^{p,q+\halb,n+1} - \tilde{h}^{p,q-\halb,n+1} (\rho v)_{2}^{p,q-\halb,n+1} \right). && 
\end{eqnarray}
Here, we have used the abbreviation $\rho E_1\left( p^{p,q,n+1} \right) = \rho^{p,q,n+1} E_1\left( p^{p,q,n+1}, \rho^{p,q,n+1} \right)$.   
Inserting the discrete momentum equations \eqref{eqn.rhou2d} into the discrete energy equation \eqref{eqn.rhoE2d.prelim} and making tilde symbols explicit via a simple Picard iteration 
(using the lower index $r$ in the following), as suggested in \cite{DumbserCasulli2016,SIMHD}, leads to the following  discrete wave equation for the unknown pressure:  
\begin{eqnarray}
\label{eqn.p2d} 
\rho^{p,q,n+1} E_1\left( p_{r+1}^{p,q,n+1}, \rho^{p,q,n+1} \right) & & \nonumber \\
- \frac{\Delta t^2}{\Delta x^2} \left(  {{h}_{r}^{p+\halb,q,n+1}} \left( p_{r+1}^{p+1,j,n+1}-p_{r+1}^{p,q,n+1} \right) 
- {{h}_{r}^{p-\halb,q,n+1}} \left( p_{r+1}^{p,q,n+1}-p_{r+1}^{p-1,q,n+1} \right) \right) & & \nonumber \\  
- \frac{\Delta t^2}{\Delta y^2}\left(  {{h}_{r}^{p,q+\halb,n+1}} \left( p_{r+1}^{p,q+1,n+1}-p_{r+1}^{p,q,n+1} \right) 
- {{h}_{r}^{p,q-\halb,n+1}} \left( p_{r+1}^{p,q,n+1}-p_{r+1}^{p,q-1,n+1} \right) \right)  
& = & b_{r}^{p,q,n}, 
\end{eqnarray} 
with the known right hand side 
\begin{eqnarray}
b_{i,j}^r  =   \rho E_{}^{p,q,*} - \rho E_{2}^{p,q,n+1} - \rho E_{3,r}^{p,q,n+1}    && 
\nonumber \\
- \frac{\Delta t}{\Delta x} \left( 
{h}_{r}^{p+\halb,q,n+1} (\rho v)_{1}^{p+\halb,q,*} - 
{h}_{r}^{p-\halb,q,n+1} (\rho v)_{1}^{p-\halb,q,*} \right) \phantom{.}  && \nonumber \\ 
- \frac{\Delta t}{\Delta y} \left( {h}_{r}^{p,q+\halb,n+1} (\rho v)_{2}^{p,q+\halb,*} - {h}_{r}^{p,q-\halb,n+1} (\rho v)_{2}^{p,q-\halb,*} \right). 
\label{eqn.rhs.2d} 
\end{eqnarray} 
The density at the new time $\rho_{}^{p,q,n+1} = \rho_{}^{p,q,*} $ is already known from \eqref{eqn.Qstar}, and also the energy 
contribution $\rho E_{2}^{p,q,n+1}$ of the distortion field $\mathbf{A}^{h,n+1}$ and of the thermal impulse $\mathbf{J}^{h,n+1}$  is already known, after averaging onto the main grid of the staggered field components that have been evolved in the vertices
via the compatible discretization \eqref{eqn.Ah} and \eqref{eqn.Jh}.  
The final system for the pressure \eqref{eqn.p2d} forms a \textit{mildly nonlinear system} of the form 
\begin{equation}
  \rho \mathbf{E}_1 \left( \mathbf{p}_{r+1}^{n+1} \right) + \mathbf{M}_r \cdot \mathbf{p}_{r+1}^{n+1} = \mathbf{b}_r^{n} 
  \label{eqn.nonlinear} 
\end{equation}
with a linear part contained in $\mathbf{M}$ that is symmetric and at least positive semi-definite. Hence, with the usual  assumptions on the nonlinearity detailed in \cite{CasulliZanolli2012}, it can be again efficiently solved with 
the nested Newton method of Casulli and Zanolli \cite{CasulliZanolli2010,CasulliZanolli2012}. Note that in the incompressible 
limit $M \to 0$, following the asymptotic analysis performed in \cite{KlaMaj,KlaMaj82,Klein2001,Munz2003,MunzDumbserRoller}, 
the pressure tends to a constant and the contribution of the kinetic energy $\rho E_3$  
can be neglected w.r.t. $\rho E_1$. Therefore, in the incompressible limit the system \eqref{eqn.p2d} tends to the usual pressure Poisson equation of incompressible flow solvers. 
In each Picard iteration, after the solution of the pressure system \eqref{eqn.p2d}, the enthalpies at the interfaces are  recomputed and the momentum is updated  by 
\begin{eqnarray}
\label{eqn.rhou2d.pic} 
(\rho v)_{1,r+1}^{p+\halb,q,n+1} &=& (\rho v)_{1}^{p+\halb,q,*} - \frac{\Delta t}{\Delta x} \left( p_{r+1}^{p+1,q,n+1} - p_{r+1}^{p,q,n+1} \right),  \\
(\rho v)_{2,r+1}^{p,q+\halb,n+1} &=& (\rho v)_{2}^{p,q+\halb,*} - \frac{\Delta t}{\Delta y} \left( p_{r+1}^{p,q+1,n+1} - p_{r+1}^{p,q,n+1} \right),  
\end{eqnarray} 
from which the new kinetic energy density $(\rho E)_{3,r+1}^{p,q,n+1}$ can be computed after averaging the momentum onto the main grid.  
At the end of the Picard iterations, the total energy is updated as 
\begin{eqnarray}
\label{eqn.rhoE2d} 
(\rho E)_{}^{p,q,n+1} &=& (\rho E)_{}^{p,q,*}  
- \frac{\Delta t}{\Delta x} \left( {h}^{p+\halb,q,n+1} (\rho v)_{1}^{p+\halb,q,n+1} - {h}_{}^{p-\halb,q,n+1} (\rho v)_{1}^{p-\halb,q,n+1} \right) \nonumber \\
& & \phantom{(\rho E)_{}^{p,q,*}} 
- \frac{\Delta t}{\Delta y} \left( {h}_{}^{p,q+\halb,n+1} (\rho v)_{2}^{p,q+\halb,n+1} - {h}_{}^{p,q-\halb,n+1} (\rho v)_{2}^{p,q-\halb,n+1} \right),   
\end{eqnarray} 
while the final momentum is averaged back onto the main grid. This completes the description of our new curl-free semi-implicit finite volume scheme for the GPR model of continuum mechanics in the two-dimensional case. In the following Section \ref{sec.analysis} we provide a detailed analysis of the properties of the new algorithm. 

\subsection{No-slip wall boundary conditions for fluids}
\begin{figure}[!htbp]
	\begin{center}
		\includegraphics[width=0.5\textwidth]{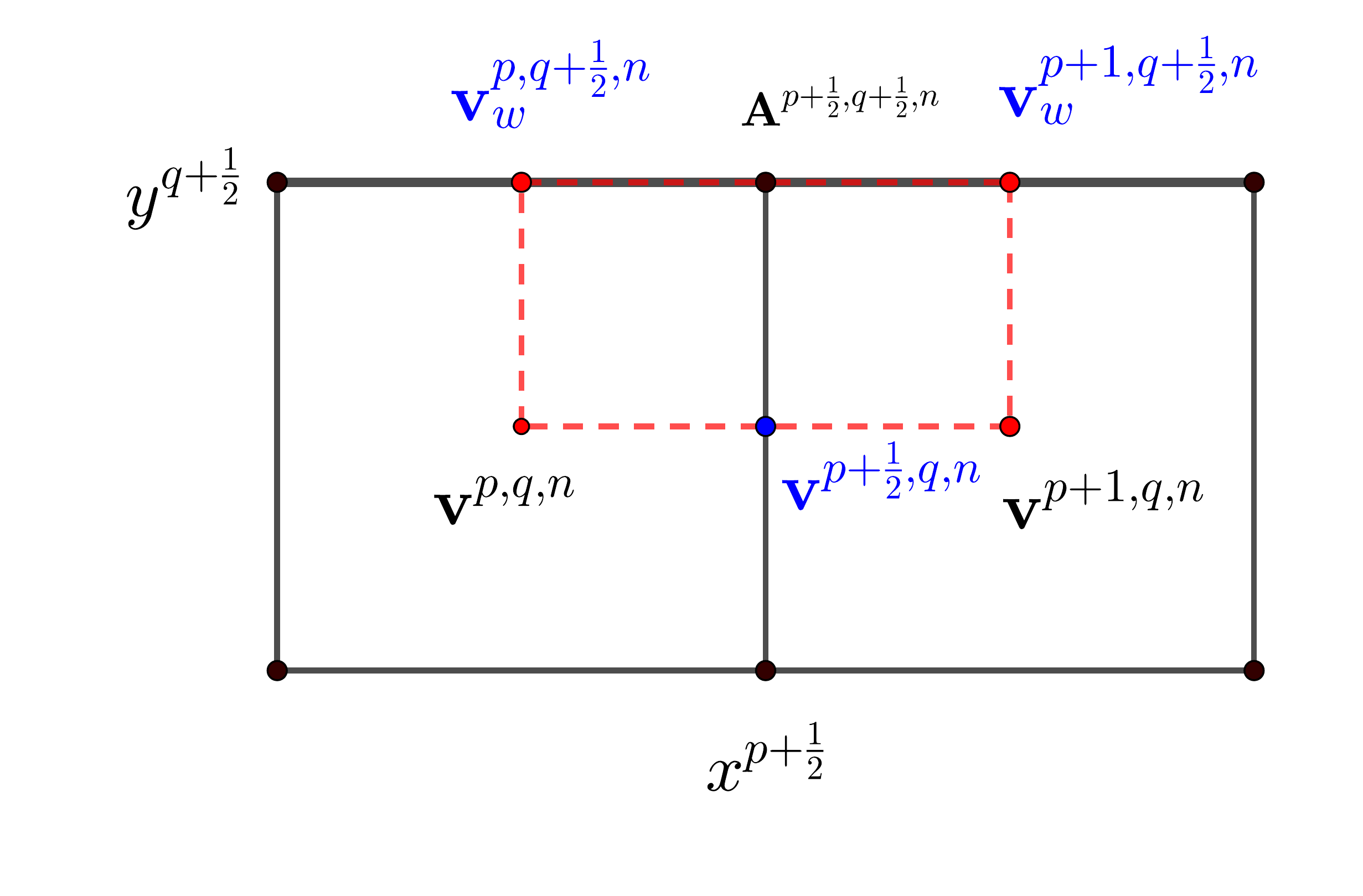}  
		\caption{Calculation of the velocity gradient for a wall boundary condition at the upper $y$ boundary of a rectangular domain $\Omega$.}  
		\label{fig.wall.bc}
	\end{center}
\end{figure}

In the case of a viscous fluid, we want to impose $\mathbf{v}=\mathbf{v}_{w}$ for all $\mathbf{x} \in \partial \Omega_w$, where $\mathbf{v}_{w}$ is a given velocity at the wall and $\partial \Omega_w$ is the part of the boundary occupied by the wall. In order to evolve the distortion field $\mathbf{A}$ properly at wall boundary points, we rewrite its governing equation as 
\begin{equation} 
 \partial_t A_{ik} + v_m \partial_m A_{ik} + A_{im} \partial_k v_m = -\frac{c_s^2 A_{ij} \mathring{G}_{jk} }{\theta_1(\tau_1)},  
 \label{eqn.A.expand} 
\end{equation} 
which can be written in matrix-notation as 
\begin{equation} 
\partial_t \mathbf{A} + \mathbf{v} \cdot \nabla \mathbf{A} + \mathbf{A} \nabla \mathbf{v} = -\frac{c_s^2  }{\theta_1(\tau_1)} \mathbf{A} \mathring{\mathbf{G}}.  
\label{eqn.A.expand2} 
\end{equation} 
In the following, we illustrate the procedure for a no-slip wall on the upper boundary, see also Fig. \ref{fig.wall.bc}. Analogous formulas can be derived also for the lower boundary as well as for a boundary on the left and right of the domain, respectively.  
At the wall, the velocity field is known as $\mathbf{v}_w^{p+\halb,q+\halb,n} = \mathbf{v}_w(x^{p+\halb},y^{q+\halb},t^n)$ and the velocity gradient at the boundary 
$\nabla \mathbf{v}^{p + \halb,q+\halb,n} = (\partial_x \mathbf{v}^{p + \halb,q+\halb,n}, \partial_y \mathbf{v}^{p + \halb,q+\halb,n})^T$ is computed as in a classical Navier-Stokes code, making
use of one-sided differences and the known velocity field at the wall: 
\begin{equation}
    \partial_x \mathbf{v}^{p+\halb,q+\halb,n} =  \frac{\mathbf{v}_w^{p+1,q+\halb,n}-\mathbf{v}_w^{p,q+\halb,n}}{\Delta x}, \qquad  
    \partial_y \mathbf{v}^{p+\halb,q+\halb,n} = \frac{\mathbf{v}_w^{p+\halb,q+\halb,n}-\mathbf{v}^{p+\halb,q,n}}{\Delta y/2},
    \label{eqn.bc.grad.y} 
\end{equation}
with $\mathbf{v}_w^{p+\halb,q+\halb,n} = \halb \left( \mathbf{v}_w^{p,q+\halb,n} + \mathbf{v}_w^{p+1,q+\halb,n} \right) $ computed from the known wall velocities on the boundary and $\mathbf{v}^{p+\halb,q,n} = \halb \left(  \mathbf{v}^{p,q,n} + \mathbf{v}^{p+1,q,n} \right)$, where $\mathbf{v}^{p,q,n}$  
and $\mathbf{v}^{p+1,q,n}$ are the known velocity vectors inside the computational domain and defined at the barycenter of a primary control volume.  

For completeness, we also give the formula for a wall boundary at the right $x$ boundary of the domain: 
\begin{equation}
\partial_x \mathbf{v}^{p+\halb,q+\halb,n} =  \frac{\mathbf{v}_w^{p+\halb,q+\halb,n}-\mathbf{v}^{p,q+\halb,n}}{\Delta x/2}, \qquad  
\partial_y \mathbf{v}^{p+\halb,q+\halb,n} = \frac{\mathbf{v}_w^{p+\halb,q+1,n}-\mathbf{v}_w^{p+\halb,q,n}}{\Delta y},
\label{eqn.bc.grad.x} 
\end{equation}
with the analogous definitions $\mathbf{v}_w^{p+\halb,q+\halb,n} = \halb \left( \mathbf{v}_w^{p+\halb,q,n} + \mathbf{v}_w^{p+\halb,q+1,n} \right) $ and $\mathbf{v}^{p,q+\halb,n} = \halb \left(  \mathbf{v}^{p,q,n} + \mathbf{v}^{p,q+1,n} \right)$. 

Once the velocity gradients have been computed the distortion field at the boundary points is evolved via the following implicit formula: 
\begin{equation} 
\mathbf{A}^{p+\halb,q+\halb,n+1} 
+ \Delta t \, \mathbf{A}^{p+\halb,q+\halb,n+1} \, \nabla \mathbf{v}^{p + \halb,q+\halb,n} 
+  \frac{ \Delta t c_s^2  }{\theta_1(\tau_1)} \mathbf{A}^{p+\halb,q+\halb,n+1} \mathring{\mathbf{G}}^{p+\halb,q+\halb,n+1} = 
\mathbf{A}^{p+\halb,q+\halb,n} - \Delta t \, \mathbf{v}_w^{p+\halb,q+\halb,n} \cdot \nabla_h^{p+\halb,q+\halb,n} \mathbf{A}_h,  
\label{eqn.A.bc} 
\end{equation} 
where $\nabla_h^{p+\halb,q+\halb,n} \mathbf{A}_h$ is a suitable upwind discretization of the convective term, based on the known velocity field $\mathbf{v}_w^{p+\halb,q+\halb,n}$ at the wall. For a wall at rest, this 
term obviously vanishes. In the interior of the domain, the formula \eqref{eqn.Ah.visc} remains valid. 

\section{Analysis of the scheme}
\label{sec.analysis} 

While the discrete curl-free property of the scheme in the case $\tau_1 \to \infty$ and $\tau_2 \to \infty$ is very easy to see directly from the structure of the discrete equations \eqref{eqn.Jh} and \eqref{eqn.Ah} together with the discrete property \eqref{eqn.rot.grad}, the behaviour of the scheme in the stiff relaxation limit $\tau_1 \to 0$ and $\tau_2 \to 0$ deserves more attention and is analyzed in the following. For the sake of simplicity, we start with the analysis of the fully discrete equation for the specific thermal impulse $\mathbf{J}$, showing that in this case the proposed method is \textit{asymptotic preserving} (AP). 
Then, we also study the stiff relaxation limit of the semi-discrete equation for $\mathbf{A}$, which is much 
more complicated.    

\subsection{Asymptotic relaxation limit for the heat flux for $\tau_2 \to 0$} 
\label{sec.arl.tau2} 

Without numerical viscosity, the fully-discrete equation for $\J$ reads (see \eqref{eqn.Jh})   
\begin{eqnarray}
J_{k}^{p+\halb,q+\halb,n+1} - J_{k}^{p+\halb,q+\halb,n} + \Delta t \partial^{p+\halb,q+\halb}_k \left( J_{m}^{h,n} v_m^{h,n}    \right) 
+ \Delta t \partial^{p+\halb,q+\halb}_k T^{h,n}  && \nonumber \\ 
+ \Delta t \frac{1}{4} \sum \limits_{r=0}^{1} \sum \limits_{s=0}^{1} v_m^{p+r,q+s,n} \left( \partial_m^{p+r,q+s} J_{k}^{h,n} - \partial_k^{p+r,q+s} J_{m}^{h,n} \right) \nonumber \\
= - \Delta t \frac{\rho^{p+\halb,q+\halb,n} \, T^{p+\halb,q+\halb,n}}{\tau_2} J_{k}^{p+\halb,q+\halb,n+1}. && 
\label{eqn.semi.Jh} 
\end{eqnarray}
Formal asymptotic expansion of the discrete solution $J^h_m$ in powers of $\tau_2$ yields 
\begin{eqnarray}
J_{k}^{p+\halb,q+\halb,n+1} = J_{k,(0)}^{p+\halb,q+\halb,n+1} + \tau_2 
J_{k,(1)}^{p+\halb,q+\halb,n+1} + \cdots 
\label{eqn.Jh.expansion}
\end{eqnarray}
Inserting \eqref{eqn.Jh.expansion} into \eqref{eqn.semi.Jh}, collecting terms of equal powers in $\tau_2$ and retaining only the leading order terms yields  
\begin{eqnarray}
J_{k,(0)}^{p+\halb,q+\halb,n+1} - J_{k,(0)}^{p+\halb,q+\halb,n} + \Delta t \partial^{p+\halb,q+\halb}_k \left( J_{m,(0)}^{h,n} v_m^{h,n}    \right) 
+ \Delta t \partial^{p+\halb,q+\halb}_k T^{h,n}  && \nonumber \\ 
+ \Delta t \frac{1}{4} \sum \limits_{r=0}^{1} \sum \limits_{s=0}^{1} v_m^{p+r,q+s,n} \left( \partial_m^{p+r,q+s} J_{k,(0)}^{h,n} - \partial_k^{p+r,q+s} J_{m,(0)}^{h,n} \right) \nonumber \\
= - \Delta t \frac{1}{\tau_2} \, \rho^{p+\halb,q+\halb} \, T^{p+\halb,q+\halb,n} J_{k,(0)}^{p+\halb,q+\halb,n+1} 
  - \Delta t \,                \rho^{p+\halb,q+\halb} \, T^{p+\halb,q+\halb,n} J_{k,(1)}^{p+\halb,q+\halb,n+1}.  && 
\label{eqn.Jh.tau} 
\end{eqnarray}
From the leading order term $\tau_2^{-1}$ one can conclude that the leading order contribution to $\mathbf{J}$ must vanish, i.e. 
\begin{equation}
 J_{k,(0)}^{p+\halb,q+\halb,n+1} = 0.   
 \label{eqn.Jh0} 
\end{equation}  
Inserting \eqref{eqn.Jh0} into \eqref{eqn.Jh.tau} and assuming that due to \eqref{eqn.Jh0} also $J_{k,(0)}^{p+\halb,q+\halb,n} = 0$ (well-prepared initial data) leads to the following result for 
the terms of order $\tau_2^0$   
\begin{equation}
J_{k,(1)}^{p+\halb,q+\halb,n+1} = - \frac{1}{\rho^{p+\halb,q+\halb,n} \, T^{p+\halb,q+\halb,n}} \, \partial^{p+\halb,q+\halb}_k T^{h,n}.   
\label{eqn.Jh1}
\end{equation}
Inserting \eqref{eqn.Jh0} and \eqref{eqn.Jh1} into \eqref{eqn.Jh.expansion} yields the following final result 
for the discrete specific thermal impulse when $\tau_2 \to 0$: 
\begin{equation}
J_{k}^{p+\halb,q+\halb,n+1} = - \frac{\tau_2}{\rho^{p+\halb,q+\halb,n} \, T^{p+\halb,q+\halb,n}} \, \partial^{p+\halb,q+\halb}_k T^{h,n}.   
\label{eqn.Jh.result}
\end{equation}
As a result, in the stiff relaxation limit $\tau_2 \to 0$ the discrete heat flux vector $\mathbf{q} = \rho T \mathbf{J}$ becomes 
\begin{equation}
q_{k}^{p+\halb,q+\halb,n} = - \tau_2 \alpha^2 \, \partial^{p+\halb,q+\halb}_k T^{h,n},   
\label{eqn.qk}
\end{equation}
which is the discrete analogue of the Fourier law \eqref{eqn.heatflux.limit}. As a consequence, for the heat flux the proposed staggered semi-implicit finite volume scheme is \textit{asymptotic preserving}, 
see \cite{Jin_Xin,Buet_Despres,Burman_Sainsaulieu,PareschiRusso2000,PareschiRusso2005,JinPareschiToscani,Naldi_Pareschi,Chen_Levermore_Liu}, i.e. in the stiff relaxation limit the classical parabolic 
Navier-Stokes heat flux is retrieved also at the fully discrete level. The same AP result is also obtained in the presence of numerical viscosity, since all extra terms with respect to \eqref{eqn.Jh} scale with $\mathbf{J}$ and thus are of order $\tau_2$, see \eqref{eqn.Jh.visc}.        

Note that in \eqref{eqn.qk}, the gradient of the temperature is computed in the vertices of the primary control volumes (corner gradient), which is a rather common choice for the discretization of the compressible Navier-Stokes equations. 

\subsection{Asymptotic relaxation limit of the stress tensor for $\tau_1 \to 0$} 
\label{sec.arl.tau1} 

The asymptotic analysis for the stress tensor is much more complex than the previous one for the heat flux. 
In the following, we consider only the following \textit{semi-discrete} scheme for $\mathbf{A}$, without numerical viscosity and restricting the discussion to the two-dimensional case, i.e. $\partial_3 = 0$
and $v_3 =0$: 
\begin{eqnarray}
\partial_t A_{ik}^{p+\halb,q+\halb,n+1} + \partial^{p+\halb,q+\halb}_k \left( A_{im}^{h} v_m^{h} \right)    + \frac{1}{4} \sum \limits_{r=0}^{1} \sum \limits_{s=0}^{1} v_m^{p+r,q+s} \left( \partial_m^{p+r,q+s} A_{ik}^{h} - \partial_k^{p+r,q+s} A_{im}^{h} \right) && \nonumber \\
 =  -  \frac{\left| A_{ij}^{p+\halb,q+\halb} \right|^{\frac{5}{3}}}{\tau_1}  
A_{ij}^{p+\halb,q+\halb} \mathring{G}_{jk}^{p+\halb,q+\halb}.  && 
\label{eqn.semi.Ah}
\end{eqnarray}
For the sake of clarity, in the following, we give the explicit expansion of several terms appearing in \eqref{eqn.semi.Ah}: 
\begin{equation}
\label{eqn.Avx} 
 \partial^{p+\halb,q+\halb}_1 \left( A_{im}^{h} v_m^{h} \right) = \halb 
 \frac{A_{im}^{p+1,q+1} v_m^{p+1,q+1} + A_{im}^{p+1,q} v_m^{p+1,q} - 
       A_{im}^{p,q+1} v_m^{p,q+1} - A_{im}^{p,q} v_m^{p,q}  }{\Delta x},  
\end{equation}
\begin{equation}
\label{eqn.Avy} 
\partial^{p+\halb,q+\halb}_2 \left( A_{im}^{h} v_m^{h} \right) = \halb 
\frac{A_{im}^{p+1,q+1} v_m^{p+1,q+1} + A_{im}^{p,q+1} v_m^{p,q+1} - 
   	  A_{im}^{p+1,q} v_m^{p+1,q} - A_{im}^{p,q} v_m^{p,q}  }{\Delta y},  
\end{equation}
with the quantity $A_{im}^{p,q}$ at the barycenter computed via averaging from the four surrounding vertices as 
\begin{equation}
   A_{im}^{p,q} = \frac{1}{4} \left( A_{im}^{p+\halb,q+\halb} + A_{im}^{p+\halb,q-\halb} + A_{im}^{p-\halb,q+\halb}+A_{im}^{p-\halb,q-\halb}\right).  
\end{equation}
We furthermore expand the second term in the double sum in $x$ and $y$ direction as follows: 
\begin{equation}
   -\frac{1}{4} \sum \limits_{r=0}^{1} \sum \limits_{s=0}^{1} v_m^{p+r,q+s} \partial_1^{p+r,q+s} A_{im}^{h} = -\frac{1}{8} \sum \limits_{r=0}^{1} \sum \limits_{s=0}^{1} v_m^{p+r,q+s} 
   \frac{ A_{im}^{p+\halb+r,q+\halb+s} + A_{im}^{p+\halb+r,q-\halb+s} 
         -A_{im}^{p-\halb+r,q+\halb+s} - A_{im}^{p-\halb+r,q-\halb+s}}{\Delta x}, 
\end{equation}
\begin{equation}
-\frac{1}{4} \sum \limits_{r=0}^{1} \sum \limits_{s=0}^{1} v_m^{p+r,q+s} \partial_2^{p+r,q+s} A_{im}^{h} = -\frac{1}{8} \sum \limits_{r=0}^{1} \sum \limits_{s=0}^{1} v_m^{p+r,q+s} 
\frac{ A_{im}^{p+\halb+r,q+\halb+s} + A_{im}^{p-\halb+r,q+\halb+s} 
      -A_{im}^{p+\halb+r,q-\halb+s} - A_{im}^{p-\halb+r,q-\halb+s}}{\Delta y}. 
\end{equation}
After some calculations one obtains the following intermediate results, where all Taylor series expansions
are carried out about the point $(x^p,y^q)$:
\begin{eqnarray}
   \partial^{p+\halb,q+\halb}_1 \left( A_{im}^{h} v_m^{h} \right) -\frac{1}{4} \sum \limits_{r=0}^{1} \sum \limits_{s=0}^{1} v_m^{p+r,q+s} \partial_1^{p+r,q+s} A_{im}^{h} &=&      
   A_{im}^{p+\halb,q+\halb} \cdot \halb \, \frac{v_m^{p+1,q+1}+v_m^{p+1,q}-v_m^{p,q+1}-v_m^{p,q}}{\Delta x} \nonumber \\
   && + \frac{1}{4} \left( A_{im}^{p+\halb,q+\frac{3}{2}} - A_{im}^{p+\halb,q+\halb} \right) 
   \frac{v_m^{p+1,q+1} - v_m^{p,q+1}  }{\Delta x} \nonumber \\ 
   && - \frac{1}{4} \left( A_{im}^{p+\halb,q+\frac{1}{2}} - A_{im}^{p+\halb,q-\halb} \right) 
   \frac{v_m^{p+1,q} - v_m^{p,q}  }{\Delta x} \nonumber \\ 
   &=& A_{im}^{p+\halb,q+\halb} \partial_1^{p+\halb,q+\halb} v^h_m + \frac{1}{4} \Delta y^2 \left( 
   \partial_y A_{im} \, \partial_{xy} v_m + \partial_{yy} A_{im} \, \partial_x v_m  \right), \nonumber \\    
   &=& A_{im}^{p+\halb,q+\halb} \partial_1^{p+\halb,q+\halb} v^h_m + \mathcal{O}\left(\Delta y^2\right),    
\end{eqnarray} 
\begin{eqnarray}
\partial^{p+\halb,q+\halb}_2 \left( A_{im}^{h} v_m^{h} \right) -\frac{1}{4} \sum \limits_{r=0}^{1} \sum \limits_{s=0}^{1} v_m^{p+r,q+s} \partial_2^{p+r,q+s} A_{im}^{h} &=&      
A_{im}^{p+\halb,q+\halb} \cdot \halb \, \frac{v_m^{p+1,q+1}+v_m^{p,q+1}-v_m^{p+1,q}-v_m^{p,q}}{\Delta y} \nonumber \\
&& + \frac{1}{4} \left( A_{im}^{p+\frac{3}{2},q+\halb} - A_{im}^{p+\halb,q+\halb} \right) 
\frac{v_m^{p+1,q+1} - v_m^{p+1,q}  }{\Delta y} \nonumber \\ 
&& - \frac{1}{4} \left( A_{im}^{p+\halb,q+\frac{1}{2}} - A_{im}^{p-\halb,q+\halb} \right) 
\frac{v_m^{p,q+1} - v_m^{p,q}  }{\Delta y} \nonumber \\ 
&=& A_{im}^{p+\halb,q+\halb} \partial_2^{p+\halb,q+\halb} v^h_m + \frac{1}{4} \Delta x^2 
\left( \partial_x A_{im} \, \partial_{xy} v_m + \partial_{xx} A_{im} \, \partial_y v_m\right), \nonumber \\ 
&=& A_{im}^{p+\halb,q+\halb} \partial_2^{p+\halb,q+\halb} v^h_m + \mathcal{O}\left(\Delta x^2\right),  
\end{eqnarray}
The convective term appears from the first term under the double sum expands as 
\begin{eqnarray}
   \frac{1}{4} \sum \limits_{r=0}^{1} \sum \limits_{s=0}^{1} \left( v_m^{p+r,q+s} \partial_m^{p+r,q+s} A_{ik}^{h} \right) &=& 
   \frac{1}{8} \sum \limits_{r=0}^{1} \sum \limits_{s=0}^{1} v_1^{p+r,q+s} \frac{ A_{im}^{p+\halb+r,q+\halb+s} + A_{im}^{p+\halb+r,q-\halb+s} 
   	-A_{im}^{p-\halb+r,q+\halb+s} - A_{im}^{p-\halb+r,q-\halb+s}}{\Delta x} + \nonumber \\ 
   &&  \frac{1}{8} \sum \limits_{r=0}^{1} \sum \limits_{s=0}^{1} v_2^{p+r,q+s} \frac{ A_{im}^{p+\halb+r,q+\halb+s} + A_{im}^{p-\halb+r,q+\halb+s} 
   	-A_{im}^{p+\halb+r,q-\halb+s} - A_{im}^{p-\halb+r,q-\halb+s}}{\Delta y} \nonumber \\ 
   &=& v_m^{p+\halb,q+\halb} \partial_m^{p+\halb,q+\halb} A_{ik}^h + \frac{1}{8} \Delta x^2 \left( 
   2 \partial_{xx} A_{ik} \partial_x v_1 + 2 \partial_{xy} A_{ik} \partial_x v_2  + \partial_x A_{ik} \partial_{xx} v_1 + \partial_y A_{ik} \partial_{xx} v_2 \right) \nonumber \\ 
   && \phantom{v_m^{p+\halb,q+\halb} \partial_m^{p+\halb,q+\halb} A_{ik}^h}  
   + \frac{1}{8} \Delta y^2 \left( 
   2 \partial_{xy} A_{ik} \partial_y v_1 + 2 \partial_{yy} A_{ik} \partial_y v_2  + \partial_x A_{ik} \partial_{yy} v_1 + \partial_y A_{ik} \partial_{yy} v_2 \right) \nonumber \\ 
   &=& v_m^{p+\halb,q+\halb} \partial_m^{p+\halb,q+\halb} A_{ik}^h + \mathcal{O}\left( \Delta x^2, \Delta y^2 \right).
\end{eqnarray}

Combining the previous results and rearranging terms, one can finally rewrite \eqref{eqn.semi.Ah} as  follows: 
\begin{equation}
\partial_t A_{ik}^{p+\halb,q+\halb} + v^{p+\halb,q+\halb}_m \partial_m^{p+\halb,q+\halb} A_{ik}^{h} 
 + A_{im}^{p+\halb,q+\halb} \partial^{p+\halb,q+\halb}_k   v_m^{h} + \mathcal{O}(\Delta x^2, \Delta y^2) =   
 - \frac{\left| A_{ij}^{p+\halb,q+\halb} \right|^{\frac{5}{3}}}{\tau_1}  
A_{ij}^{p+\halb,q+\halb} \mathring{G}_{jk}^{p+\halb,q+\halb}. 
\label{eqn.semi.Ah.final}
\end{equation}
Using the definition of the material derivative $D q/Dt = \partial q/\partial t + \mathbf{v} \cdot \nabla q$ we can rewrite the previous equation in the following semi-discrete form based on the material derivative:  
\begin{equation}
\frac{D}{Dt} A_{ik}^{p+\halb,q+\halb}  
+ A_{im}^{p+\halb,q+\halb} \partial^{p+\halb,q+\halb}_k   v_m^{h} + \mathcal{O}(\Delta x^2, \Delta y^2) =  
- \frac{\left| A_{ij}^{p+\halb,q+\halb} \right|^{\frac{5}{3}}}{\tau_1}  
A_{ij}^{p+\halb,q+\halb} \mathring{G}_{jk}^{p+\halb,q+\halb}. 
\label{eqn.semiL.Ah}
\end{equation}
Multiplication of the above equation with $\mathbf{A}^T$ from the left and adding the transposed equation multiplied by $\mathbf{A}$ from the right yields the semi-discrete Lagrangian evolution equation for the metric tensor $\mathbf{G}$: 
\begin{equation}
\frac{D}{Dt} G_{ik}^{p+\halb,q+\halb}  
+ G_{im}^{p+\halb,q+\halb} \partial^{p+\halb,q+\halb}_k   v_m^{h} 
+ G_{mk}^{p+\halb,q+\halb} \partial^{p+\halb,q+\halb}_i   v_m^{h} 
+ \mathcal{O}(\Delta x^2, \Delta y^2) =  
- 2 \frac{\left| A_{ij}^{p+\halb,q+\halb} \right|^{\frac{5}{3}}}{\tau_1}  
G_{ij}^{p+\halb,q+\halb} \mathring{G}_{jk}^{p+\halb,q+\halb}. 
\label{eqn.semiL.Gh}
\end{equation}
Formal asymptotic expansion of the metric tensor in terms of $\tau_1$ provides the ansatz 
\begin{eqnarray}
G_{ij}^{p+\halb,q+\halb} = G_{ij,(0)}^{p+\halb,q+\halb} + \tau_1 G_{ij,(1)}^{p+\halb,q+\halb} + \cdots 
\label{eqn.Gh.expansion}
\end{eqnarray}
which can now be inserted into \eqref{eqn.semiL.Gh}. The leading order term $\tau_1^{-1}$ leads to 
the first result  
\begin{equation}
 \mathring{G}_{ij,(0)}^{p+\halb,q+\halb} = 0, 
\end{equation}
i.e. at leading order the discrete metric tensor $\mathbf{G}^{p+\halb,q+\halb}$ becomes trace-free. 
Applying the ``$\textnormal{dev}$'' operator, with $\textnormal{dev}(\mathbf{G}) = \mathbf{G} - 
\frac{1}{3} \textnormal{tr} ( \mathbf{G}) \mathbf{I}$, to \eqref{eqn.semiL.Gh} and repeating the 
calculations 
already presented in \cite{GPRmodel} for the continuous case, leads to the following asymptotic 
result for the discrete stress tensor in the absence of heat conduction ($\alpha=0$), from which we 
can conclude that for $\tau_1 \to 0$ the compressible Navier-Stokes stress tensor is retrieved up 
to second order of accuracy in space: 
%
\begin{eqnarray}
\sigma_{ik}^{p+\halb,q+\halb} = \rho c_s^2 G_{ij}^{p+\halb,q+\halb} \mathring{G}_{jk}^{p+\halb,q+\halb} = \frac{1}{6} \rho_0 c_s^2 \tau_1 
\left( \partial_i^{p+\halb,q+\halb} v^h_k + \partial_k^{p+\halb,q+\halb} v^h_i 
- \frac{2}{3} \delta_{ik} \, \partial_m^{p+\halb,q+\halb} v^h_m  \right)  + \mathcal{O}(\Delta x^2, \Delta y^2). 
\label{eqn.sigma.final}
\end{eqnarray}
Note that in \eqref{eqn.sigma.final} the gradient of the velocity field is computed in the vertices of the primary control volumes (corner gradient), which is a very common choice for the discretization of the compressible Navier-Stokes equations.  

Unfortunately, the second order error terms $\mathcal{O}(\Delta x^2, \Delta y^2)$ remain in the asymptotic relaxation limit of the discrete stress tensor \eqref{eqn.sigma.final} and do \textit{not} vanish for $\tau_1 \to 0$.  Hence, the scheme is \textit{not} rigorously asymptotic  preserving (AP) for the viscous stress when $\tau_1 \to 0$, in contrast to the discrete heat flux discussed in the previous section, which reduces to a perfect discrete analogue of the Fourier law for $\tau_2 \to 0$. Instead, for $\boldsymbol{\sigma}$ the method proposed in this paper is only \textit{quasi} asymptotic preserving, up to the second order error terms in \eqref{eqn.sigma.final}. 
However, when inserting the discrete stress tensor \eqref{eqn.sigma.final} into the discrete momentum equation \eqref{eqn.rhou2d} with \eqref{eqn.Qstar} the errors $\mathcal{O}(\Delta x^2, \Delta y^2)$ will add to the numerical errors already made in the discretization of the nonlinear convective terms, see \eqref{eqn.Qstar}, which, in general, are of second order in space and time when a second order MUSCL-Hancock-type upwind scheme is used. Since the numerical errors of the discretization of the nonlinear convective terms also do \textit{not} scale with $\tau_1$, overall the above result can still be considered as satisfactory for our practical purposes, despite its obvious limitations and shortcomings.  

The difference in the asymptotic behaviour of the stress tensor compared to the one of the heat 
flux is mainly due to the highly nonlinear nature of the equation for $\mathbf{A}$ due to the 
nonlinearity of its relaxation source term, compared to the simple quasi linear relaxation source 
term in the equation for $\mathbf{J}$. 
Further research on this topic needs to be carried out in the future, investigating the possibility to find 
a compatible curl-free discretization of the PDE for $\textbf{A}$ that is also rigorously asymptotic preserving with the Navier-Stokes limit of $\boldsymbol{\sigma}$.

\subsection{Behavior of the scheme at low Mach numbers}
\label{sec.lowMach} 

Assuming no viscosity and no heat conduction (e.g. by simply setting $c_s = \alpha = 0$ and thus 
$E_2=0$), in the low Mach number limit the pressure tends to a constant in space and the velocity 
field will asymptotically satisfy the classical divergence-free condition of incompressible 
inviscid flows, see 
\cite{KlaMaj,KlaMaj82,MunzDumbserRoller}. As a result of the divergence-free condition of the velocity field,
the mass conservation equation can be rewritten as
\begin{equation}
   \frac{\partial \rho}{\partial t} + \mathbf{v} \cdot \nabla \rho = 0,  
\end{equation}
which is a simple scalar transport equation for the density. If the initial condition satisfies 
$\rho(\mathbf{x},0) = \rho_0 = \textnormal{const}$ then $\rho(\mathbf{x},t) = \rho_0 = 
\textnormal{const}$ 
for all times. 
Assuming therefore $\rho$ and $p$ constant in space and time at the leading order and the contribution of the kinetic energy negligible, 
we have $h = \textnormal{const}$ and $\rho E_1 = \textnormal{const}$ and therefore with $E_2=0$ the 
pressure equation \eqref{eqn.p2d} tends to the classical pressure Poisson equations for incompressible flows:  
\begin{eqnarray}
- \frac{\Delta t^2}{\Delta x^2} \left(  
\left( p_{}^{p+1,q,n+1}-p_{}^{p,q,n+1} \right) -  
\left( p_{}^{p,q,n+1}-p_{}^{p-1,q,n+1} \right) \right)  
- \frac{\Delta t^2}{\Delta y^2}\left( 
\left( p_{}^{p,q+1,n+1}-p_{}^{p,q,n+1} \right) - 
\left( p_{}^{p,q,n+1}-p_{}^{p,q-1,n+1} \right) \right)  
 =  && \nonumber \\ 
 - \frac{\Delta t}{\Delta x} \left( 
 (\rho v)_{1}^{p+\halb,q,*} - 
 (\rho v)_{1}^{p-\halb,q,*} \right)   
- \frac{\Delta t}{\Delta y} \left(  (\rho v)_{2}^{p,q+\halb,*} - (\rho v)_{2}^{p,q-\halb,*} \right). && 
\end{eqnarray}  
This behavior was expected, since the method presented in this paper is a staggered pressure-based scheme.

\section{Numerical results}
\label{sec.results} 

In all the following test problems, whenever a viscosity coefficient $\mu$ is specified together with a shear  sound speed $c_s$ the corresponding relaxation time $\tau_1$ in the GPR model is computed from \eqref{eqn.mu.kappa} as $\tau_1 = 6 \mu / (\rho_0 c_s^2)$, according to the results of the asymptotic  
analysis carried out in  \cite{GPRmodel} for $ E_2(A_{ik},J_k) $ given by \eqref{eqn.E2}. 
If another form of the energy $ E_2 $ is used the expression for the viscosity may change, while staying of the form $ \mu\sim\tau c_s^2 $. 

\subsection{Solid rotor problem and numerical verification of the discrete curl-free property} 

\begin{figure}[!htbp]
	\begin{center}
		\begin{tabular}{cc} 
			\includegraphics[width=0.47\textwidth]{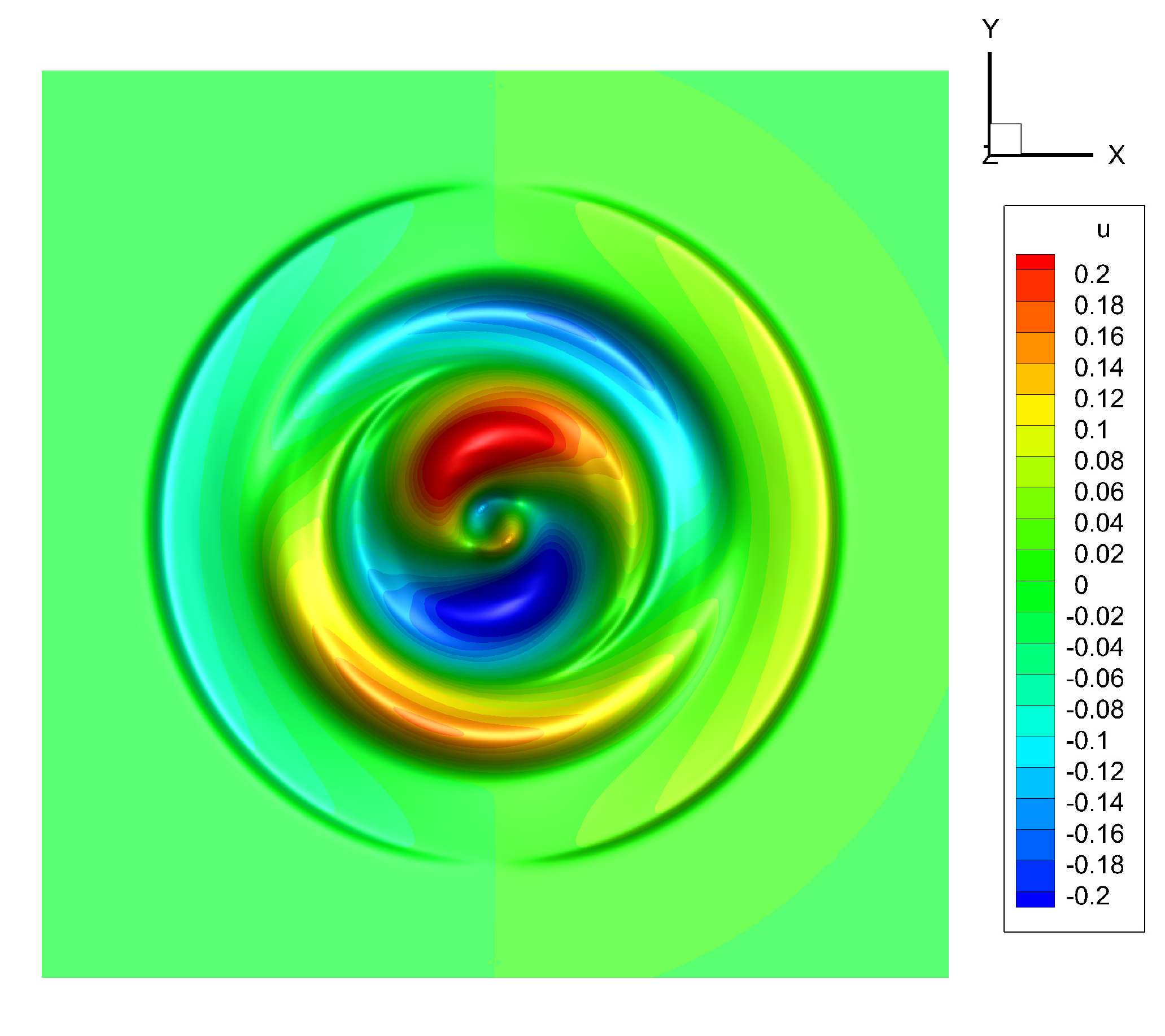}  &
            \includegraphics[width=0.47\textwidth]{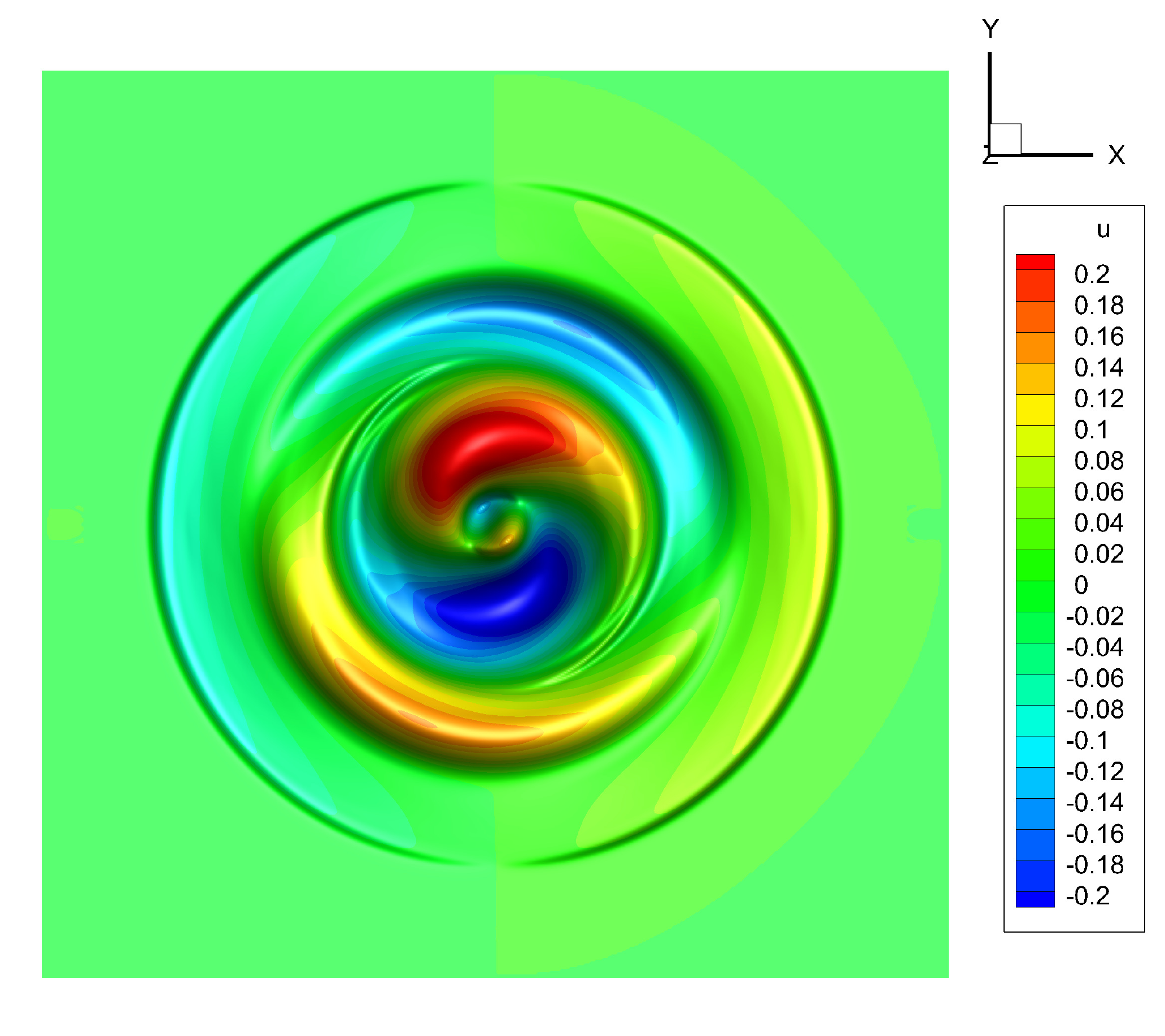}  \\ 
			\includegraphics[width=0.47\textwidth]{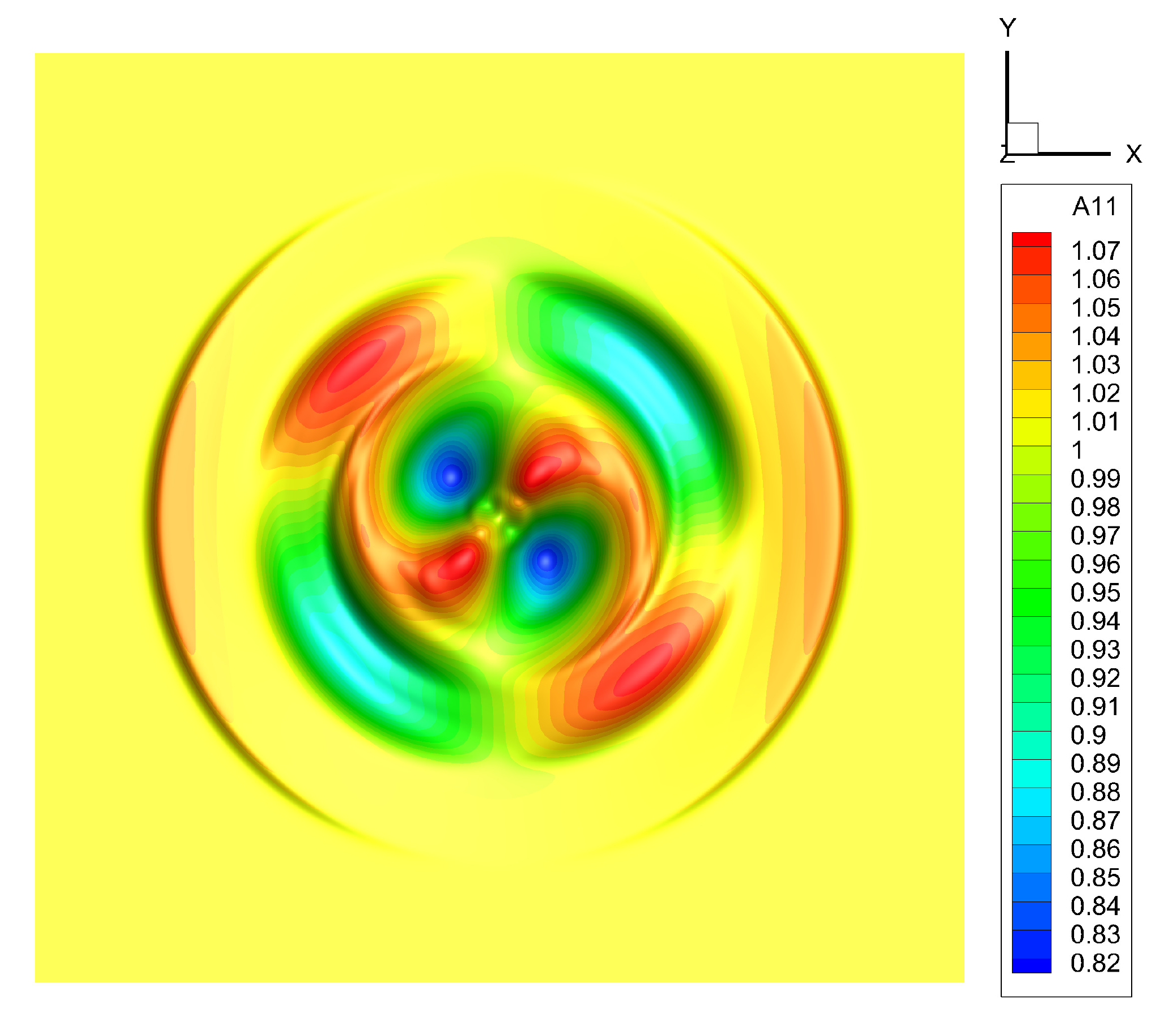}  &
			\includegraphics[width=0.47\textwidth]{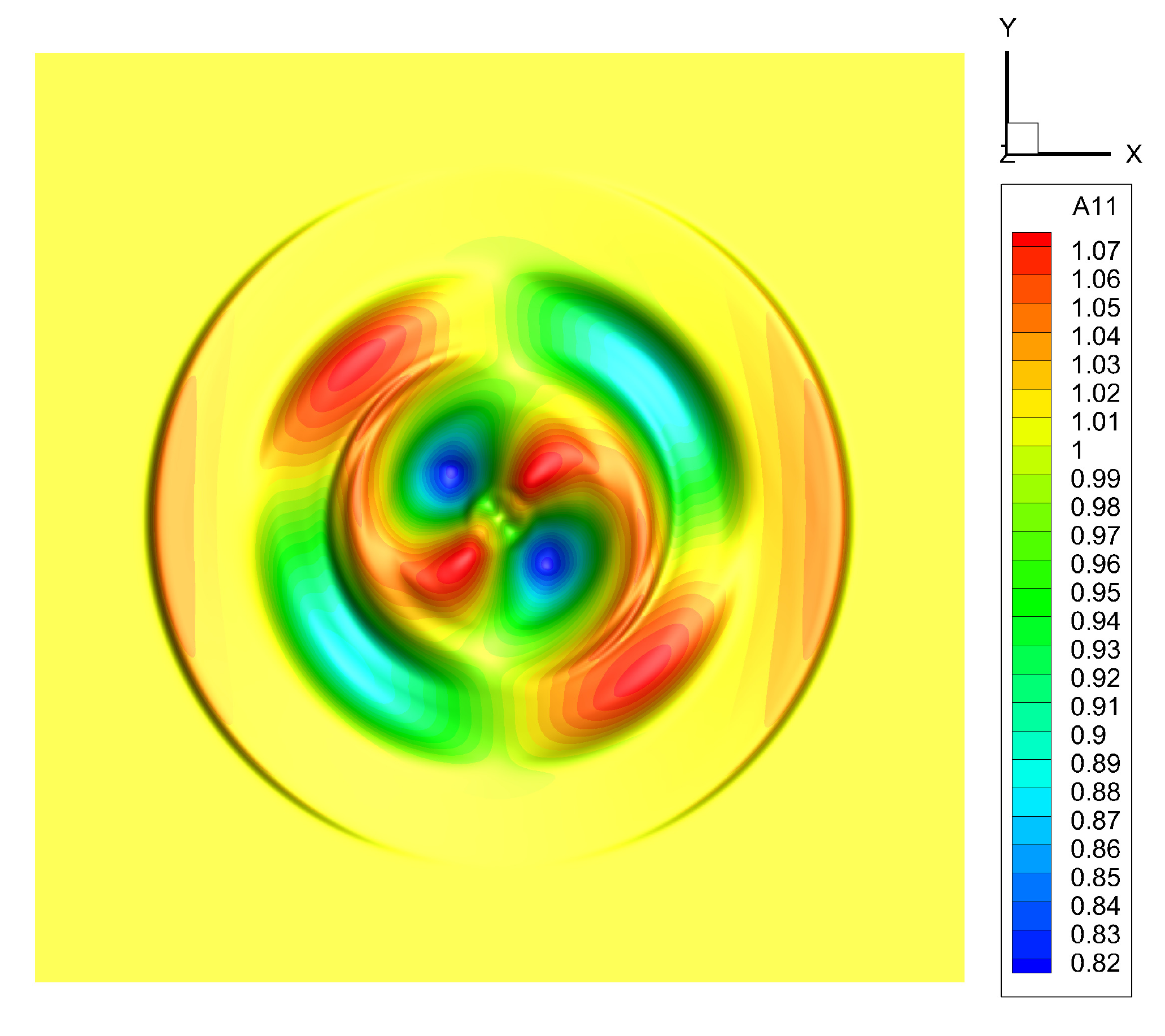}   
		\end{tabular} 
		\caption{2D solid rotor problem for the homogeneous GPR model with $\tau_1 = \tau_2 = 10^{20}$ at time $t=0.30$. Color contours
			for the velocity component $u$ (top) and the distortion field component $A_{11}$ (bottom) computed with a classical explicit
			MUSCL TVD scheme (right) and with the new structure-preserving semi-implicit FV scheme (left).    } 
		\label{fig.solidrotor}
	\end{center}
\end{figure}

\begin{figure}[!htbp]
	\begin{center}
		\includegraphics[width=0.7\textwidth]{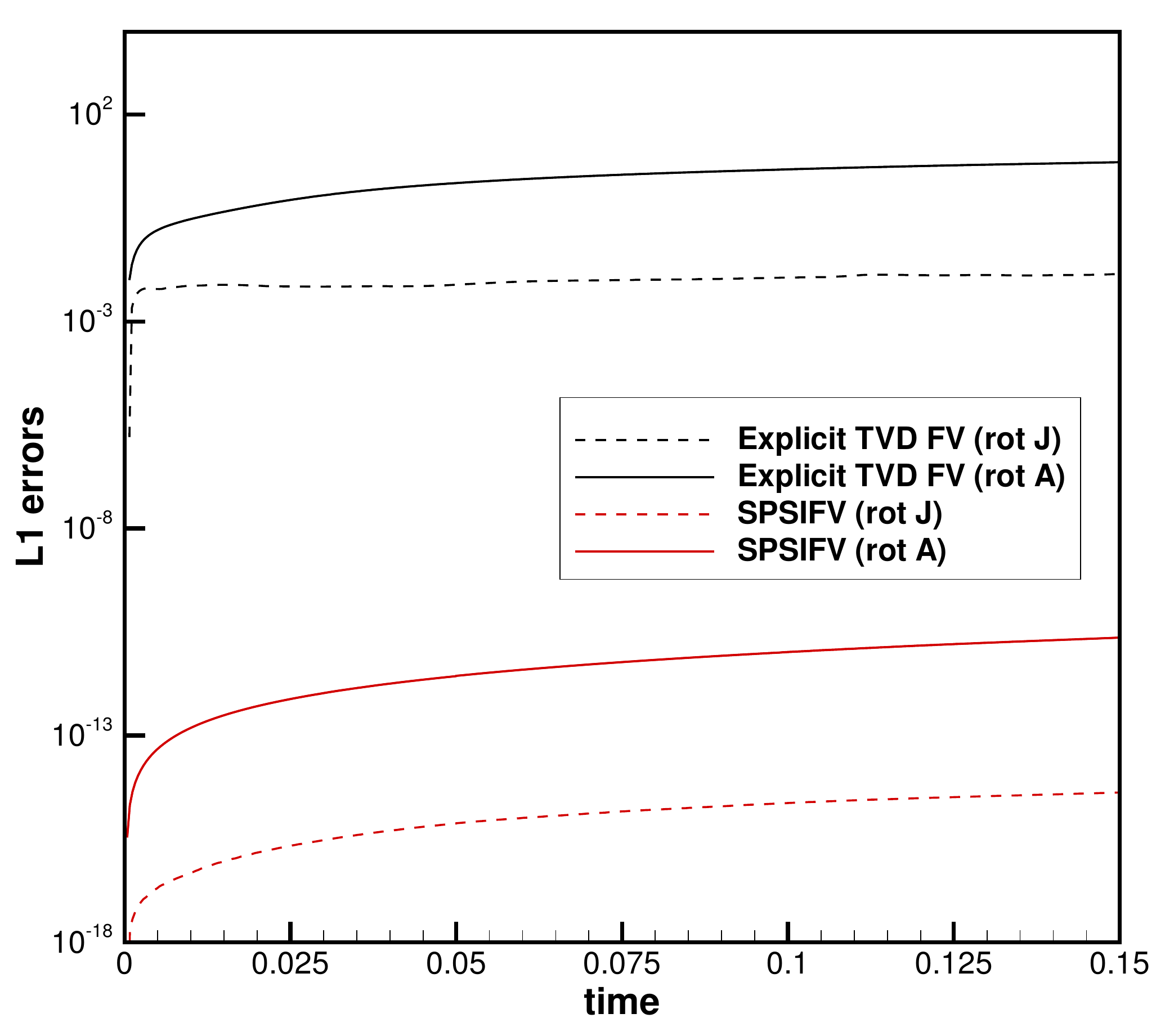}  
		\caption{Time series of the $L_1$ error norms of the curl of $\AAA$ and $\mathbf{J}$ for the 2D solid rotor problem until time $t=0.3$ using a standard second order MUSCL-Hancock-type TVD finite volume scheme (black) and the new structure-preserving  semi-implicit finite volume scheme (red). The new structure-preserving method is able to preserve all curl-free conditions of the  GPR model essentially up to machine precision.   } 
		\label{fig.solidrotor.curlerrors}
	\end{center}
\end{figure}

In this first test problem we carry out a numerical verification of the discrete curl-free property of the new structure-preserving semi-implicit finite volume scheme proposed in this paper. The main objective here is to check whether the practical implementation of the new method is correct and if the scheme properly achieves the curl-free property that was proven in the previous section \ref{sec.analysis}. 
For this purpose we solve the homogeneous system \eqref{eqn.GPR} by setting $\tau_1 = \tau_2 = 
10^{20}$ which corresponds to a hyperelastic solid. The computational domain  
is chosen as $\Omega = [-1,+1]^2$ with periodic boundary conditions everywhere. The initial data is given by 
$\rho = 1$, $p = 1$, $\mathbf{A} = \mathbf{I}$ and $\mathbf{J}=0$. The initial condition for the velocity field is 
chosen as $u = -y/R$, $v = +x/R$ and $w=0$ within the circular region $r \leq R$, where $r = \left\| \mathbf{x} \right\|_2$ and $R=0.2$, while $\mathbf{v}=0$ for $r > R$. The remaining parameters of the model are chosen as $\gamma = 1.4$, $c_s = 1.0$ and
$\alpha = 0.5$. The setup of the test problem is similar, but not identical, to the one proposed in 
\cite{Dumbser2008}.  We solve the problem with the new structure-preserving semi-implicit finite 
volume scheme until $t=0.3$ on a grid  composed of $500 \times 500$ elements. For comparison, the 
same problem is also solved on the same grid with a standard second-order accurate explicit 
MUSCL-Hancock finite volume scheme, see \cite{toro-book} for details. The results are compared 
with each other in Fig.,\ref{fig.solidrotor},  where the contour colors of the velocity component 
$u$ and the distortion field  component $A_{11}$ are shown. In the color contours of 
Fig.\,\ref{fig.solidrotor}, there are essentially no differences between 
the two solutions. However, if we compare the time evolution of the $L_1$ errors of the curl of $\mathbf{A}$ and $\mathbf{J}$ of the two schemes, 
see Fig.\,\ref{fig.solidrotor.curlerrors}, we observe that the SPSIFV scheme is able to maintain 
the curl errors close to  
machine zero, while the $L_1$ norms of the curl errors produced by the standard MUSCL-Hancock method are more than ten orders 
of magnitude larger. 
These results for a non-trivial test case exhibiting all types of waves (shear and pressure waves) 
confirm the correct implementation of the new SPSIFV method, for which the curl-free property has 
been proven
in the previous section \ref{sec.analysis}. 

\subsection{Simple shear motion in solids and fluids} 
\label{sec.shearlayer} 

In this section we verify the new structure-preserving scheme for simple shear motion of fluids and solids. 
For this test we use the computational domain $\Omega = [-0.5;+0.5]^2$, with periodic boundary conditions in $y$ direction and 
fixing the initial condition at the boundaries in $x$ direction. The initial condition of the problem is given by  
$\rho=1$, $u=0$, $p=1/\gamma$, $\AAA=\mathbf{I}$, $\mathbf{J}=0$, while the velocity component $v$ is $v=-v_0$ for $x<0$
and $v=+v_0$ for $x\geq 0$. The parameters of this test are $v_0=0.1$, $\gamma=1.4$, $c_v=1$, $\rho_0=1$, $c_s=1$ 
and $\alpha=0$. The simulations are performed with the SPSIFV scheme on a grid composed of $1000 
\times 20$ control 
volumes up to a final time of $t=0.4$. 
In fluid mechanics, such an isolated, unsteady and infinitely long shear layer is also known as the first problem of Stokes, 
which admits an exact solution of the incompressible Navier-Stokes equations, see e.g. \cite{BLTheory}.  
The exact solution of the incompressible Navier-Stokes equations for the velocity component $v$ 
is given by  
\begin{equation}
\label{eqn.stokes} 
v(x,t) = v_0 \, \textnormal{erf}\left( \halb \frac{x}{\sqrt{\nu t}} \right),  
\end{equation} 
with $\nu = \mu / \rho_0$ and which can be used as a reference solution for the GPR model in the stiff relaxation limit.  
For solid materials, this initial data leads to two shear waves traveling outward with speed $c_s$. 

\begin{figure}[!htbp]
	\begin{center}
		\begin{tabular}{cc} 
			\includegraphics[width=0.45\textwidth]{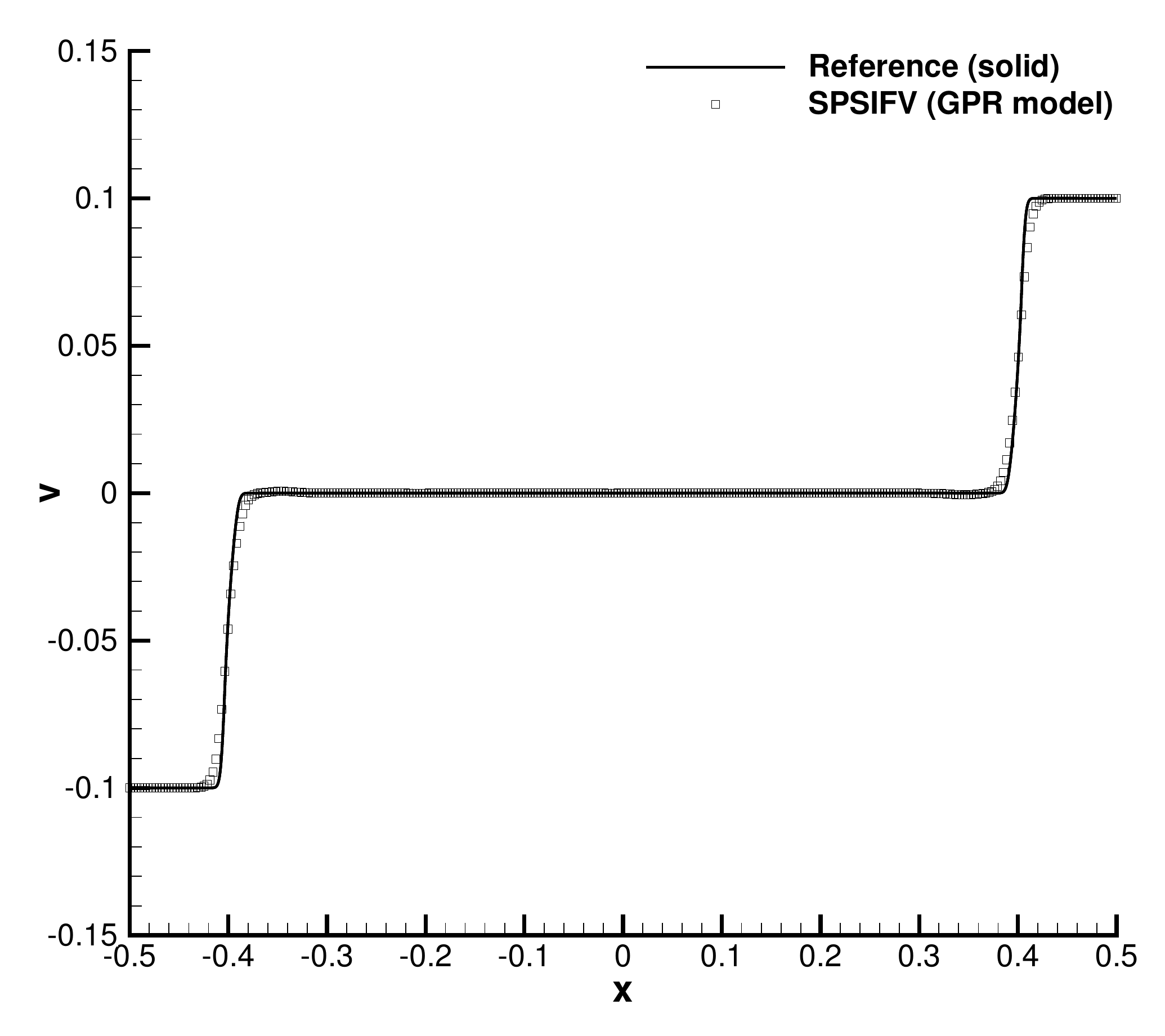}   & 
			\includegraphics[width=0.45\textwidth]{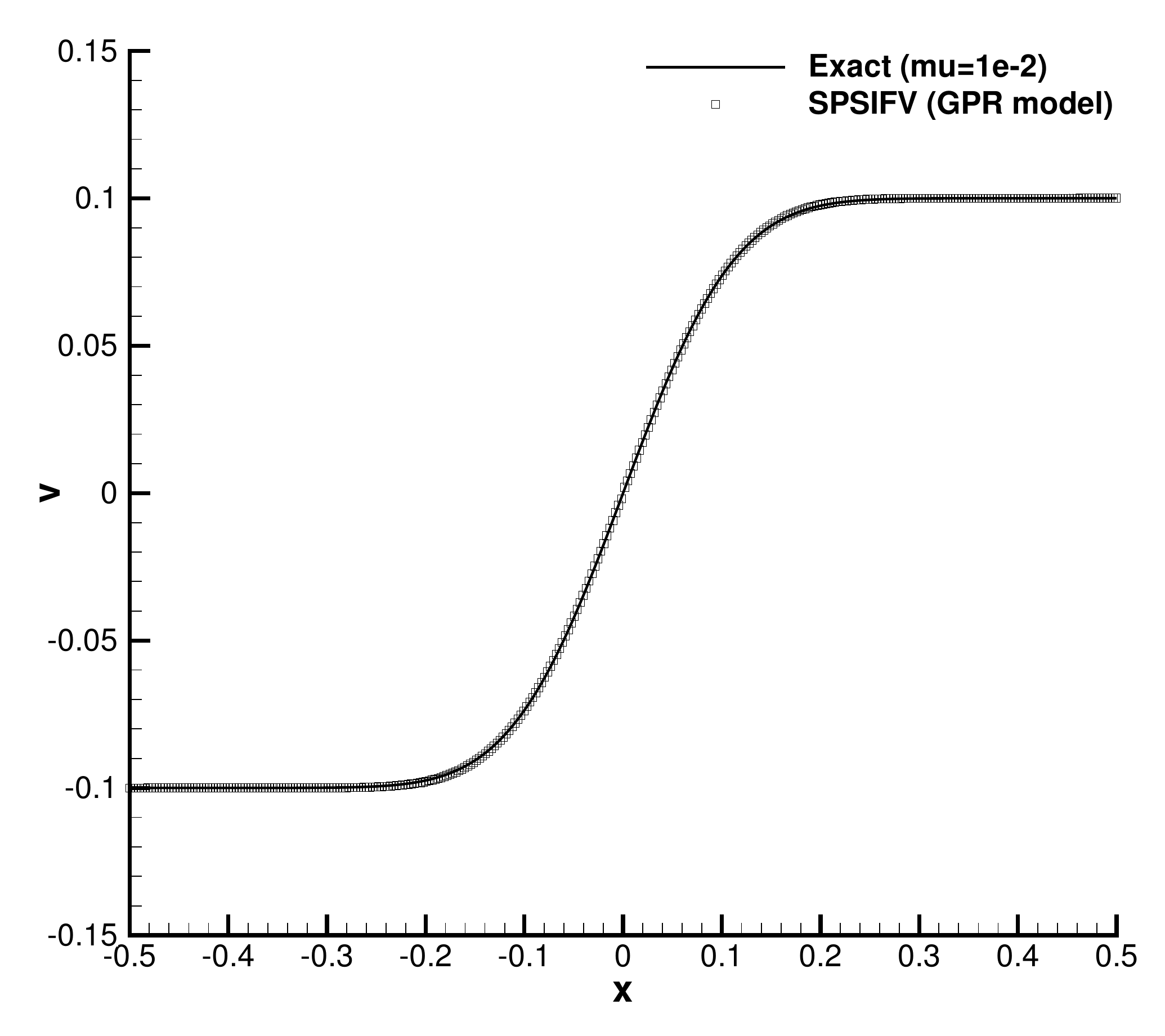}  \\  
			\includegraphics[width=0.45\textwidth]{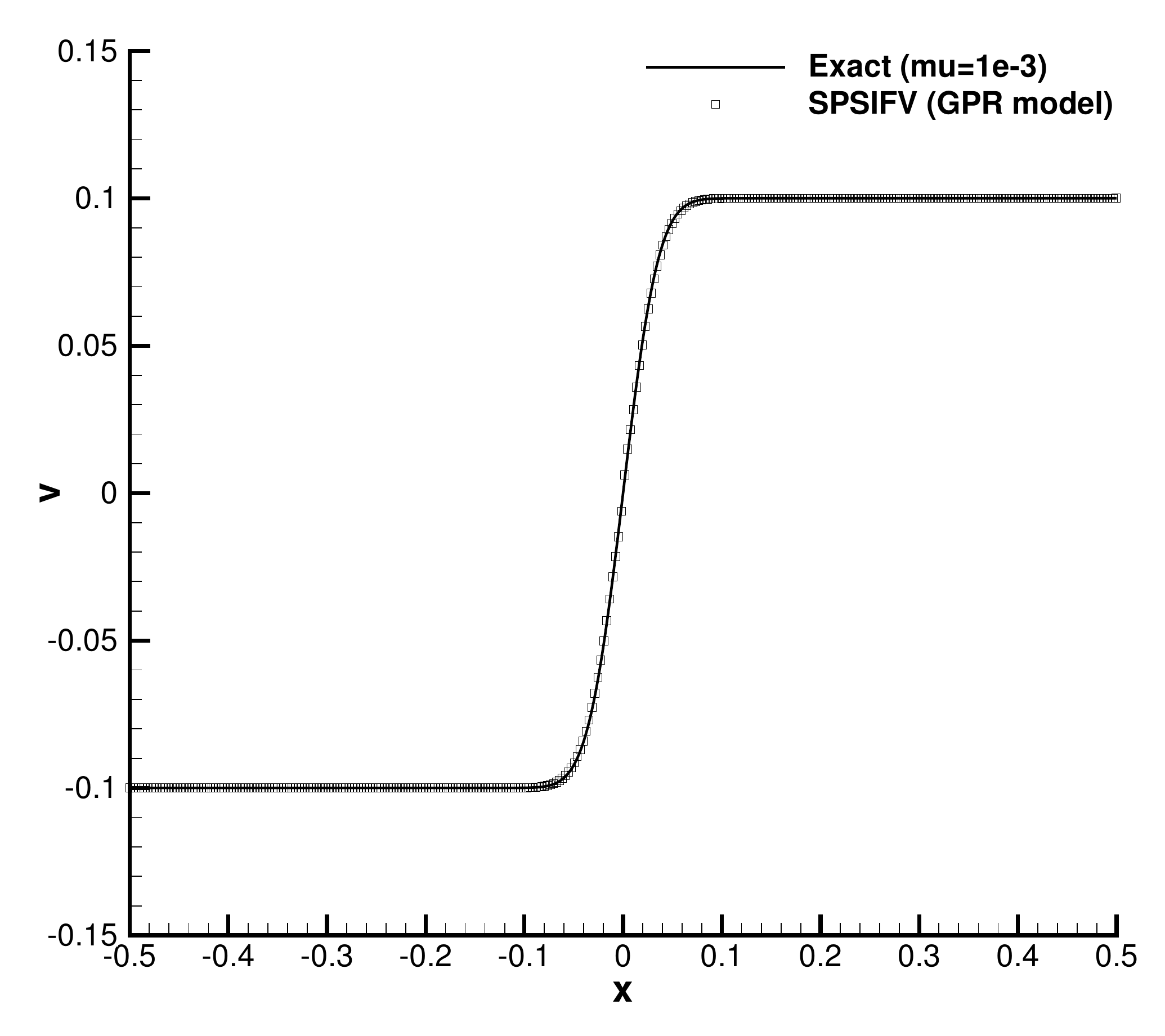}   & 
			\includegraphics[width=0.45\textwidth]{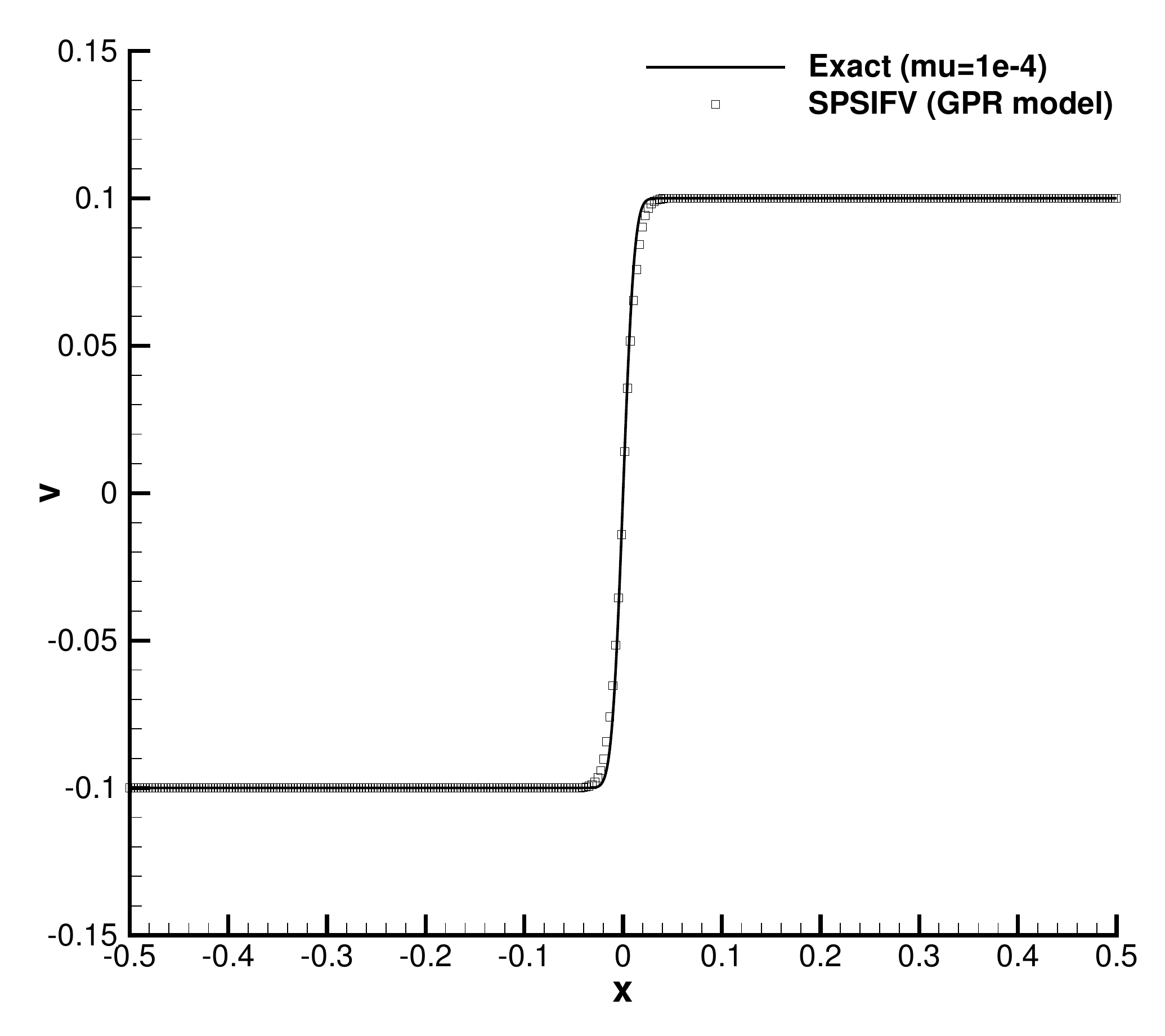}  
		\end{tabular} 
		\caption{Numerical solution obtained with the structure-preserving semi-implicit finite volume (SPSIFV) scheme for the GPR model for a simple shear flow in fluids and in an elastic solid at time $t=0.4$. Results for the solid (top left) and for fluids with different viscosities: $\mu=10^{-2}$ (top right), $\mu=10^{-3}$ (bottom left) and $\mu=10^{-4}$ (bottom right). 
		For fluids, this test corresponds to the first problem of Stokes, which has an exact solution.  } 
		\label{fig.shear}
	\end{center}
\end{figure}

The comparison between the Navier-Stokes reference solution \eqref{eqn.stokes} and the numerical results obtained with the new 
SPSIFV scheme for the GPR model are presented in Fig. \ref{fig.shear}, where one can observe an excellent agreement between the 
two for various viscosities $\mu$. In the same Figure \ref{fig.shear} we also present numerical results and a reference solution 
for the case of an ideal elastic solid, i.e. for the case when $\tau_1 \to \infty$ and thus the strain relaxation source term 
vanishes. Also in this case we obtain a very good agreement between the numerical solution and the reference solution, 
which was computed with an explicit second order MUSCL TVD finite volume scheme on a very fine mesh of 5000 cells.

\subsection{Riemann problems for nearly inviscid fluids and ideal elastic solids} 
\label{sec.RP} 

In this section we solve a set of one-dimensional Riemann problems in the fluid and solid limit of the GPR
model, i.e. for $\tau \to 0$ and $\tau \to \infty$, respectively. In the case of the inviscid fluid limit 
($\tau \to 0$) the source terms in the PDE for $\AAA$ and $\J$ become stiff and therefore an implicit treatment 
is mandatory. In this limit of the GPR model, the reference solution is given  by the exact solution of the 
Riemann problem of the Euler equations of compressible gas dynamics, see \cite{toro-book} for details.  
In the case of an ideal elastic solid ($\tau \to \infty$), the governing PDE system becomes homogeneous and 
we simply compute a reference solution by solving the governing PDE system with a classical explicit second 
order accurate shock capturing TVD finite volume scheme on a very fine mesh composed of 10000 elements. While 
the Riemann problem of the 1D Euler equations contains only five waves (two acoustic waves, one $x-y$ shear wave,
one $x-z$ shear wave and one entropy wave), 
the homogeneous GPR model contains much more waves (two fast thermo-acoustic 
waves, two slow thermo-acoustic waves, 
two $x-y$ shear waves, two $x-z$ shear waves, one entropy wave, and a set of waves associated with advection at speed $u$, see 
Appendix\,\ref{sec.appendix.sound.speed}). 
For the discussion of Riemann solvers and the exact and approximate solution of the Riemann problem for 
nonlinear hyperelasticity, the reader is referred to \cite{Bartonetal,TitarevRomenskiToro}. 

The setup of all Riemann problems is described in the following. The computational domain is given by 
$\Omega = [-0.5,+0.5]^2$ and is discretized with a computational grid of $1000 \times 20$ elements. We use periodic 
boundary conditions in $y$ direction and impose the initial condition as Dirichlet boundary condition in the $x$ 
direction. 
The parameters of the GPR model are set to $\rho_0 = 1$, $\gamma = 1.4$ and $c_s = \alpha = 1$ for 
all Riemann problems 
apart from RP2, where we use $c_s = \alpha = 2$. The initial data for density, pressure and 
velocity are given in 
Table \ref{tab.ic.ideal}, where also the relaxation times $\tau_1$ and $\tau_2$ are provided. The remaining state 
variables are set to $w=0$, $\AAA = \sqrt[3]{\rho} \mathbf{I}$ and $\J = 0$. 

\begin{figure}[!htbp]
	\begin{center}
		\includegraphics[width=0.32\textwidth]{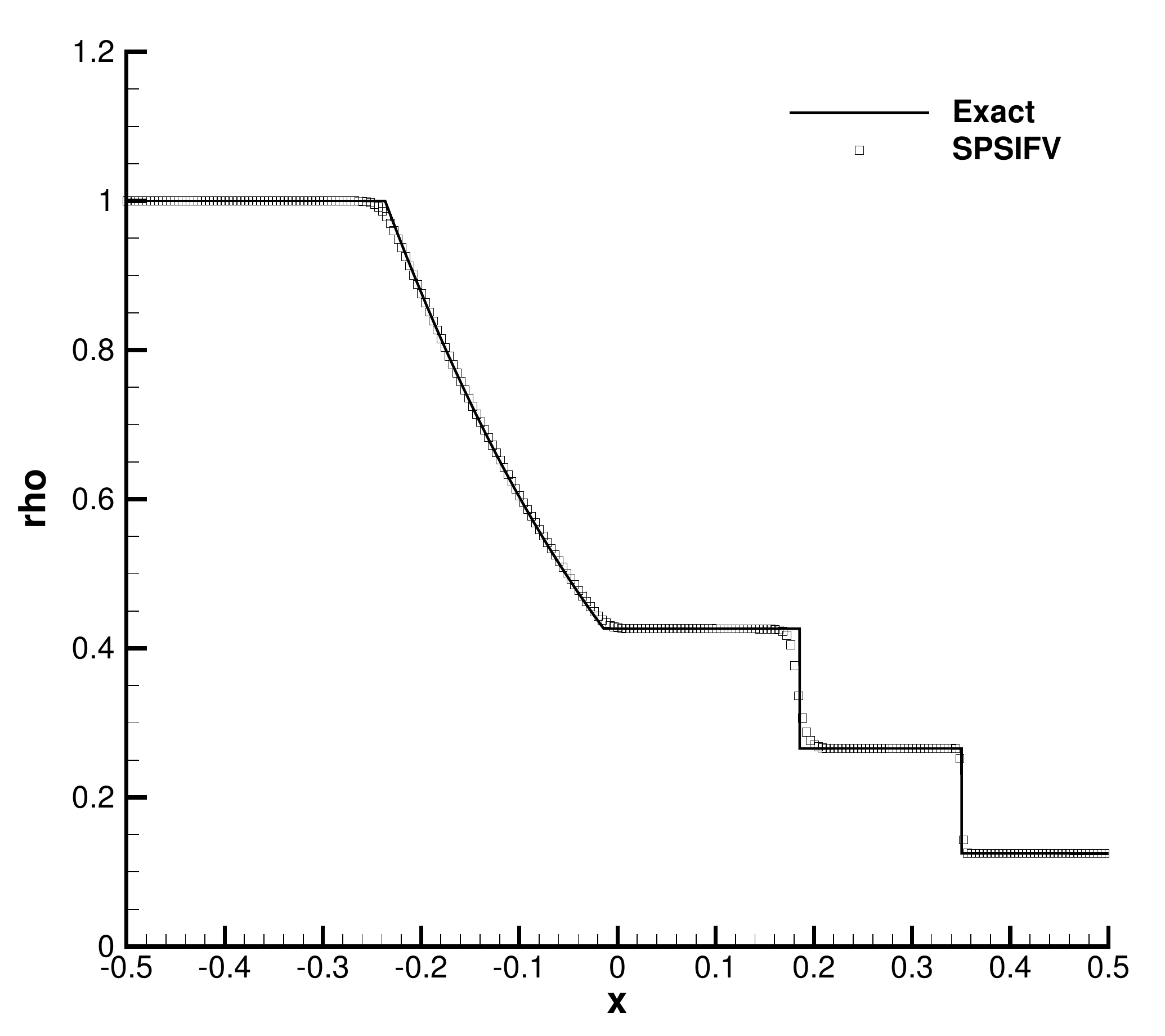}   
		\includegraphics[width=0.32\textwidth]{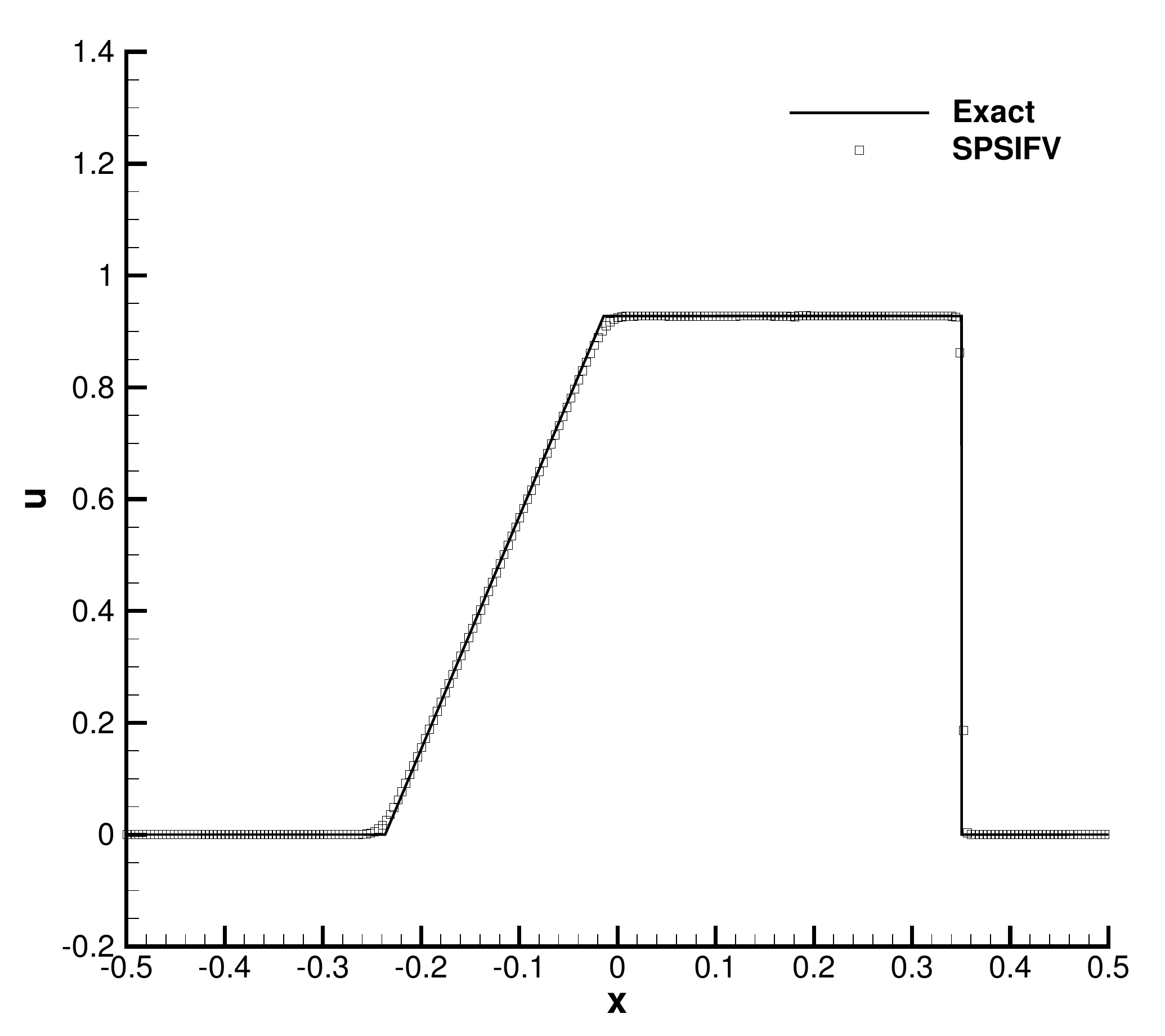}   
		\includegraphics[width=0.32\textwidth]{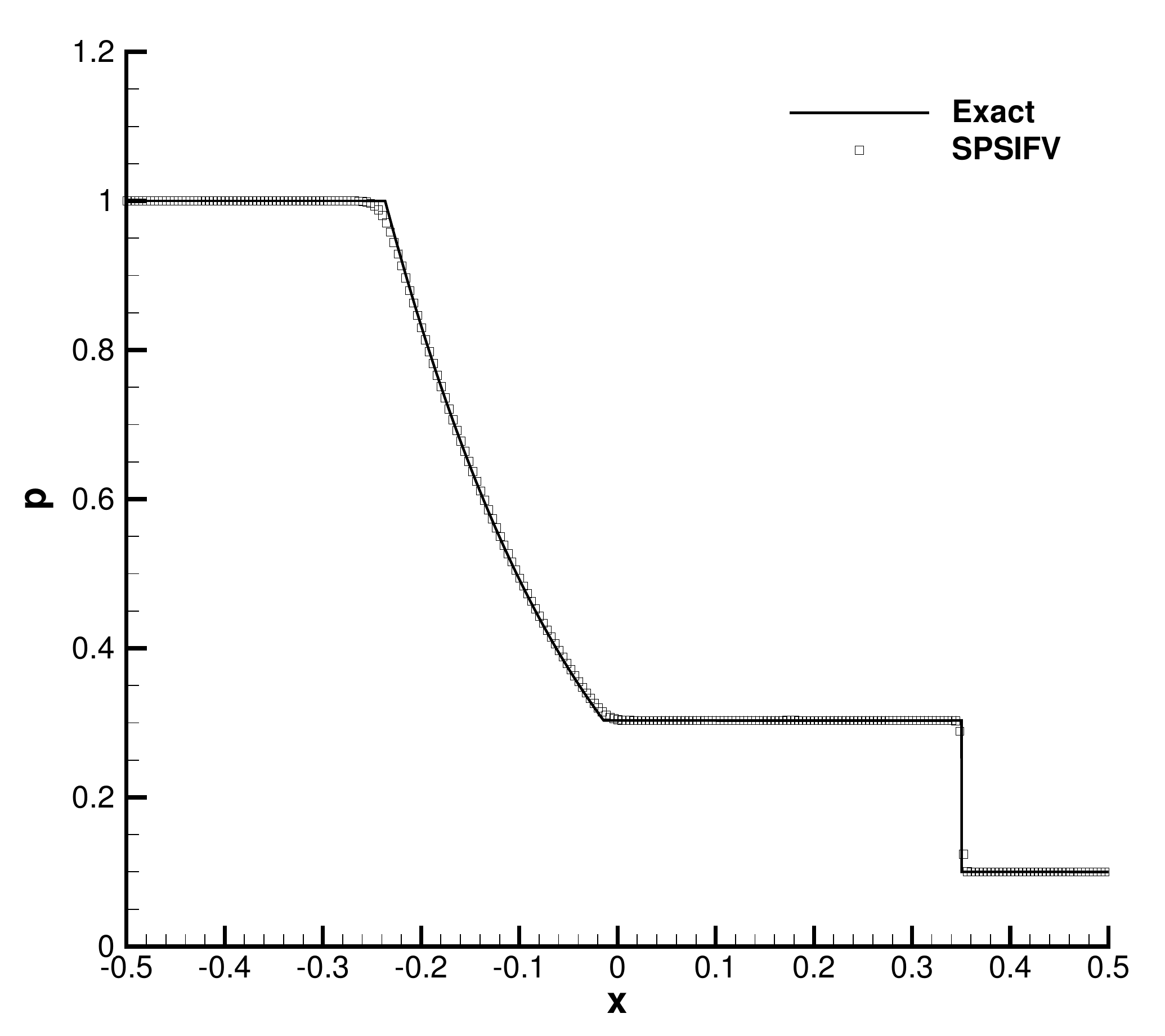}   
		\caption{Exact solution of the Euler equations and numerical solution of the GPR model in the stiff relaxation limit ($\tau_1 = \tau_2 = 10^{-6}$) for Riemann problem RP1 (Sod shock tube). The density $\rho$, the velocity component $u$ and the pressure $p$ are shown at a final time of $t=0.2$.} 
		\label{fig.rp1}
	\end{center}
\end{figure}

\begin{figure}[!htbp]
	\begin{center}
		\includegraphics[width=0.32\textwidth]{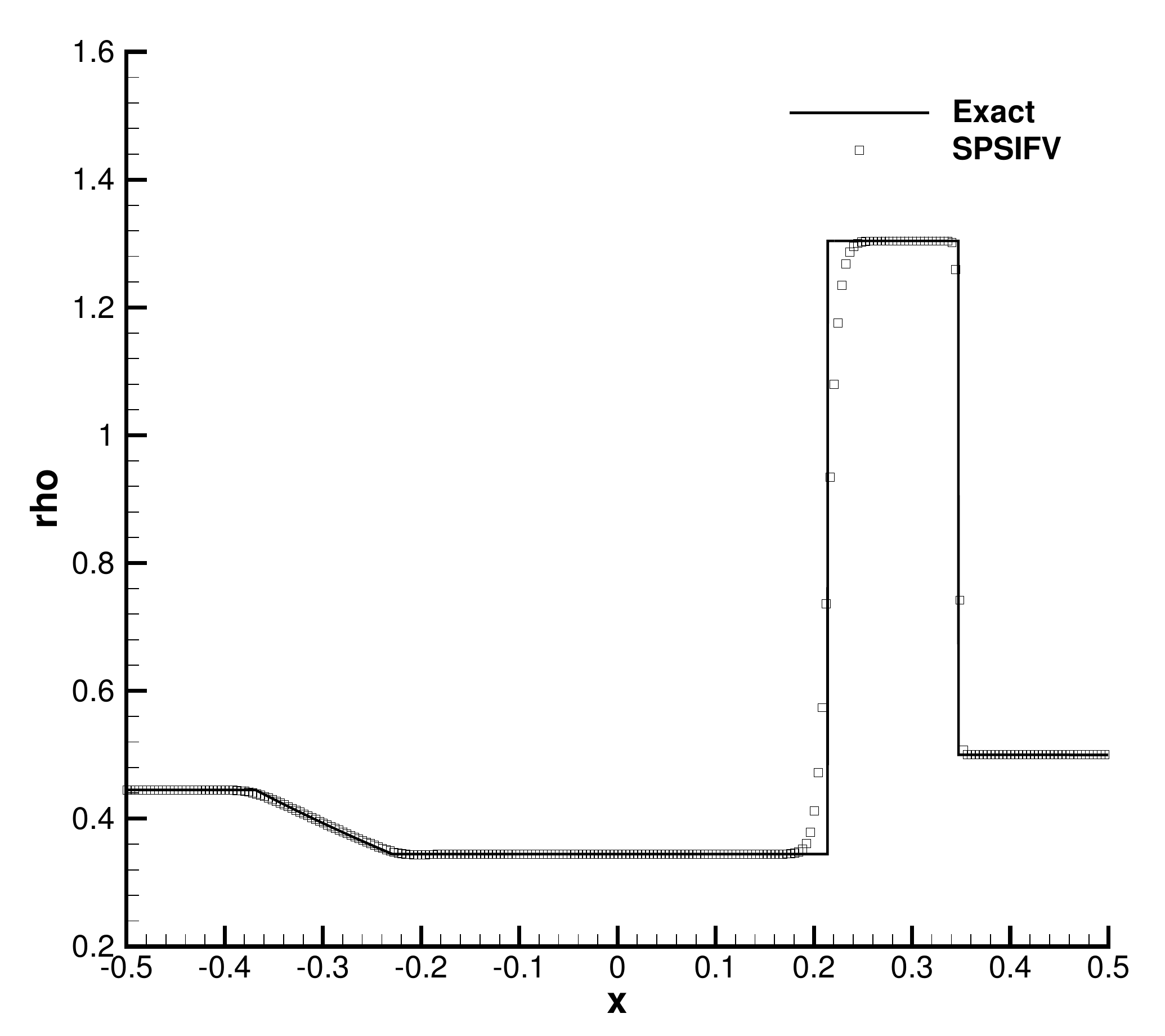}   
		\includegraphics[width=0.32\textwidth]{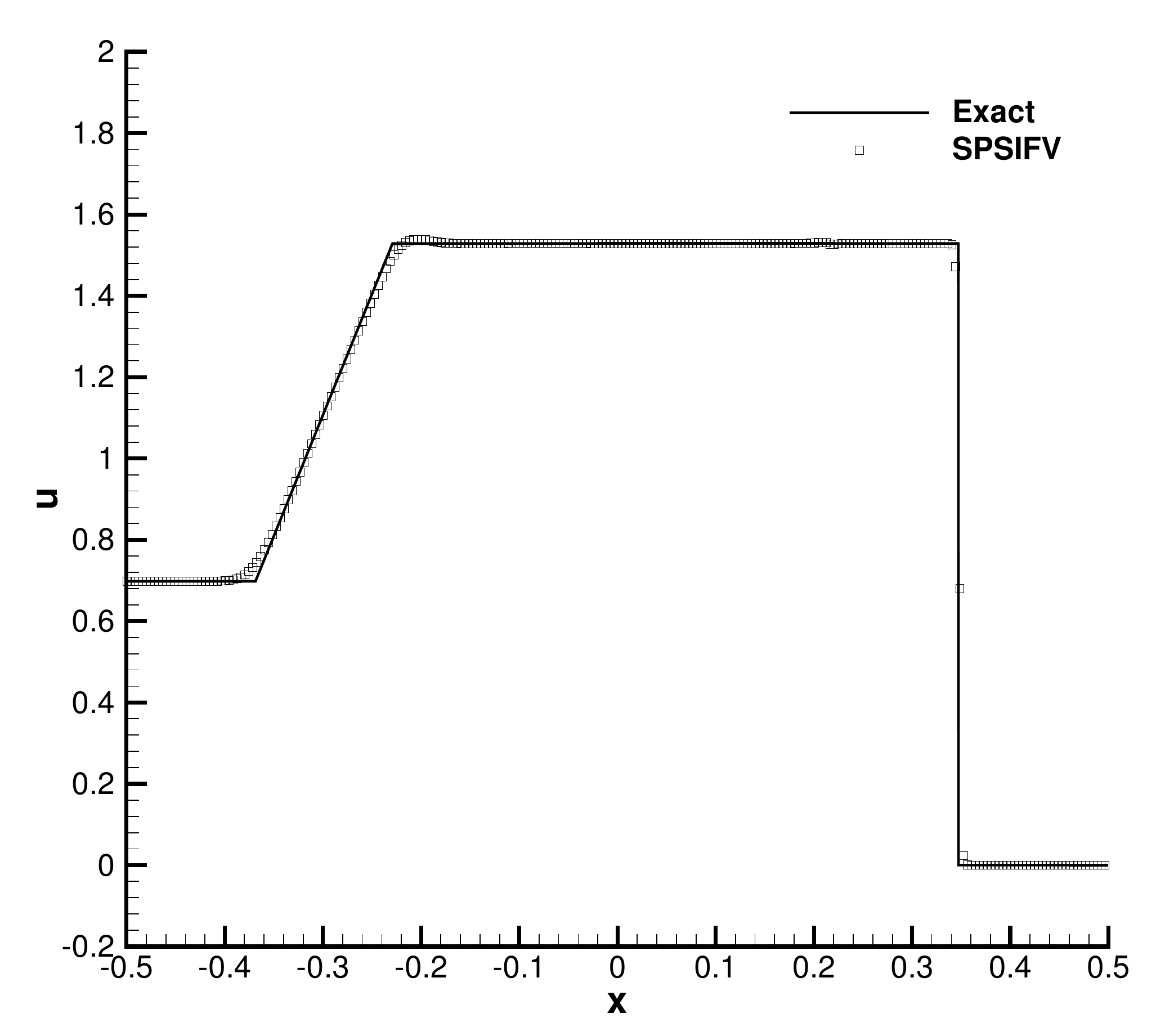}   
		\includegraphics[width=0.32\textwidth]{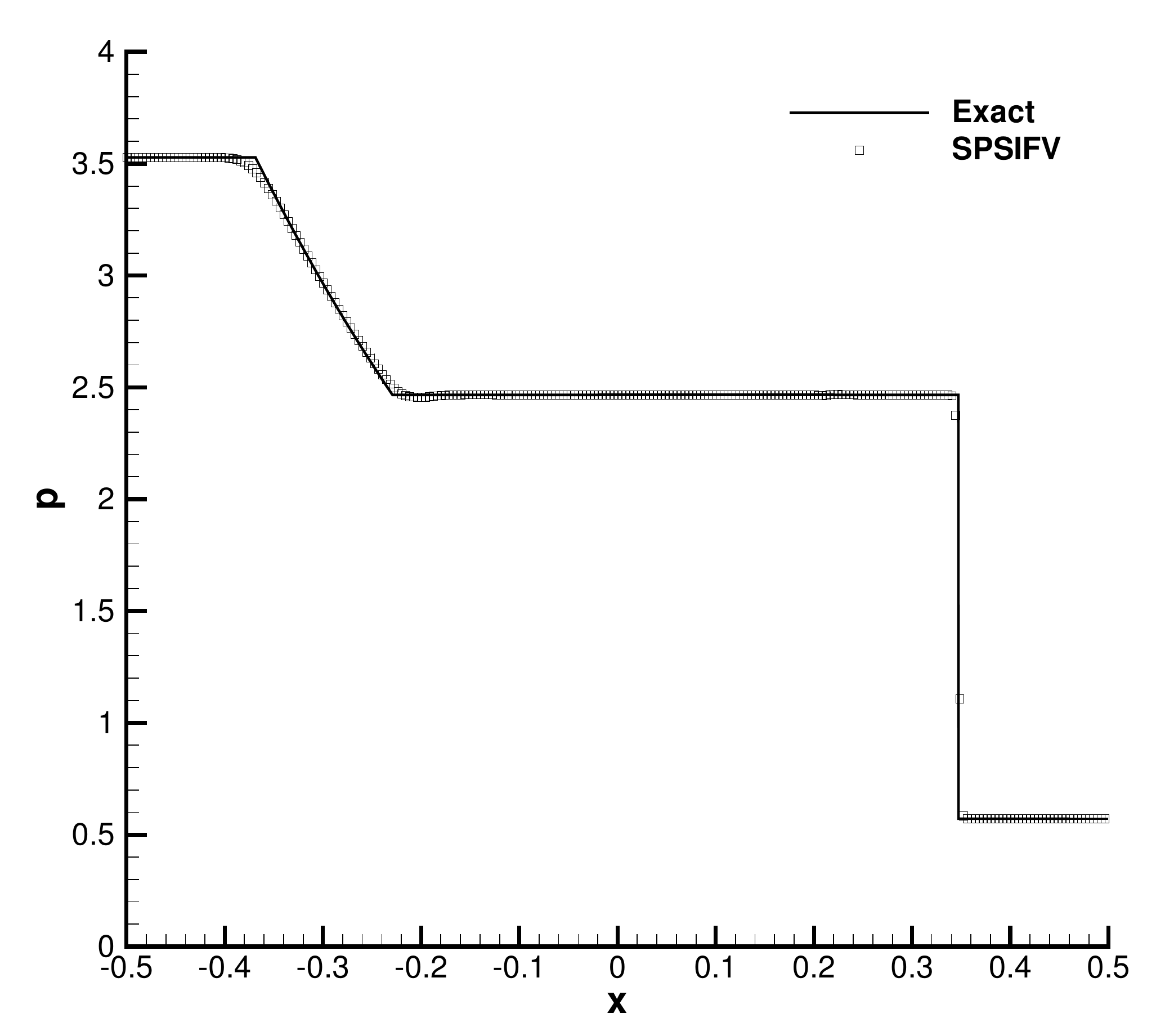}   
		\caption{Exact solution of the Euler equations and numerical solution of the GPR model in the stiff relaxation limit ($\tau_1 = \tau_2 = 10^{-6}$) for Riemann problem RP2 (Lax shock tube). The density $\rho$, the velocity component $u$ and the pressure $p$ are shown at a final time of $t=0.14$.} 
		\label{fig.rp2}
	\end{center}
\end{figure}

RP1 and RP2 are the classical Sod and Lax shock tube problem, respectively, which are well-known in the context of 
compressible gas dynamics. The respective computational results obtained with the new SPSIFV scheme are presented 
in Figures \ref{fig.rp1} and \ref{fig.rp2}, where we can in general observe a good agreement between the numerical 
solution of the GPR model and the exact solution of the Riemann problem of the compressible Euler equations.

\begin{figure}[!htbp]
	\begin{center}
		\includegraphics[width=0.32\textwidth]{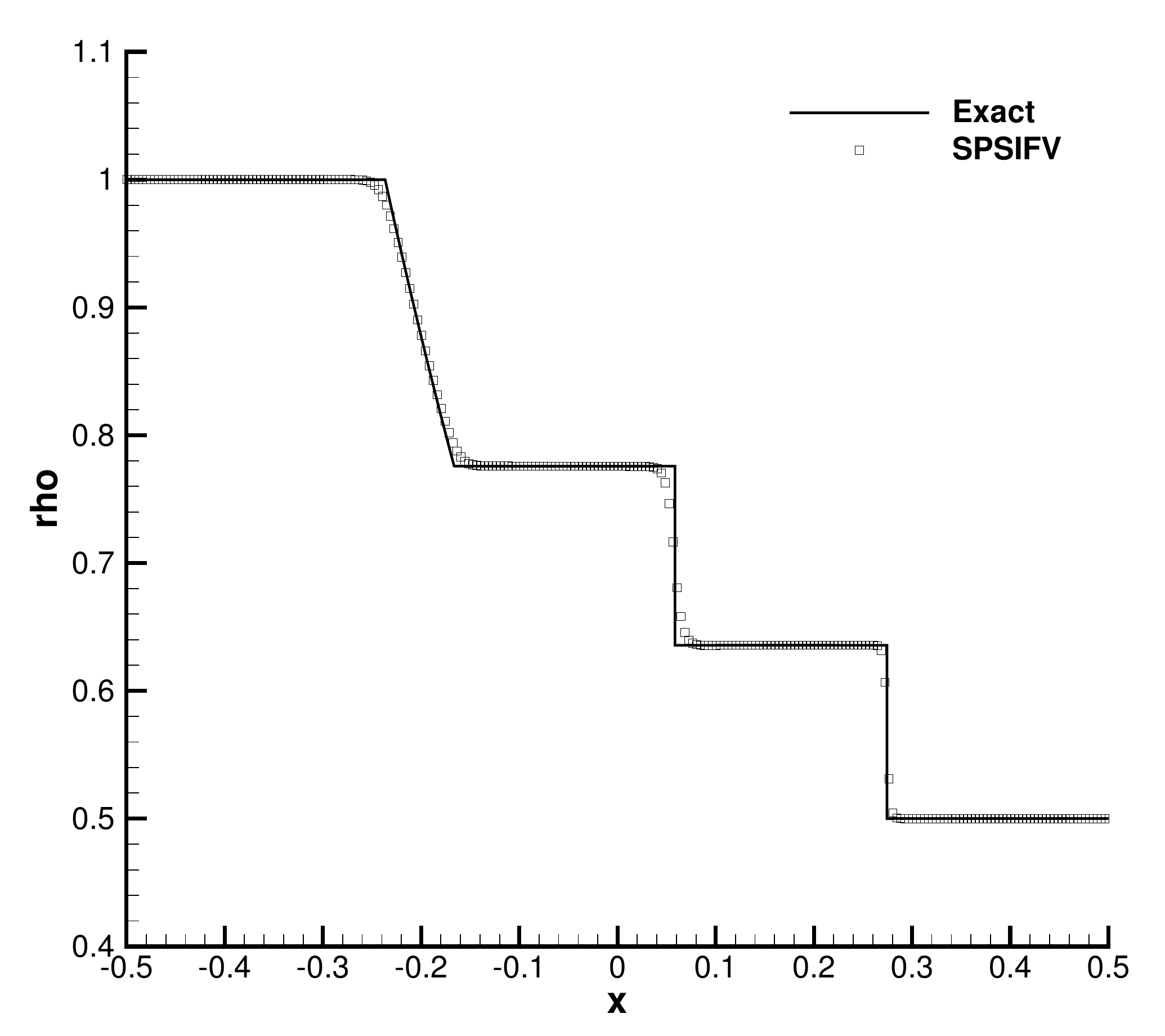}   
		\includegraphics[width=0.32\textwidth]{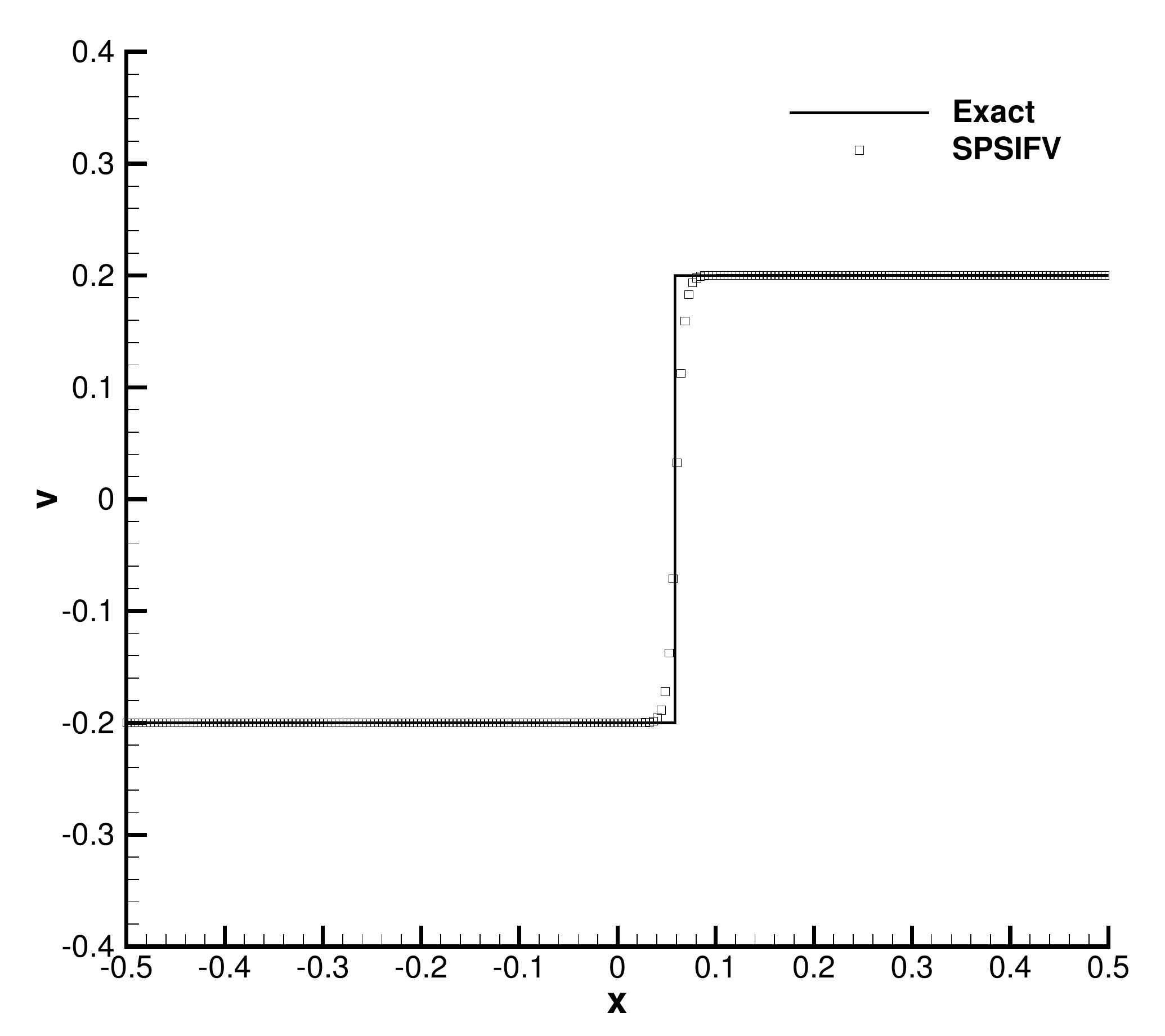}   
		\includegraphics[width=0.32\textwidth]{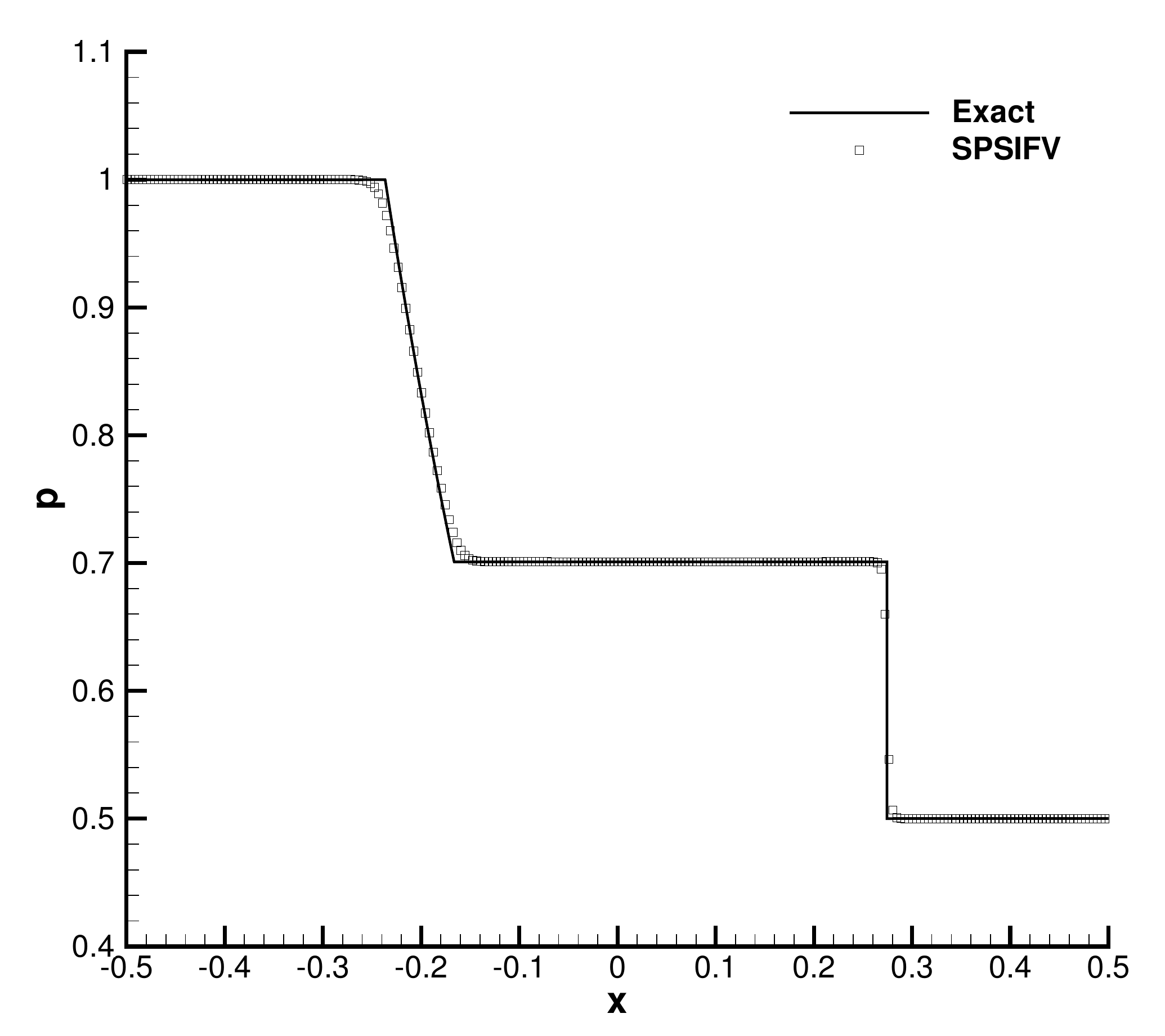}   
		\caption{Exact solution of the Euler equations and numerical solution of the GPR model in the stiff relaxation limit ($\tau_1 = \tau_2 = 10^{-6}$) for Riemann problem RP3. The density $\rho$, the velocity component $u$ and the pressure $p$ are shown at a final time of $t=0.2$.} 
		\label{fig.rp3}
	\end{center}
\end{figure}

\begin{figure}[!htbp]
	\begin{center}
		\includegraphics[width=0.32\textwidth]{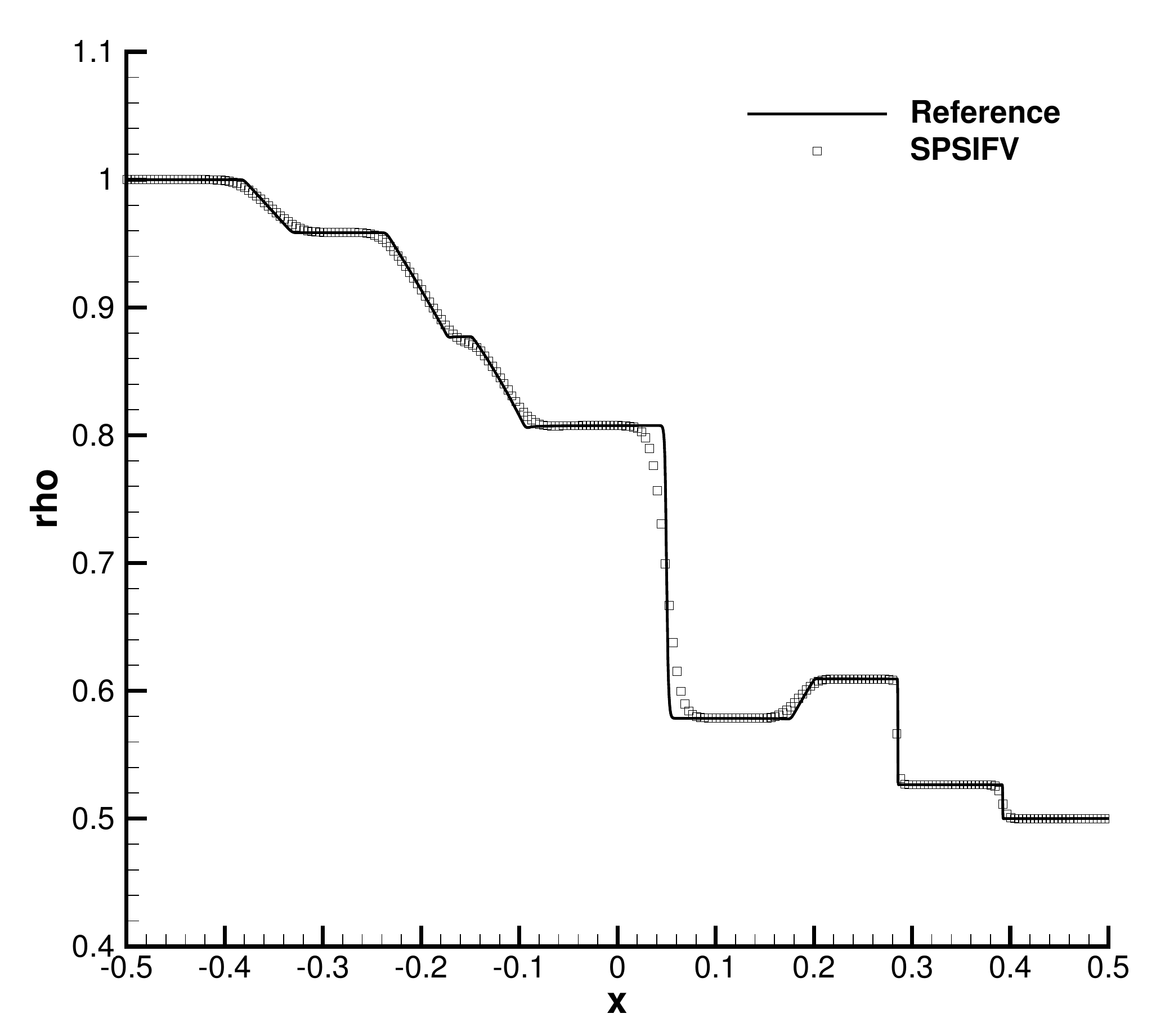}   
		\includegraphics[width=0.32\textwidth]{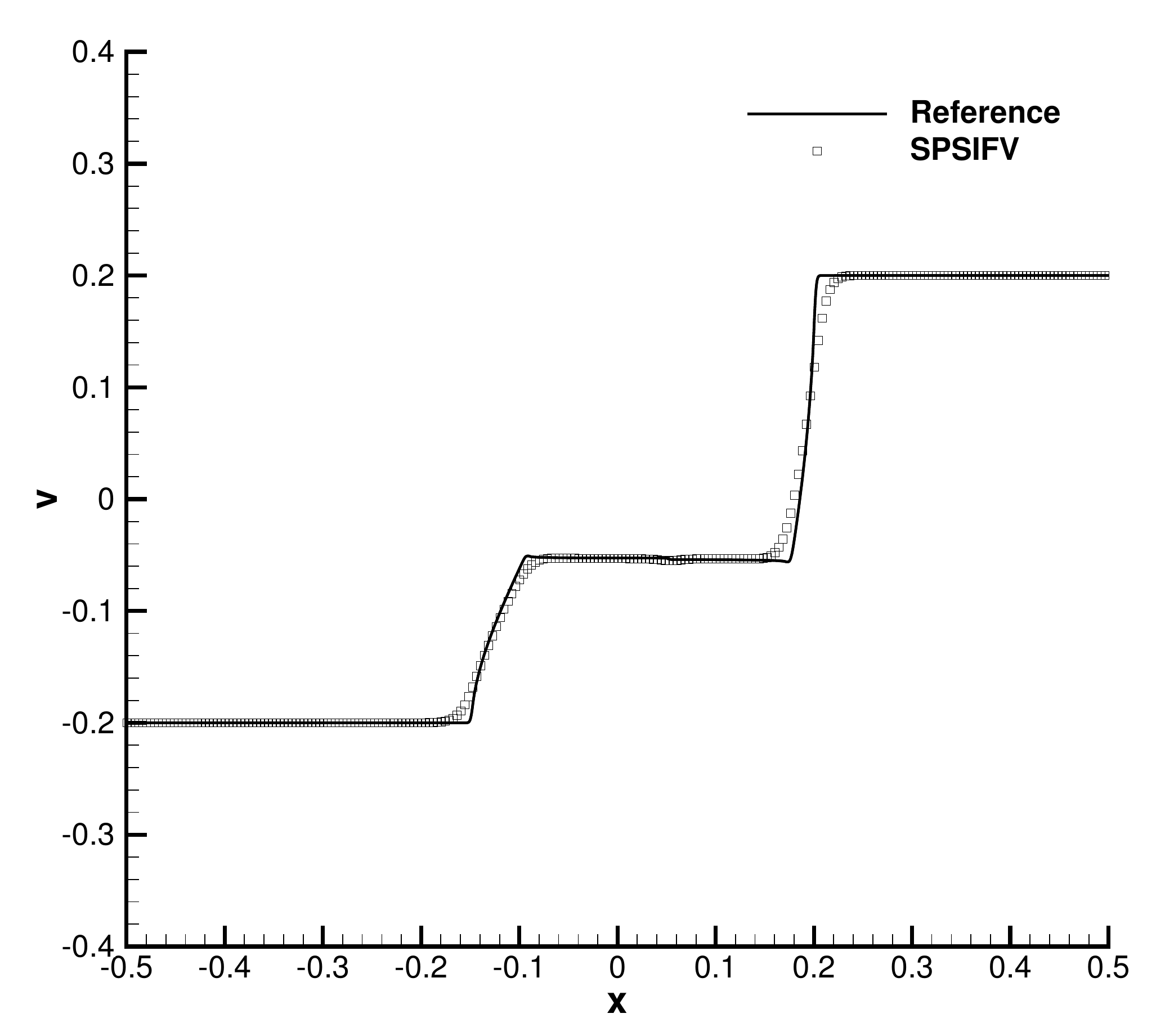}   
		\includegraphics[width=0.32\textwidth]{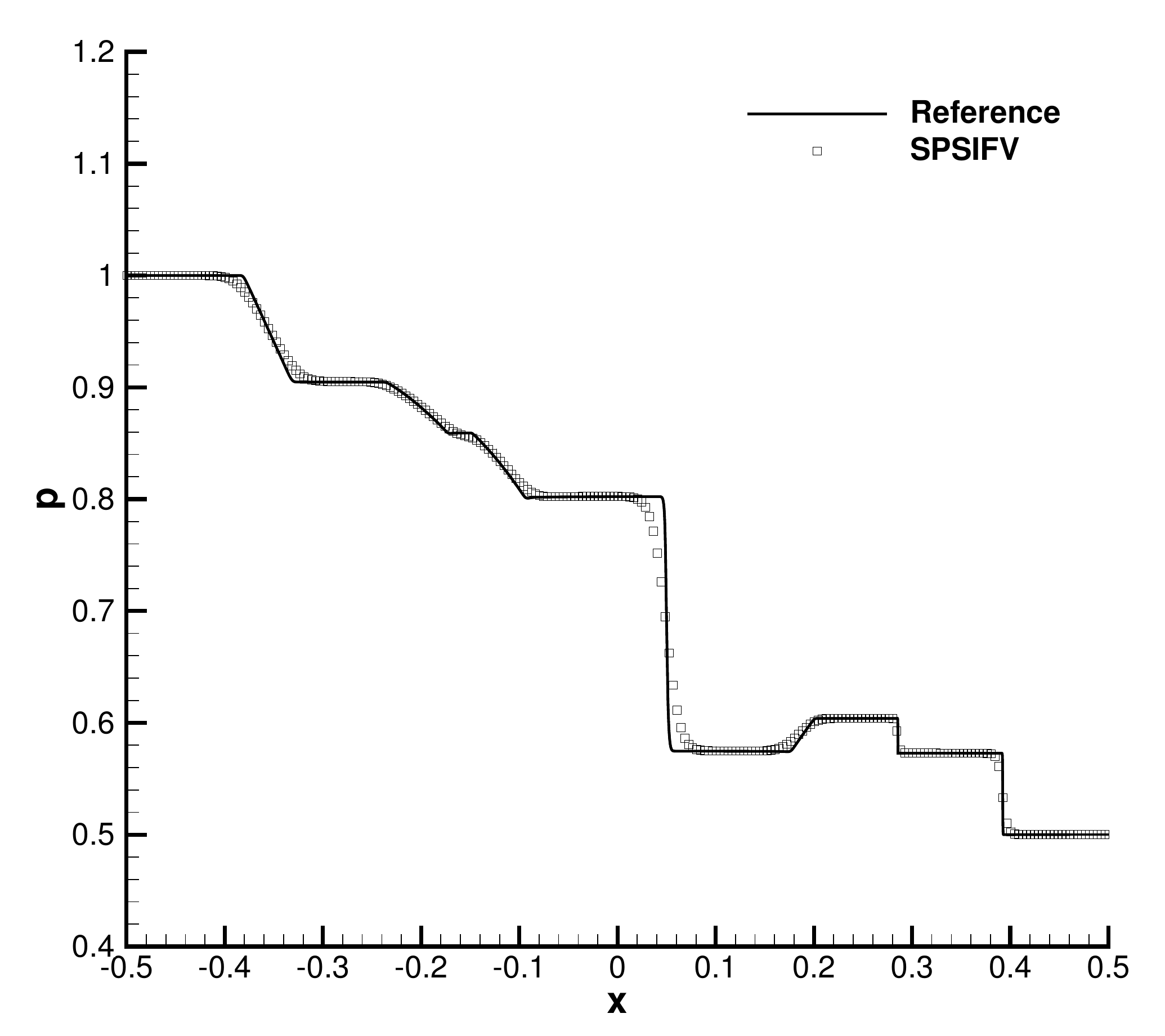}   
		\caption{Reference solution and numerical solution of the homogeneous GPR model without source terms ($\tau_1 = \tau_2 = 10^{20}$) for Riemann problem RP4. The density $\rho$, the velocity component $v$ and the pressure $p$ are shown at a final time of $t=0.2$.
		One can note seven waves that are contained in the homogeneous part of the GPR model.}  
		\label{fig.rp4}
	\end{center}
\end{figure}

Instead, RP3 and RP4 are the same Riemann problem with shear, once computed in the fluid limit (RP3) and once solved 
in the solid limit (RP4). The computational results are shown in Figures \ref{fig.rp3} and \ref{fig.rp4}. In the fluid 
limit (Fig. \ref{fig.rp3} we can indeed observe four waves, namely one contact discontinuity, one shear wave and 
two acoustic waves (a rarefaction moving to the left and a shock wave moving to the right). The agreement with the
exact solution of the Euler equations is excellent. In case of the homogeneous system (RP4), one can observe 7 waves
in Figure \ref{fig.rp4}, namely two fast thermo-acoustic waves (a right-moving shock and a left-moving rarefaction), 
two slow thermo-acoustic waves (again a right-moving shock and a left-moving rarefaction), two $x-y$ shear waves
and one contact discontinuity. The agreement with the reference solution is very good also in this case. 

Last but not least, we also report the time needed for one single element update as measured on one single CPU core of
an Intel i9-7900X CPU with 3.3 GHz nominal clock speed and 32 GB of RAM. Since the first three Riemann problems contain
stiff source terms, the necessary CPU time is higher than the one for RP4, where no source terms are present: 
The CPU time needed for one element update was 6.6 $\mu$s for RP1, 6.9 $\mu$s for RP2, 6.4 $\mu$s for RP3 and 3.1 $\mu$s 
for RP4. 

\begin{table}[!htbp]
	\renewcommand{\arraystretch}{1.25}
	\caption{Initial states left (L) and right (R) for density $\rho$, velocity $\mathbf{v}=(u,v,0)$ and pressure $p$  
		for a set of Riemann problems solved on the domain $\Omega=[-\frac{1}{2},+\frac{1}{2}]$ using the new structure-preserving 
		semi-implicit FV scheme. 
		The remaining variables of the GPR model are initialized as $\mathbf{A}=\sqrt[3]{\rho} \, \mathbf{I}$ and  $\mathbf{J}=0$. The Riemann problems include the fluid limit (RP1-RP3) as well as the solid limit (RP4). The relaxation times $\tau_1$ 
		and $\tau_2$ are also specified. In all cases we set $\gamma=1.4$. } 
	\begin{center} 
		\begin{tabular}{ccccccccccc} 
			\hline
			RP & $\rho_L$ & $u_L$ & $v_L$ & $p_L$ & $\rho_R$ & $u_R$ & $v_R$ & $p_R$ & $\tau_1$ & $\tau_2$  \\ 
			\hline
			RP1 &  1.0      &  0.0       &  0.0 & 1.0     & 0.125      &  0.0        &  0.0 & 0.1      &  $10^{-6}$ & $10^{-6}$ \\
			RP2 &  0.445    &  0.698     &  0.0 & 3.528   & 0.5        &  0.0        &  0.0 & 0.571    &  $10^{-6}$ & $10^{-6}$ \\ 
			RP3 &  1.0      &  0.0       & -0.2 & 1.0     & 0.5        &  0.0        & +0.2 & 0.5      &  $10^{-6}$ & $10^{-6}$ \\ 
			RP4 &  1.0      &  0.0       & -0.2 & 1.0     & 0.5        &  0.0        & +0.2 & 0.5      &  $10^{20}$ & $10^{20}$ \\ 
			\hline
		\end{tabular}
	\end{center} 
	\label{tab.ic.ideal}
\end{table}

\subsection{Viscous shock wave} 

Here we simulate the problem of a stationary viscous shock wave at a shock Mach number of $M_s=2$. 
For the special case of Prandtl number Pr$=0.75$, there exists an exact solution of the compressible  
Navier-Stokes equation that was first found by Becker \cite{Becker1923} in 1923, see 
\cite{Becker1923,BonnetLuneau,GPRmodel} for details. 

The computational domain $\Omega=[0,1]^2$ is discretized with $400 \times 10$ cells and the shock wave 
is centered at $x=0.5$. We suppose that the pre-shocked fluid is moving from right to left into the
shock wave.

\begin{figure}[!htbp]
	\begin{center}
		\includegraphics[width=0.32\textwidth]{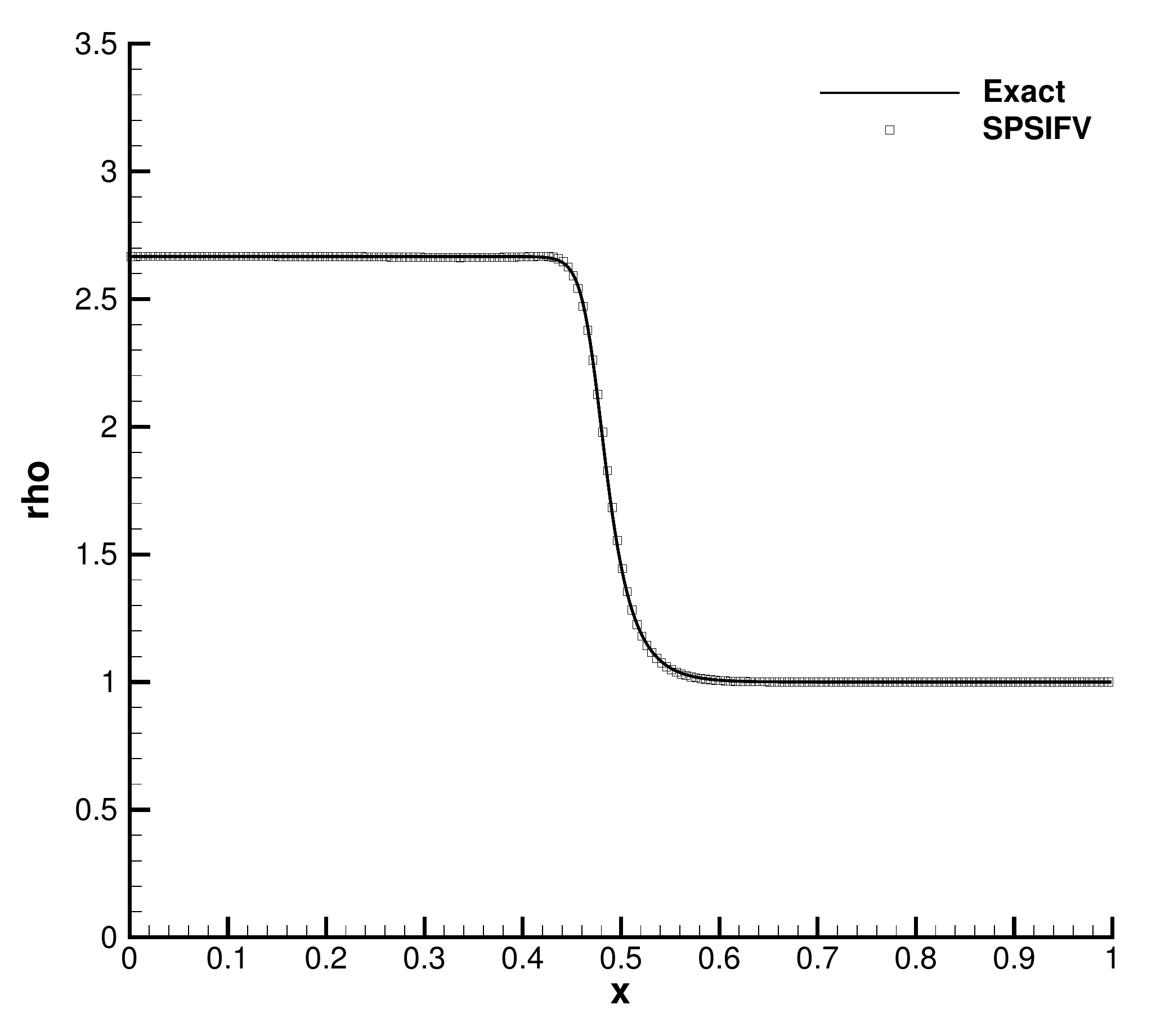}   
		\includegraphics[width=0.32\textwidth]{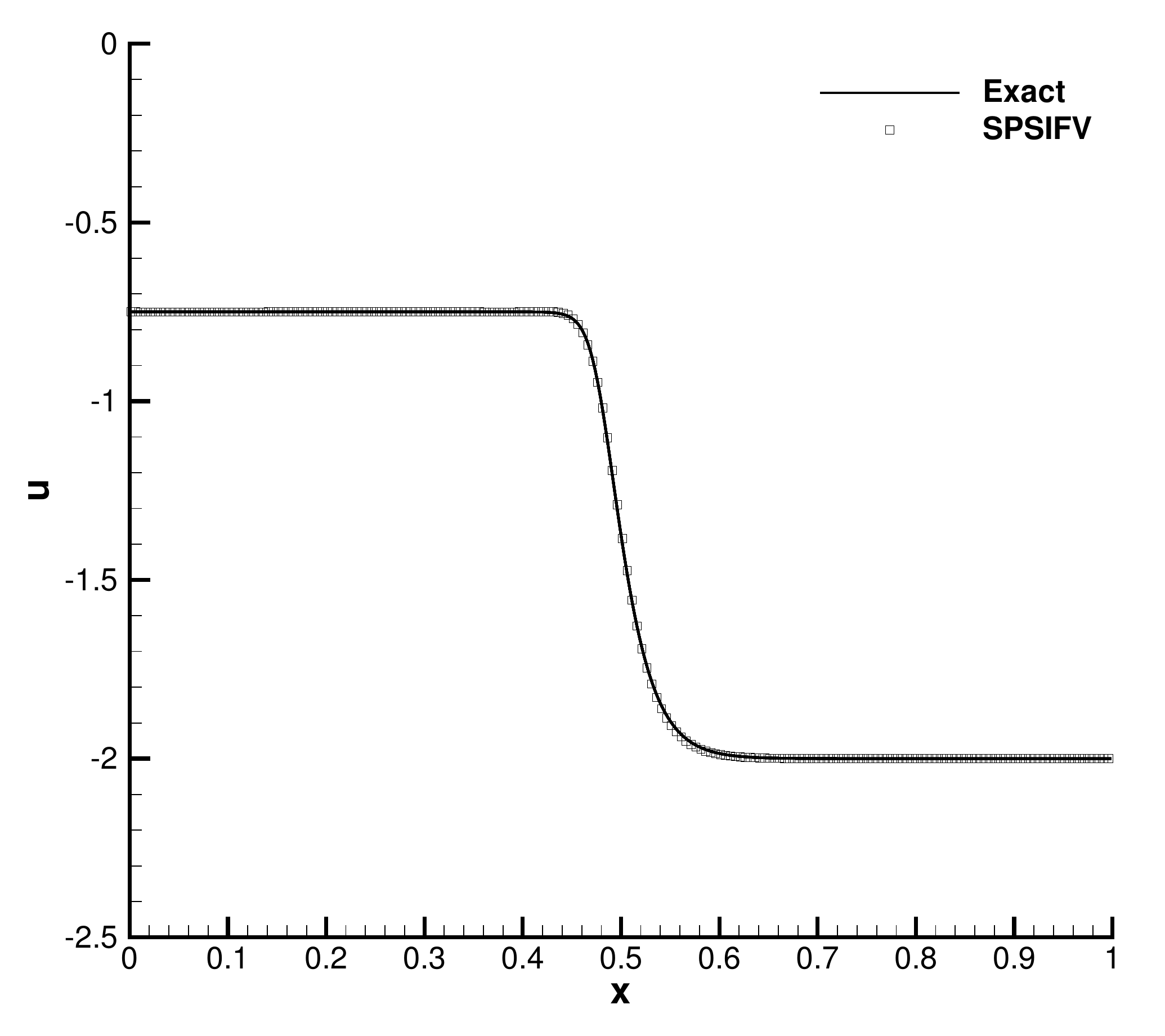}   
		\includegraphics[width=0.32\textwidth]{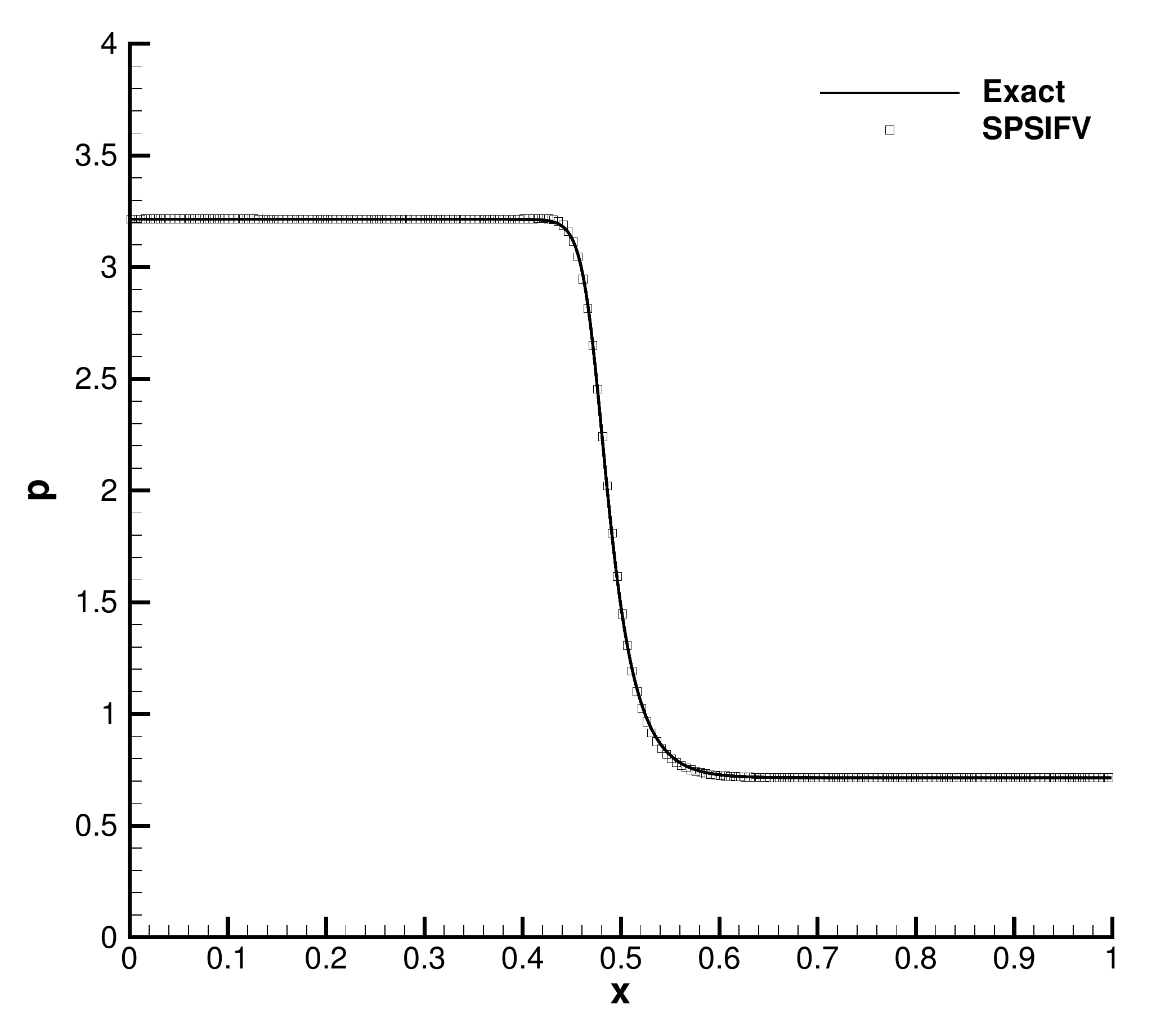}   
		\caption{Exact solution of the compressible Navier-Stokes equations and numerical solution of the GPR model for a viscous 
			shock profile at $M=2$, $Re=100$ and $Pr=0.75$.} 
		\label{fig.vshock}
	\end{center}
\end{figure}

The values of the fluid in front of the shock wave are chosen as $\rho_0 =1$, $u_0=-2$, $v_0=w_0=0$ and $p_0=1/\gamma$, 
hence $c_0 = 1$. The Reynolds number based on the flow speed $u_0$ and a unitary reference length ($L=1$) 
is given by $Re_s=\frac{\rho_0 \, c_0 \, M_s \, L }{\mu}$. 
The parameters of the GPR model are chosen as $\gamma = 1.4$, $c_v = 2.5$, $\alpha=c_s=50$, $\mu=2 
\cdot 10^{-2}$  
and $\lambda = 9 \frac{1}{3} \cdot 10^{-2}$. The shock Reynolds number is therefore $Re_s=100$. The distortion field 
and the thermal impulse are initially set to $\AAA=\sqrt[3]{\rho} \, \mathbf{I}$ and 
$\mathbf{J}=0$, respectively. The simulation with the new SPSIFV scheme is run until a final time of $t=0.2$. 
The comparison between the numerical solution of the first order hyperbolic GPR model and the exact solution of the  
compressible Navier-Stokes equations is shown for density $\rho$, velocity $u$ and pressure $p$ in Fig. \ref{fig.vshock}. 
One can note an excellent agreement for all quantities.

\subsection{2D Taylor-Green vortex at low Mach number}

Here we solve the two-dimensional Taylor-Green vortex at very low Mach numbers, which has an exact solution in the incompressible Navier-Stokes limit (i.e. for $M \to 0$) that reads 
\begin{eqnarray}
u(x,y,t)&=&\sin(x)\cos(y)e^{-2\nu t}, \label{eq:TG_0} \\
v(x,y,t)&=&-\cos(x)\sin(y)e^{-2\nu t}, \label{eq:TG_1} \\
p(x,y,t)&=& C + \frac{1}{4}(\cos(2x)+\cos(2y))e^{-4\nu t}.
\label{eq:TG_2}
\end{eqnarray}
In our numerical simulations, the computational domain is chosen as $\Omega=[0,2\pi]^2$ with periodic boundaries in $x$ and 
$y$ direction. 
The GPR model is solved with the new structure-preserving semi-implicit scheme up to a final time of $t=10$ using a computational
grid composed of $200 \times 200$ cells. The following set of parameters is used for this test case: $\gamma=1.4$, 
$\rho_0=1$, $\mu=10^{-2}$, $c_v=1$, $c_s=10$, $\alpha=1$, $Pr = 1$. The initial conditions for the 
velocity and the pressure are given by  \eqref{eq:TG_0}--\eqref{eq:TG_2}, where the additive 
constant in the pressure field is set to $C=10^5$, so that the maximal Mach number 
in this test problem is $M=0.0027$. The distortion field and the heat flux are initialized as usual with $\mathbf{A}=\mathbf{I}$ and $\mathbf{J}=0$. 

\begin{figure}[!htbp]
	\begin{center}
		\begin{tabular}{cc} 
			\includegraphics[width=0.45\textwidth]{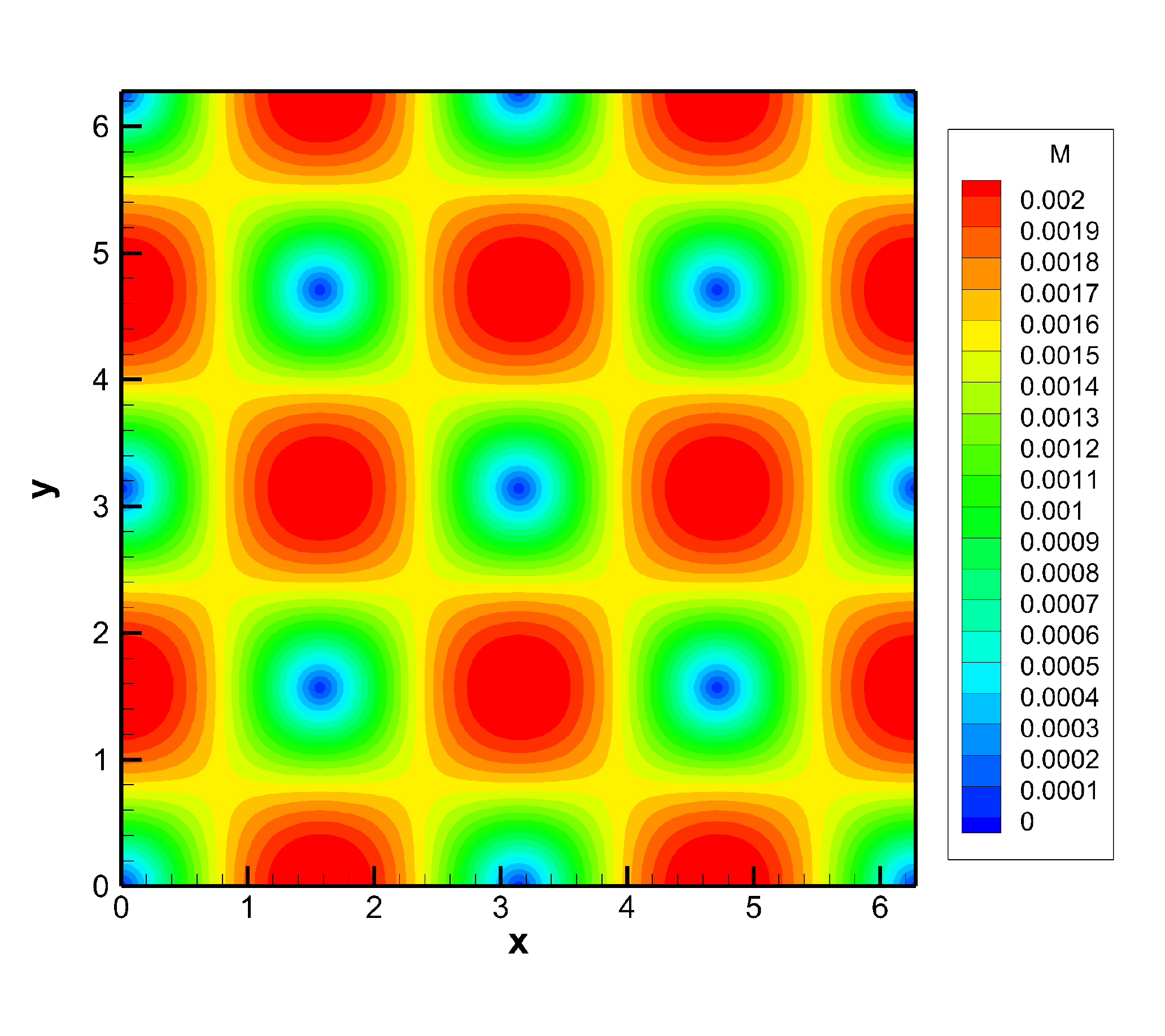}   & 
			\includegraphics[width=0.45\textwidth]{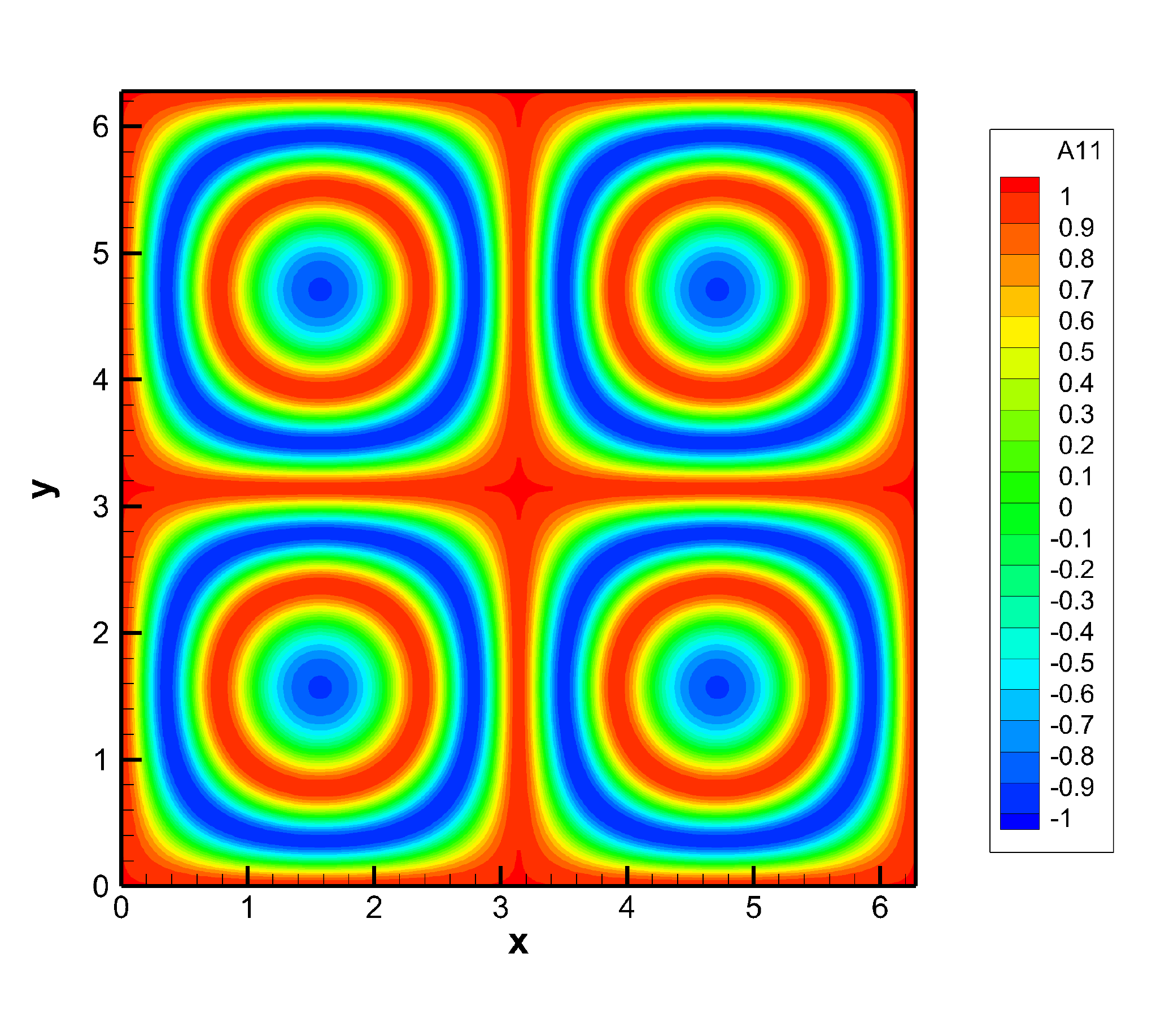} \\  
			\includegraphics[width=0.45\textwidth]{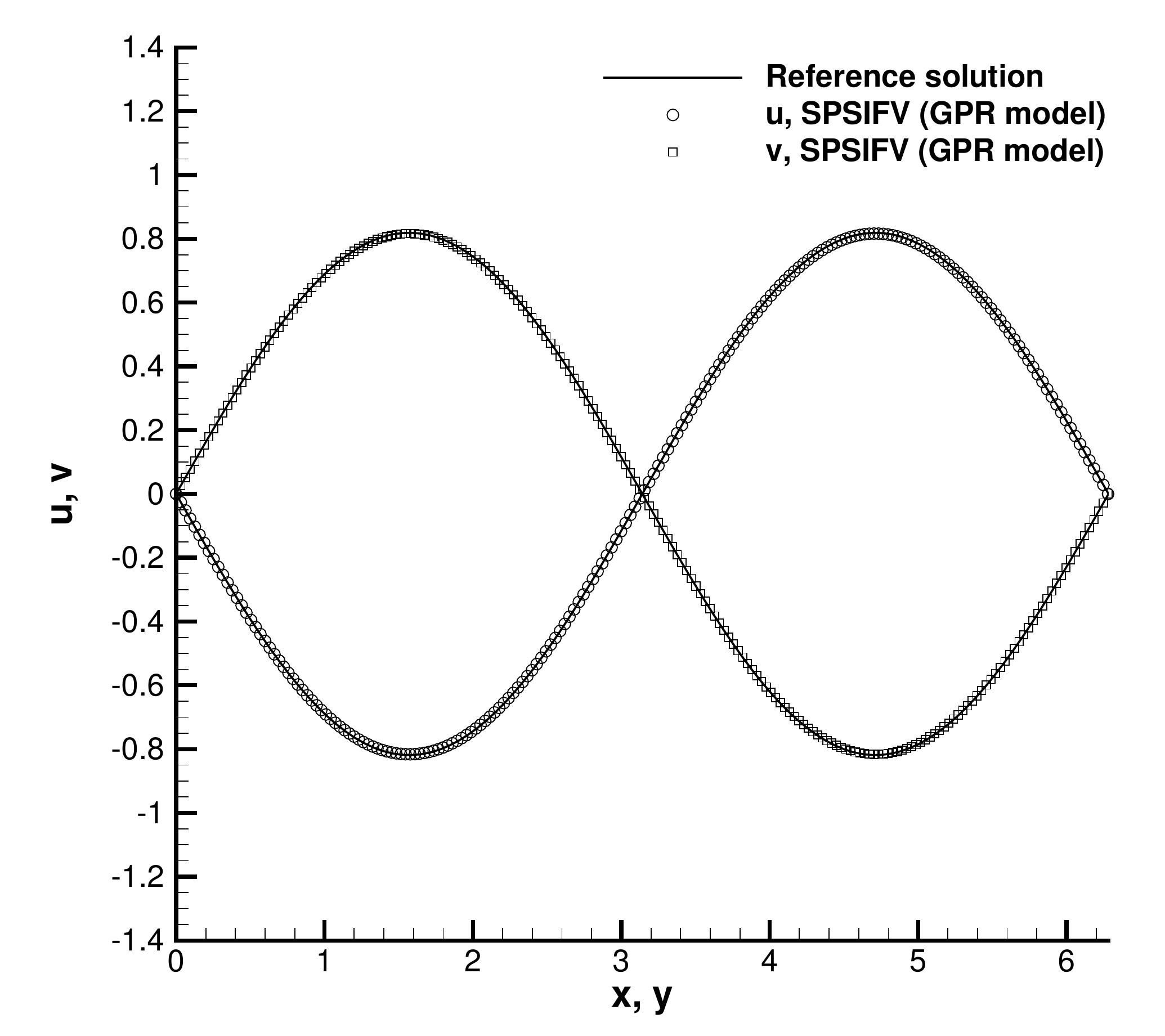}   & 
			\includegraphics[width=0.45\textwidth]{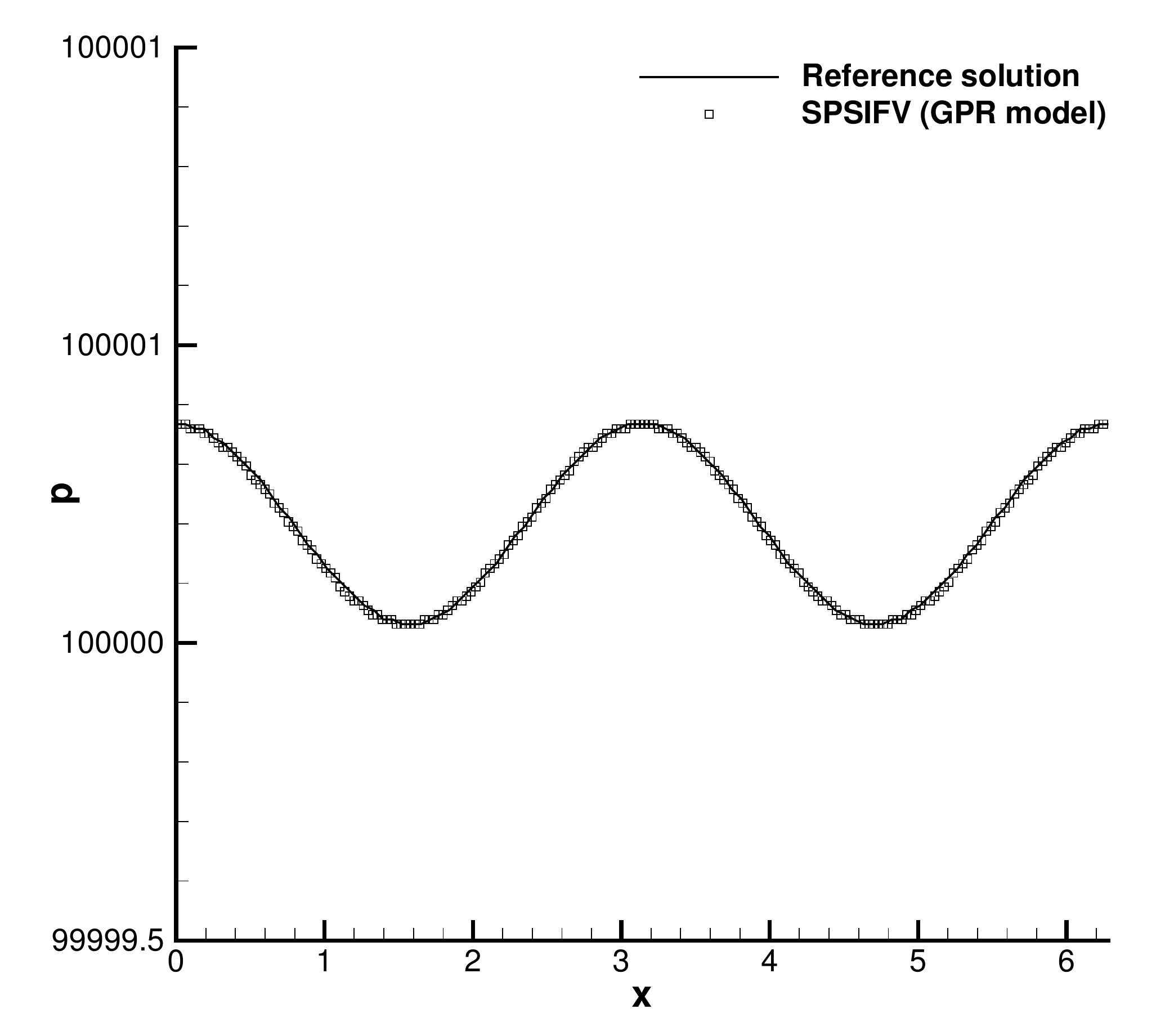}  
		\end{tabular} 
		\caption{Numerical solution of the GPR model for the Taylor-Green vortex with an effective viscosity of $\nu=10^{-2}$ at a final time of $t=10$ using the pressure-based structure-preserving semi-implicit finite volume scheme. Mach number contours (top left)  and 21 color contours in the interval [-1,1] of the distortion field component $A_{11}$ (top right). 1D cuts along the $x$ and the $y$ axis and comparison with the exact solution of the incompressible Navier-Stokes equations for the velocity components $u$ and $v$ (bottom left) and 1D cut along the $x$-axis for the pressure $p$ (bottom right).  } 
		\label{fig.tgv}
	\end{center}
\end{figure}

The $L_p$ error norm of a quantity $q$ with respect to the exact solution $q_e$ at time $t^n$ is in the following defined as 
\begin{equation}  
L_p(q,t^n) = \sqrt[p]{ \frac{1}{|\Omega|} \sum_{r,s} \Delta x \Delta y \left| q^{n,r,s} - q_e(x^r,y^s,t^n) \right|^p }, 
\end{equation}
while the $L_\infty$ norm is given as usual by
\begin{equation}
 L_\infty(q,t^n) = \max \limits_{r,s} \left| q^{n,r,s} - q_e(x^r,y^s,t^n) \right|. 
\end{equation}
The computational results obtained with the new structure-preserving semi-implicit scheme are shown 
in Fig.\,\ref{fig.tgv} at time $t=10$, where also a comparison with the exact solution of the 
incompressible Navier-Stokes equations is provided. An excellent agreement between the numerical 
results and the reference solution can be observed. The distortion field component $A_{11}$ shown 
in Fig. \ref{fig.tgv} 
matches the one presented in \cite{GPRmodel} very well and reveals the vortex structures of the flow. The total CPU time required by
the semi-implicit scheme for this low Mach number test problem on one single CPU core of an Intel i9-7940X CPU was 8607.7s. The entire simulation required 8466 time steps. The time needed for one degree of freedom (element) update (TDU) with the SPSIFV scheme was therefore 25.4 $\mu s$. The $L_\infty$ errors measured for the velocity component $u$ and for 
the pressure $p$ at the final time $t=10$ were $L_{\infty}(u,10)=3.6 \cdot 10^{-3}$ and $L_{\infty}(p,10)=3.7 \cdot 10^{-2}$, respectively. 

For comparison, we have rerun the same test problem also with an explicit second order accurate MUSCL-Hancock-type TVD finite volume scheme \cite{toro-book}, which required 111126.8 s of wall clock time on one core of an Intel i9-7940X CPU. In order to reach the final time a total number of
265418 time steps was necessary, i.e. the time needed to update one degree of freedom with the explicit scheme was TDU = 10.5 $\mu$s. 
We can conclude that in terms of total wall clock time the new SPSIFV scheme is a factor of \textbf{12.9} faster than the explicit scheme for this low Mach number test case. However, in terms of CPU time per degree of freedom update, the semi-implicit finite volume method is only a factor of 2.4 times more expensive than the explicit scheme,
despite the need to solve a linear system for the pressure in each Picard iteration in each time step. Last but not least, we also report
the $L_\infty$ error norms obtained with the explicit scheme at the final time. They were $L_{\infty}(u,10)=0.49$ and $L_{\infty}(p,10)=0.34$, respectively, which are up to two orders of magnitude larger than the errors obtained with the structure-preserving semi-implicit scheme. 
It is indeed very well-known that the low Mach number problem affects explicit density-based schemes in a very negative way not only 
in terms of computational efficiency, but also in terms of accuracy.

We finally use this test problem also in order to perform a numerical convergence study of our scheme in the stiff relaxation limit at low Mach numbers. For this purpose, we set the effective viscosity in the GPR model to $\mu=10^{-5}$ and run the 2D Taylor-Green vortex on a sequence of successively refined meshes until a final time of $t=0.2$. We measure the error norms for the velocity component $u$ and  compute the resulting convergence order, see Table \ref{tab.convGPR}. As expected, the obtained results indicate that the scheme 
achieves second order of accuracy in space.

\begin{table}  
		\caption{Numerical convergence results of the structure-preserving semi-implicit FV scheme for the GPR model in the stiff relaxation limit using the 2D Taylor-Green vortex problem. The $L_1$, $L_2$ and $L_\infty$ error norms refer to the velocity component $u$ at a final time of $t=0.2$. } 	
		\begin{center} 
			\begin{small}
				\renewcommand{\arraystretch}{1.3}
				\begin{tabular}{ccccccc} 
					\hline
					$N_x = N_y$ & ${L_1}(u,0.2)$ & $\mathcal{O}(u)$  & ${L_2}(u,0.2)$ & $\mathcal{O}(u)$  
					& ${L_\infty}(u,0.2)$ & $\mathcal{O}(u)$  
					\\ 
					\hline
					25  & 9.9081E-03 &      & 1.2214E-02 &       & 2.4335E-02 &       \\ 
					50  & 2.1079E-03 & 2.2  & 2.5905E-03 &  2.2  & 5.1097E-03 & 2.3   \\ 
					100 & 5.1362E-04 & 2.0  & 6.2444E-04 &  2.1  & 1.1784E-03 & 2.1   \\ 
					200 & 1.3577E-04 & 1.9  & 1.6400E-04 &  1.9  & 2.9661E-04 & 2.0   \\ 
					\hline 
				\end{tabular}
			\end{small}
		\end{center}
		\label{tab.convGPR}
\end{table}

\subsection{Double shear layer at low Mach number} 

\begin{figure}[!htbp]
	\begin{center}
		\begin{tabular}{cc} 
			\includegraphics[width=0.47\textwidth]{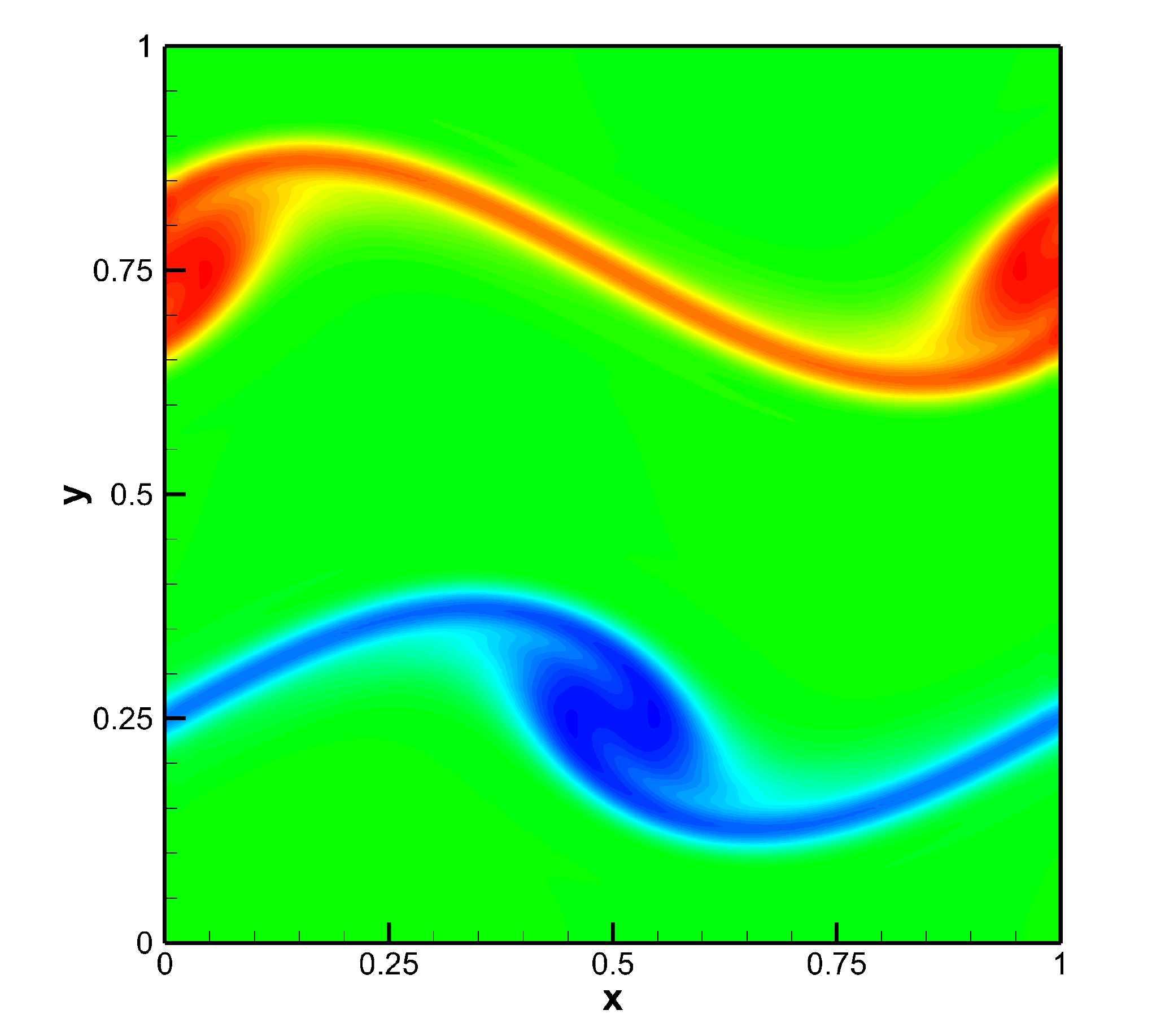}  & 
			\includegraphics[width=0.47\textwidth]{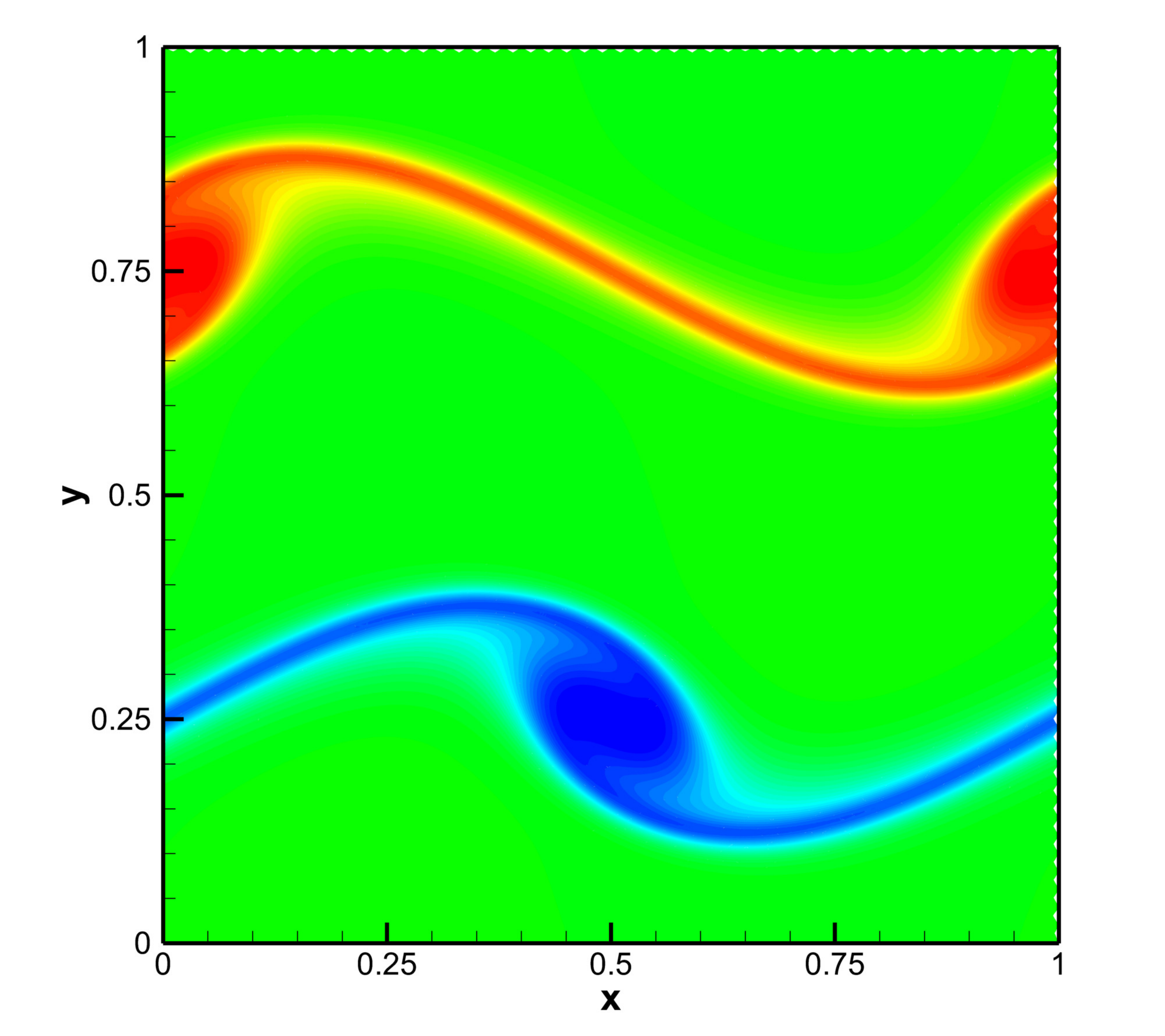}  \\ 
			\includegraphics[width=0.47\textwidth]{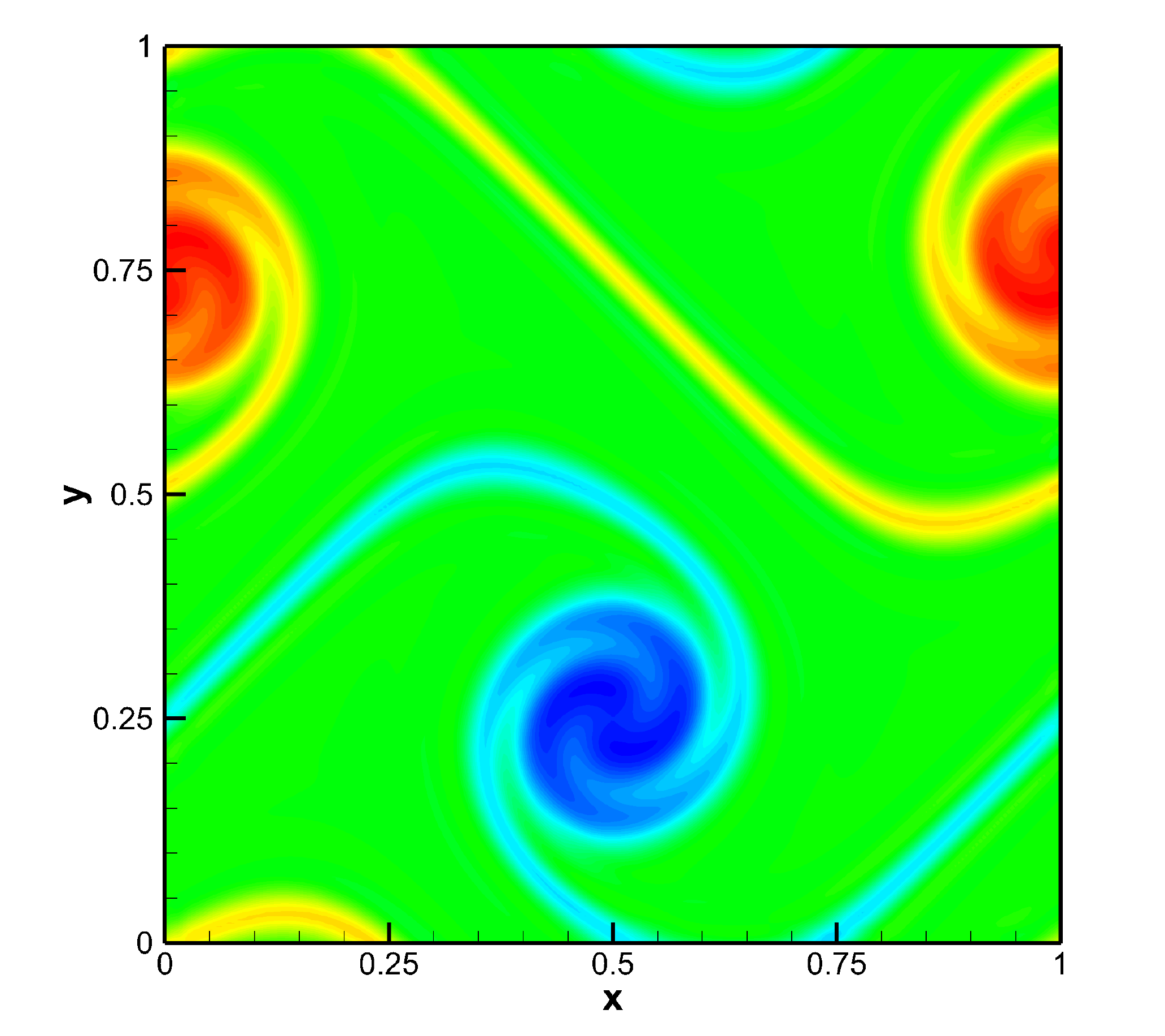}  & 
			\includegraphics[width=0.47\textwidth]{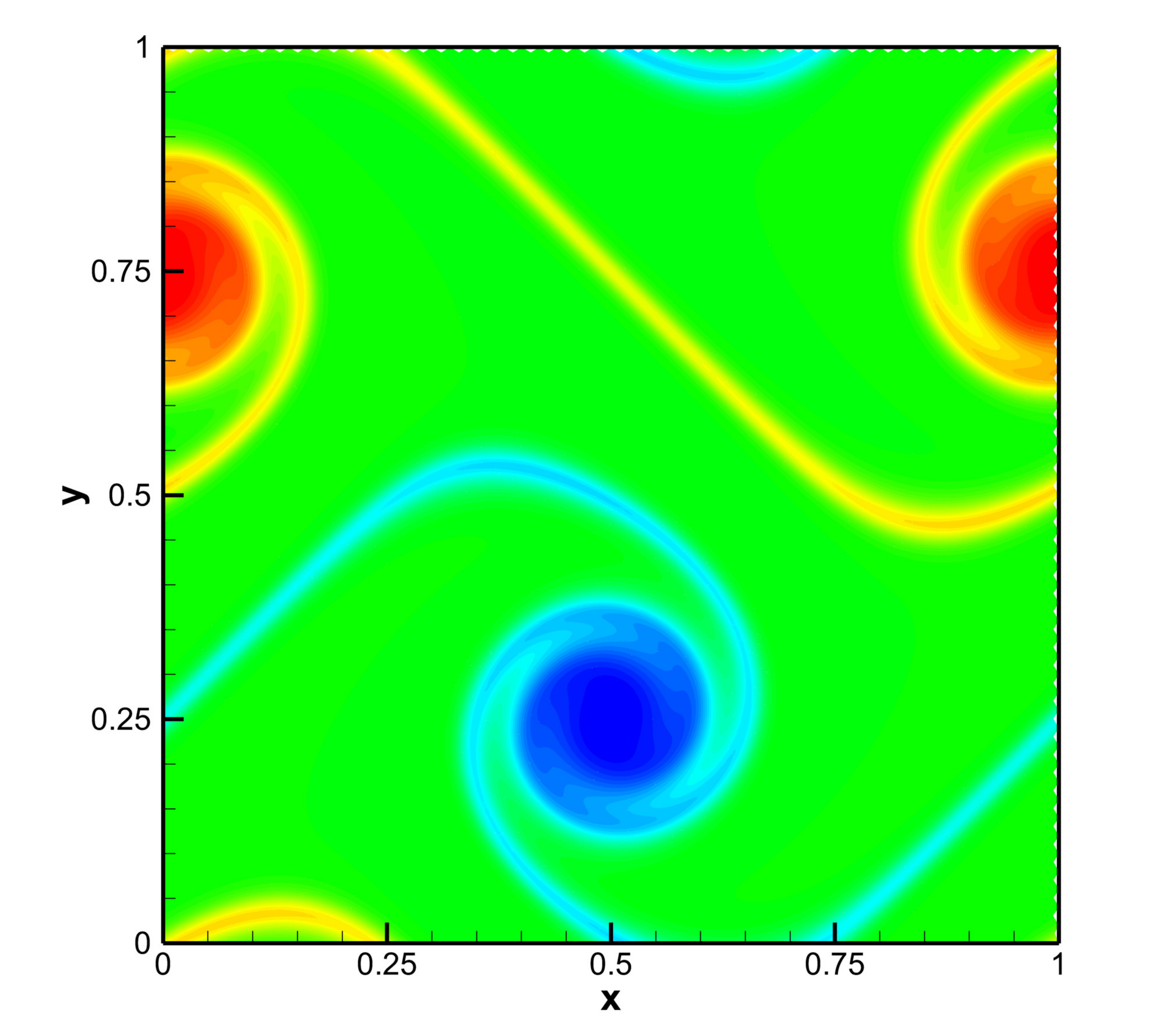}  \\ 
			\includegraphics[width=0.47\textwidth]{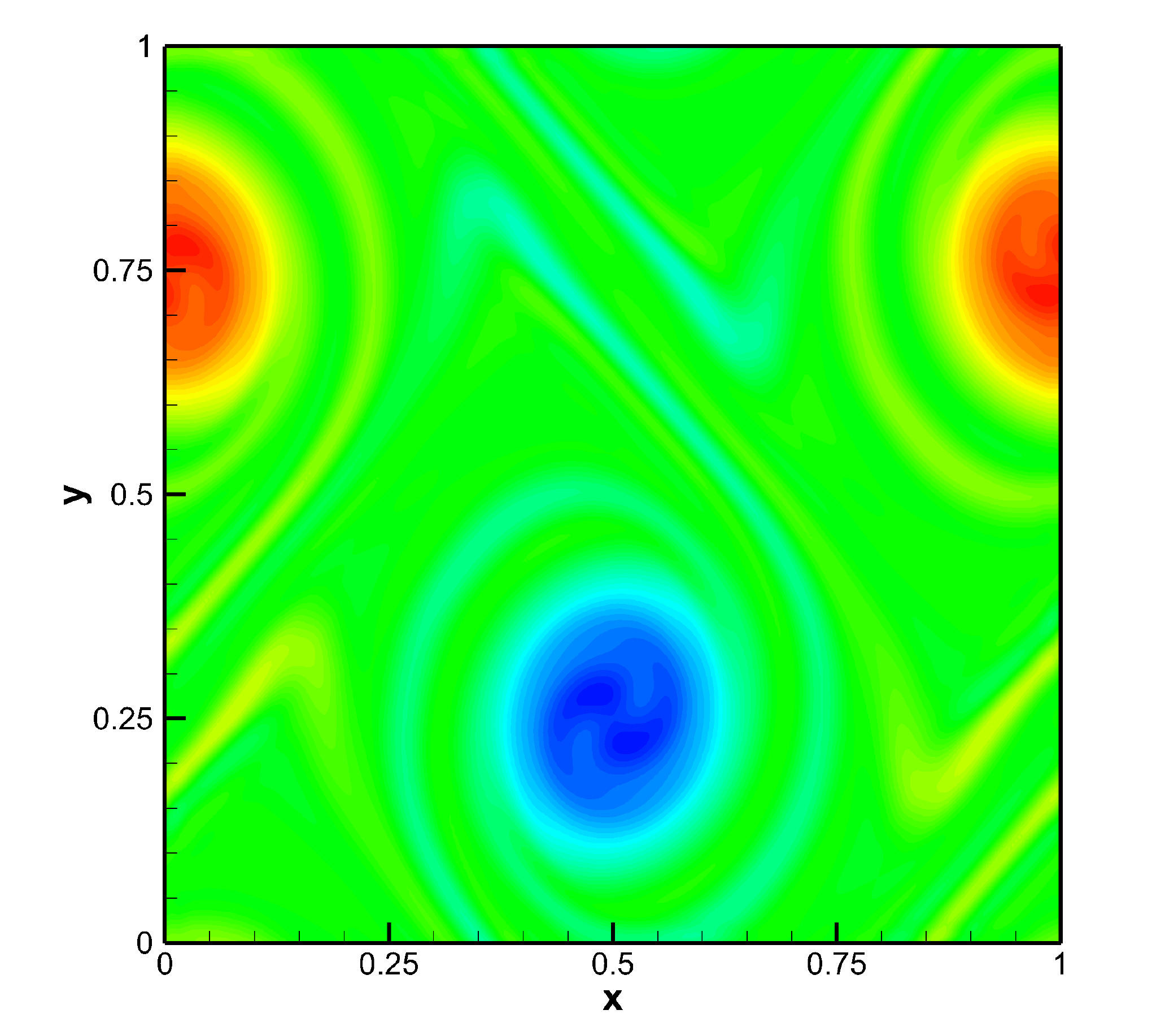}  & 
			\includegraphics[width=0.47\textwidth]{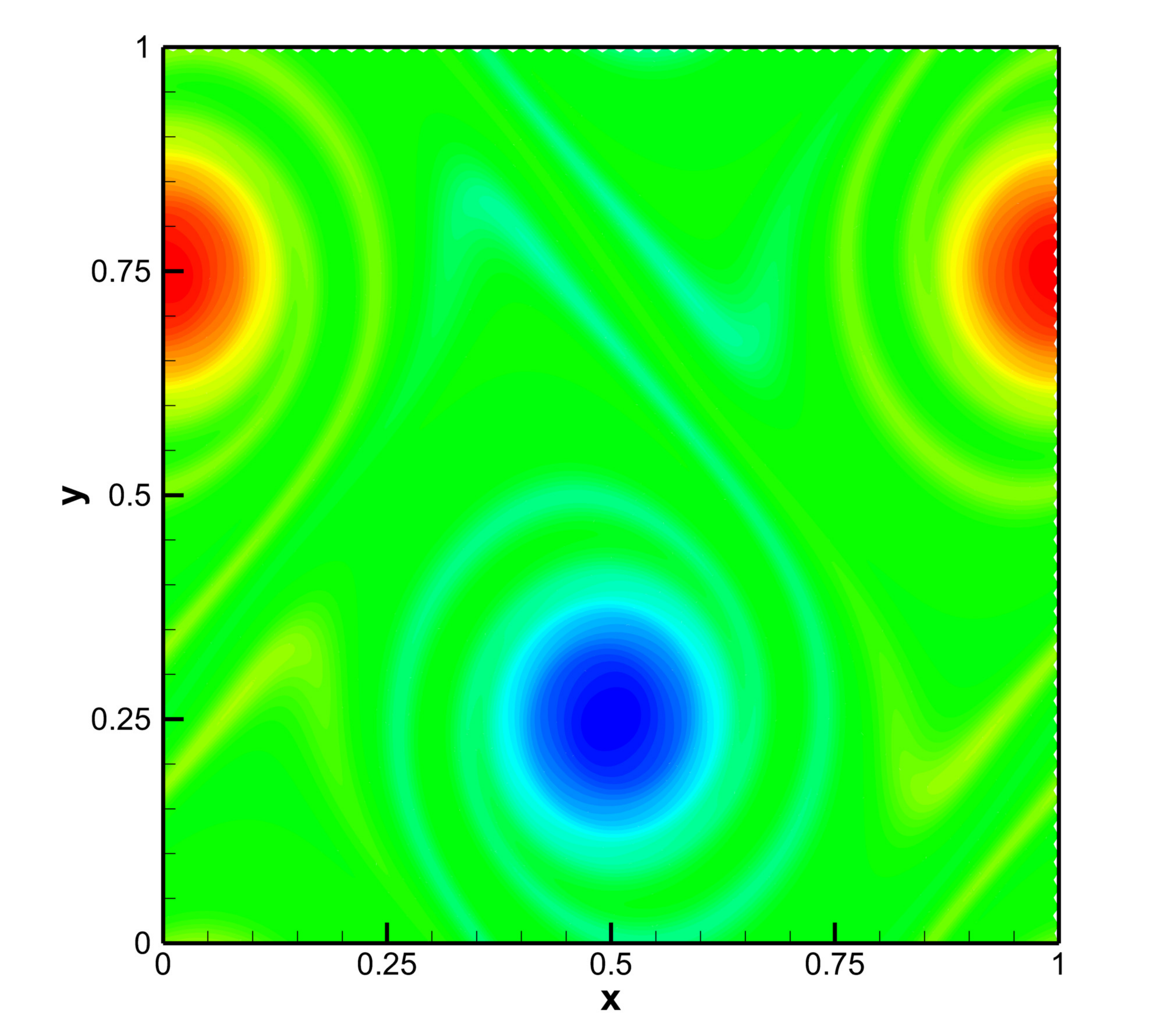}   
		\end{tabular} 
		\caption{Vorticity contours for the double shear layer with a viscosity of $\mu=2 \cdot 10^{-4}$ 
			at times $t=0.8$, $t=1.2$ and $t=1.8$ (from top to bottom). 
			Right: reference solution obtained by solving the incompressible Navier-Stokes equations with the 
			staggered semi-implicit space-time DG scheme of Tavelli and Dumbser \cite{TavelliNS,TavelliDumbser2016}. 
			Left: numerical solution of the GPR model obtained with the new structure-preserving semi-implicit finite volume scheme. } 
		\label{fig.dslrot}
	\end{center}
\end{figure}

Here we solve the double shear layer test problem, see \cite{Bell1989,Tavelli2015,GPRmodel}. 
The computational domain is $\Omega=[0,1]^2$ with periodic boundary conditions everywhere. The initial 
condition reads 
\begin{equation}
u=\left\{
\begin{array}{l}
\tanh\left( \tilde{\rho} (y-0.25) \right), \qquad \textnormal{ if } y \leq 0.5, \\
\tanh\left( \tilde{\rho} (0.75-y) \right), \qquad \textnormal{ if } y > 0.5,
\end{array}
\right.
\label{eq:DSL_0} 
\end{equation} 
\begin{equation} 		
v= \delta \sin(2\pi x), \qquad w = 0, \qquad \rho  = \rho_0 = 1,  \qquad p = 10^5. 
\label{eq:DSL_2}
\end{equation}
With this initial data, the maximum Mach number of the problem is $M=2.67 \cdot 10^{-3}$, which is about two orders of magnitude 
less than the Mach number used in \cite{GPRmodel}. 
For this test case we set the parameters that determine the the shape of the velocity field to 
$\delta=0.05$ and  $\tilde{\rho}=30$. Furthermore, we set the viscosity coefficient to $\nu= \mu / 
\rho_0 = 2 \cdot 10^{-4}$. The other parameters of the GPR model are  chosen as $\gamma = 1.4$, 
$\rho_0=1$, $c_v=1$, $c_s=8$ and $\alpha=0$. The initial condition for the distortion field is 
$\AAA=\mathbf{I}$ and furthermore we initialize the thermal impulse with $\mathbf{J}=0$. Simulations are carried out with the new structure-preserving semi-implicit finite volume scheme up to a final time of $t=1.8$. The computational mesh is composed of $250 \times 250$ control volumes. In Fig. \ref{fig.dslrot} the computational results obtained with the SPSIFV scheme applied to the GPR model are  compared with a numerical reference solution that is based on the solution of the incompressible Navier-Stokes equations and that has been obtained by a high order staggered semi-implicit space-time discontinuous Galerkin scheme, see \cite{Tavelli2015} for details. 
The flow dynamics has already been described in \cite{Bell1989} and \cite{Tavelli2015} and can be summarized by the development of  
several vortices from the initially perturbed shear layers. 
We can note an excellent agreement between the Navier-Stokes reference solution and the numerical solution of the GPR model obtained  with the new SPSIFV scheme. 

\begin{figure}[!htbp]
	\begin{center}
		\begin{tabular}{cc} 
		\includegraphics[width=0.47\textwidth]{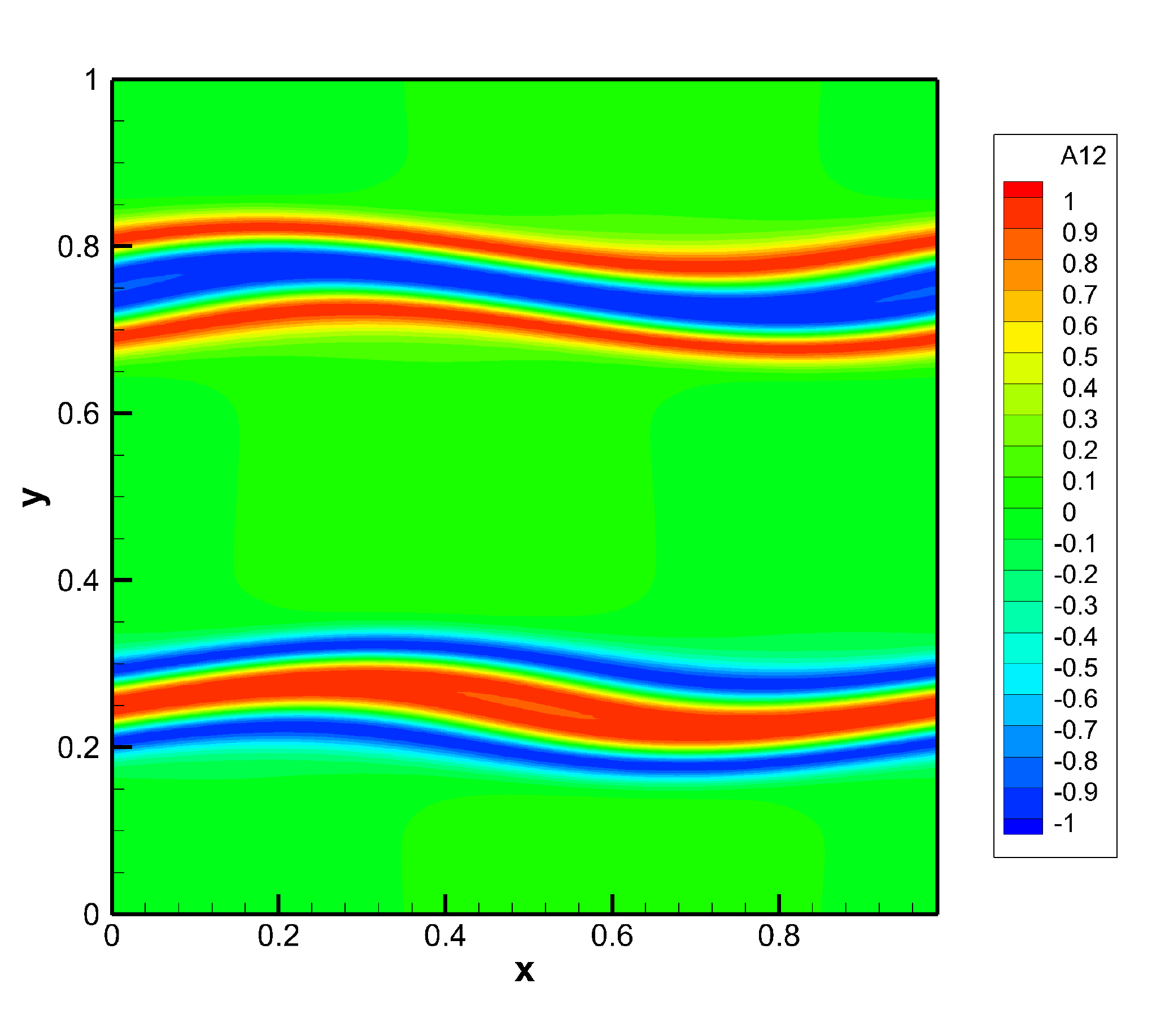}  & 
		\includegraphics[width=0.47\textwidth]{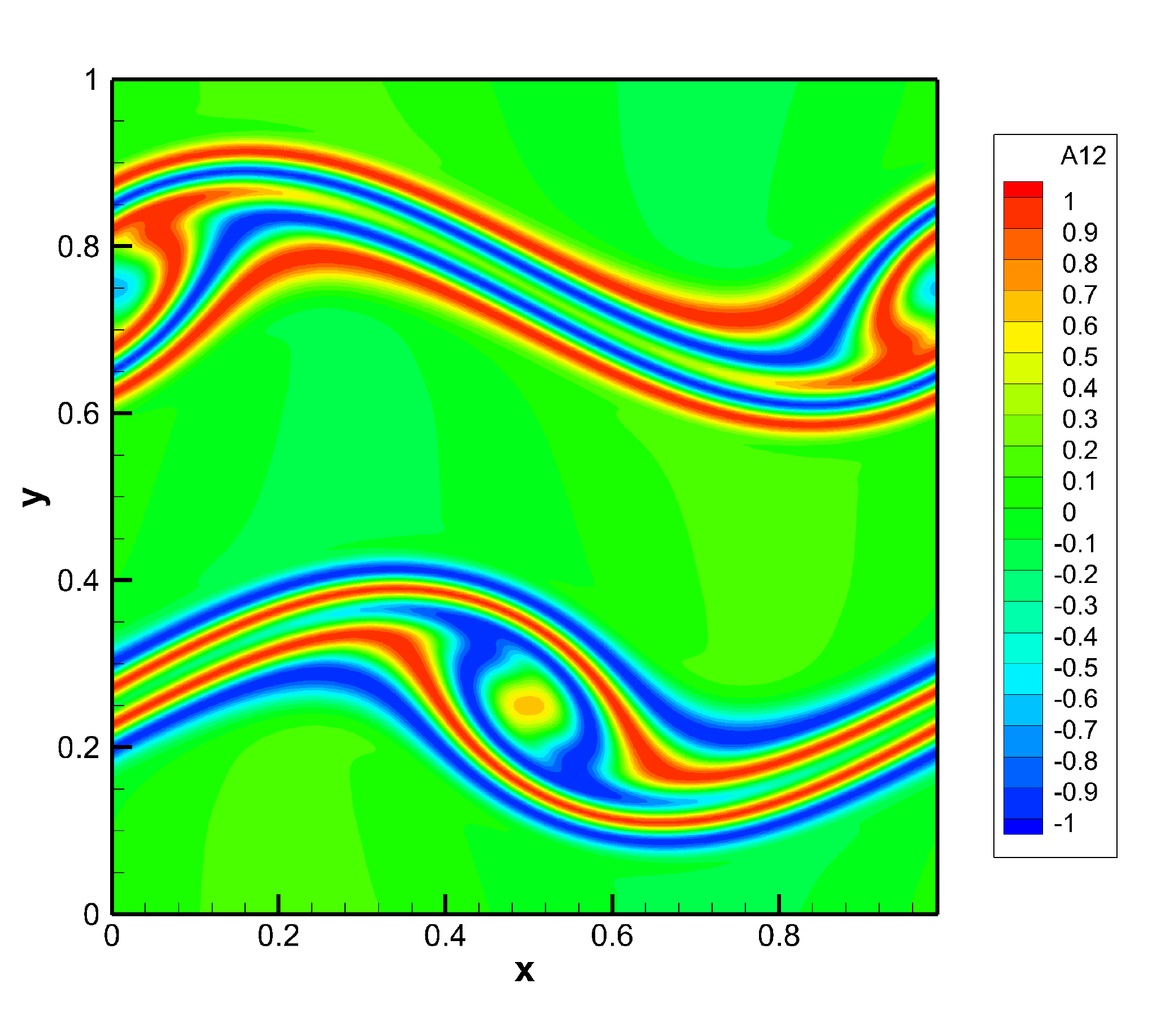}  \\ 
		\includegraphics[width=0.47\textwidth]{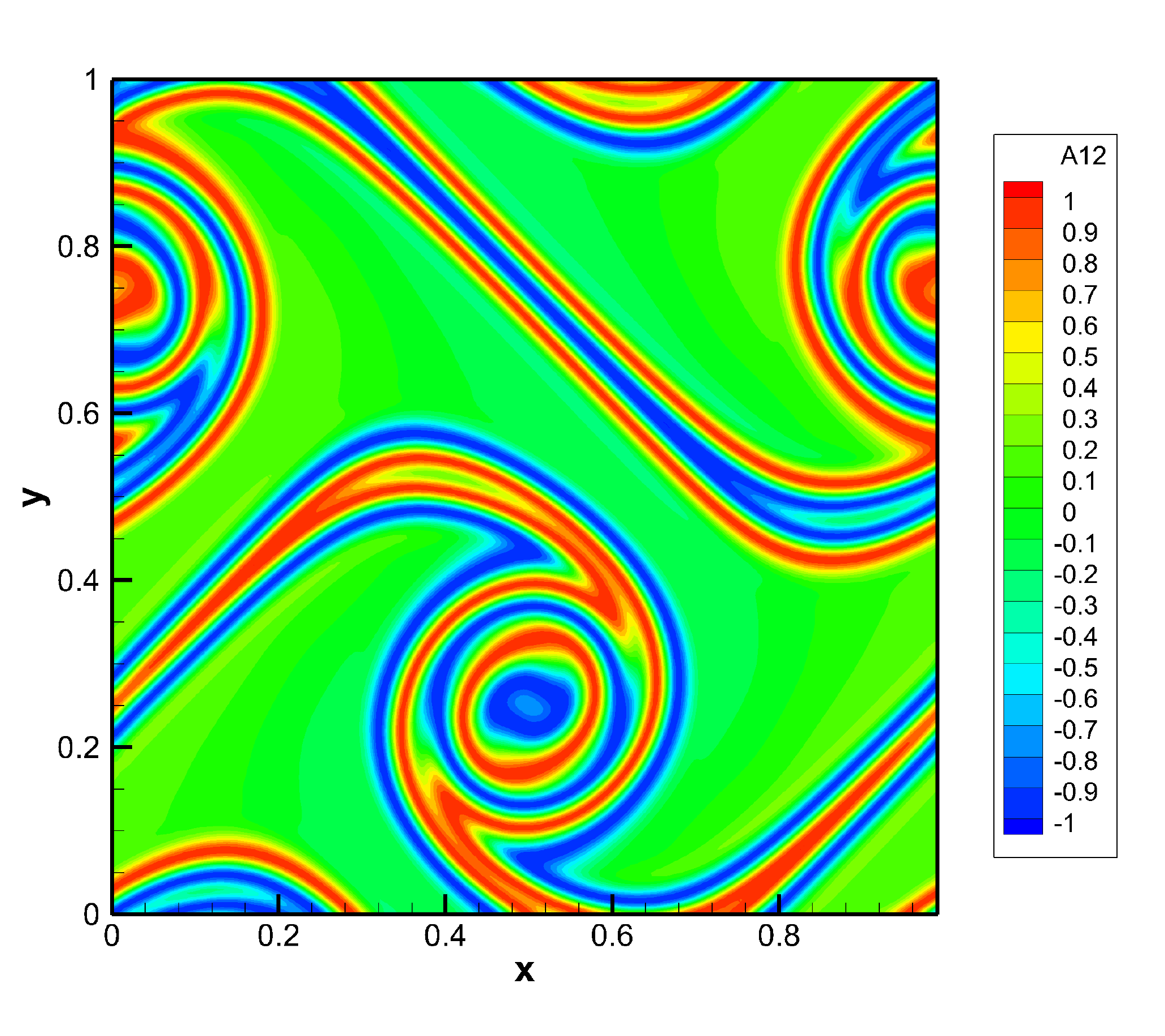}  & 
		\includegraphics[width=0.47\textwidth]{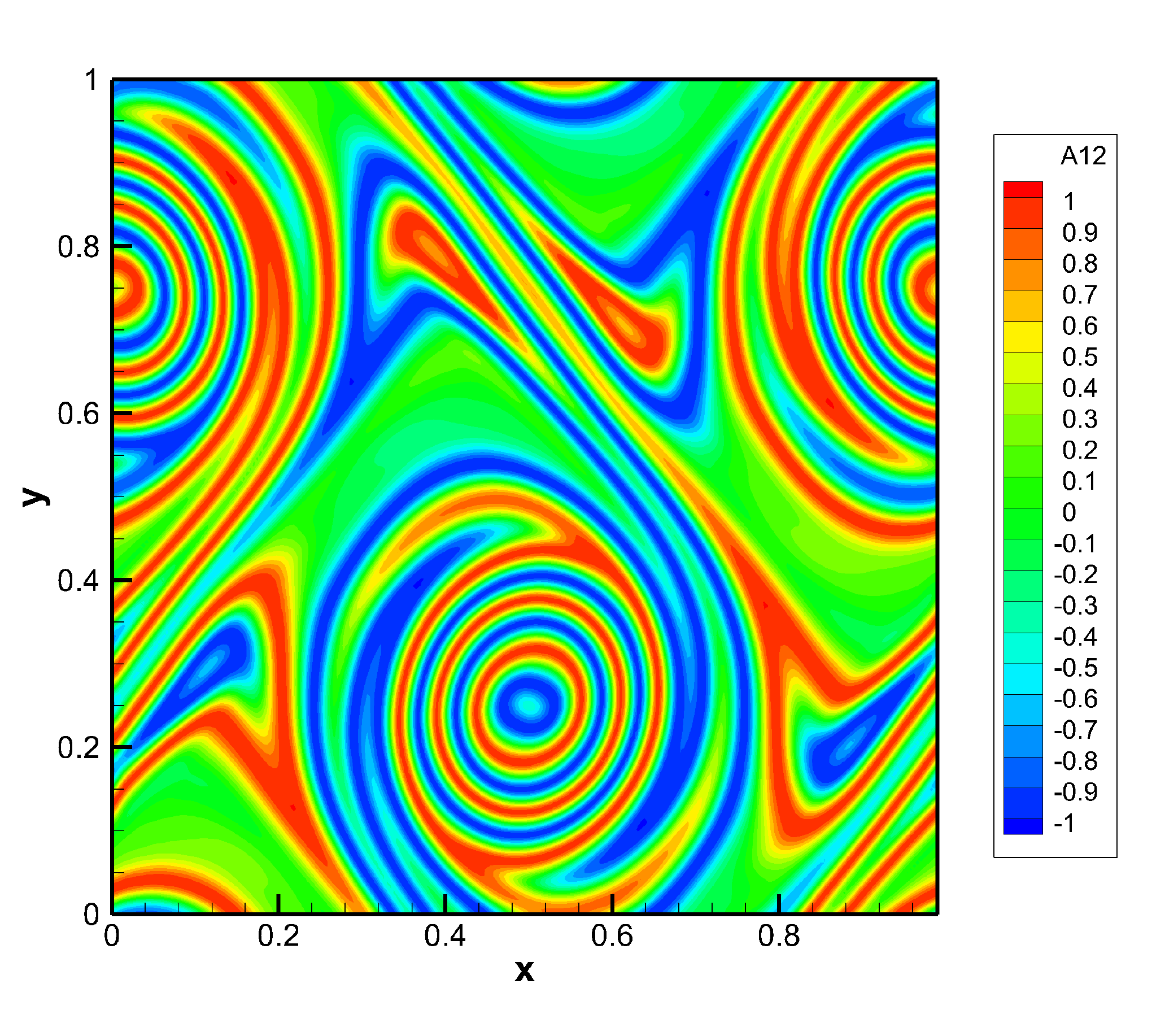}   
		\end{tabular} 
		\caption{Distortion field component $A_{12}$ obtained for the double shear layer problem at times $t=0.4$, $t=0.8$, $t=1.2$ and $t=1.8$ by solving the GPR model with the structure-preserving semi-implicit finite volume scheme 
		in the stiff relaxation limit ($\mu = 2 \cdot 10^{-4}$). } 
		\label{fig.dslA}
	\end{center}
\end{figure}

In Fig. \ref{fig.dslA} we show the temporal evolution of the distortion field component $A_{12}$, which matches well with the results shown in \cite{GPRmodel}. As already observed in \cite{GPRmodel}, the components of the field $\AAA$ seem to be excellent candidates for \textit{flow visualization}, since they reveal even more details of the flow structures than the vorticity plotted in the previous Figure \ref{fig.dslrot}. Another major advantage is that the field 
$\AAA$ is part of the state vector $\mathbf{Q}$ that is directly evolved in time via the governing PDE. It is \textit{not} a derived
flow quantity as the vorticity, which needs to be computed from the velocity field via appropriate post-processing techniques.


\subsection{Couette flow at low Mach number} 

\begin{figure}[!htbp]
	\begin{center}
		\begin{tabular}{cc} 
			\includegraphics[width=0.4\textwidth]{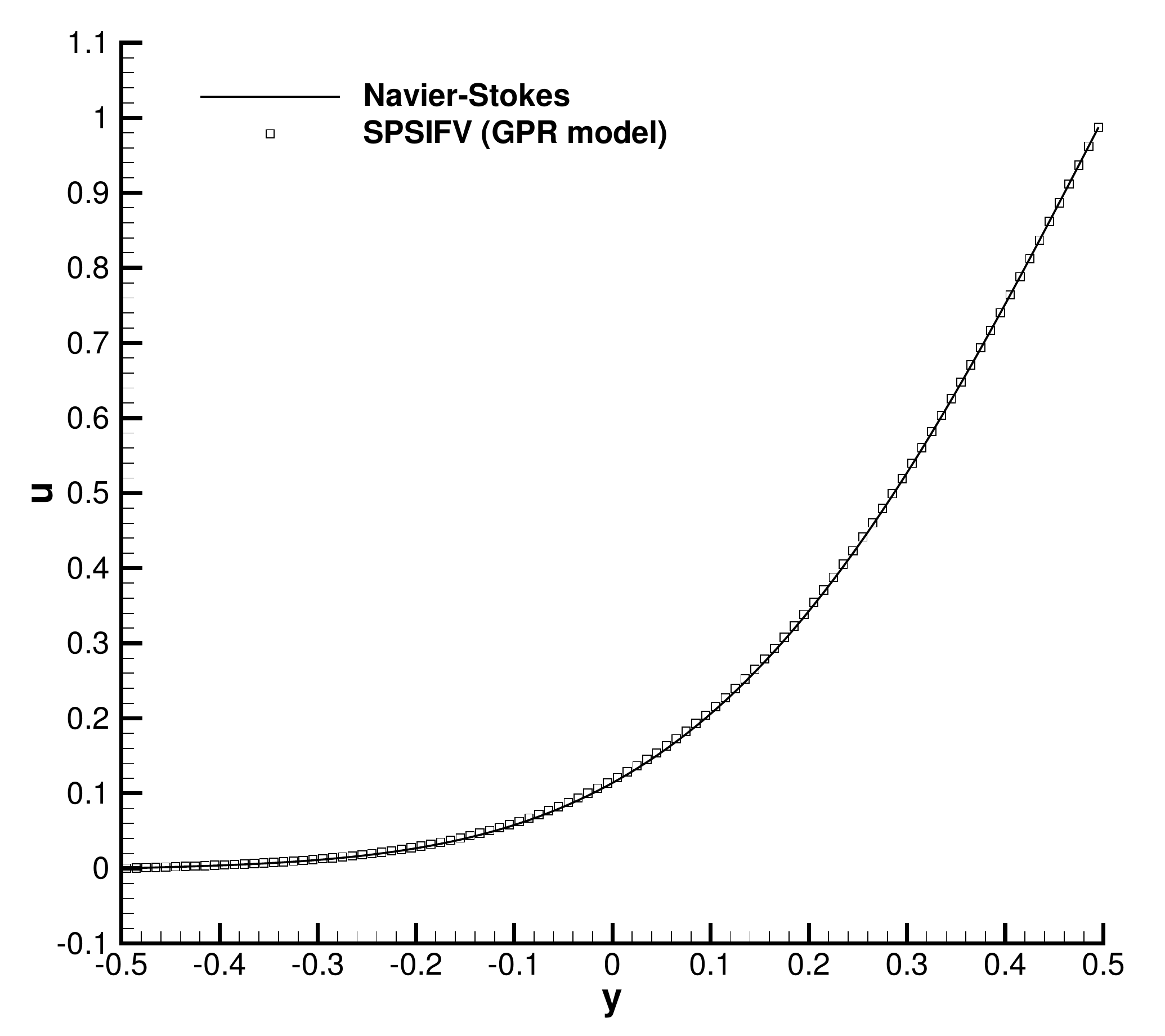}    & 
			\includegraphics[width=0.4\textwidth]{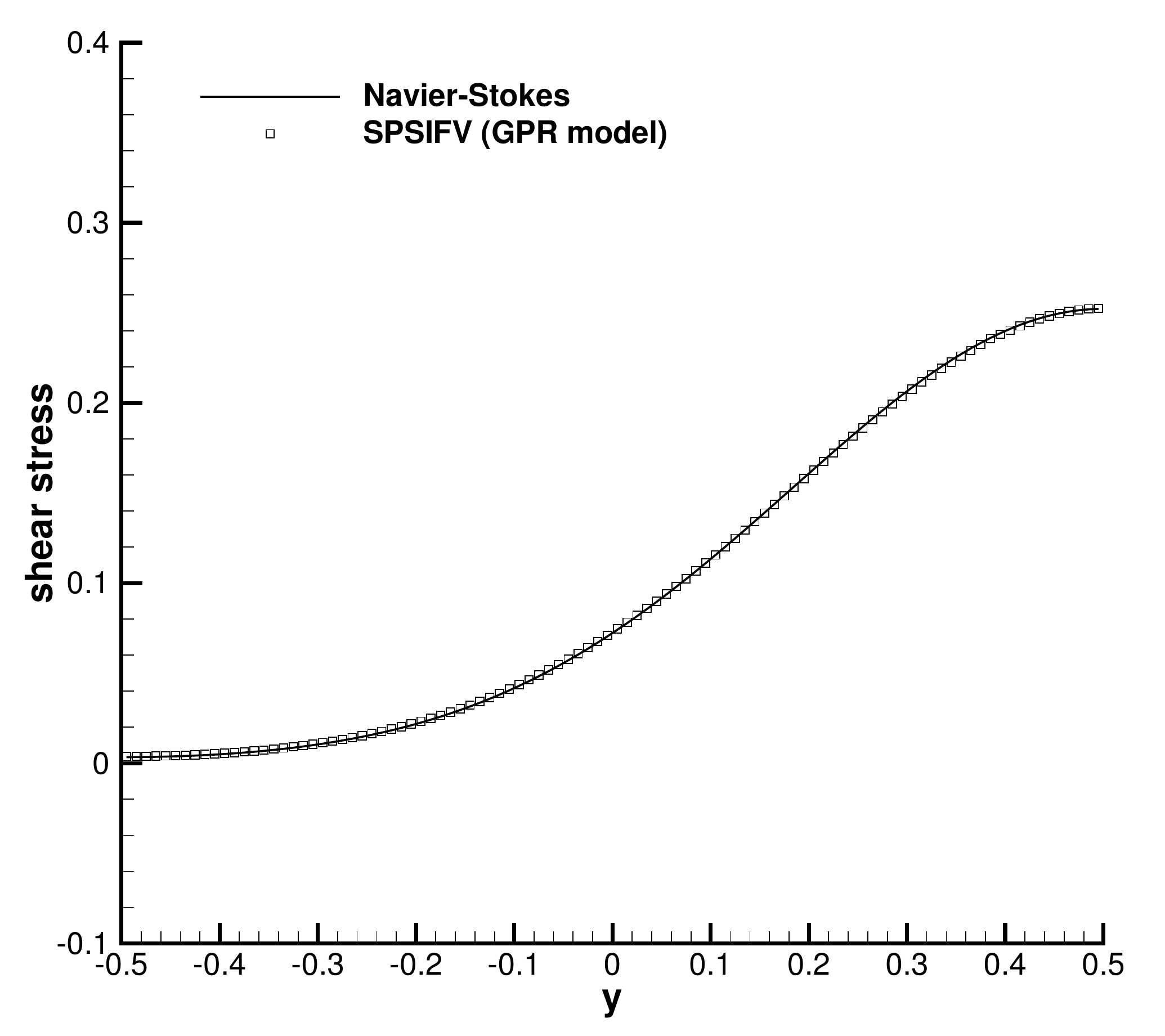}  \\  
			\includegraphics[width=0.4\textwidth]{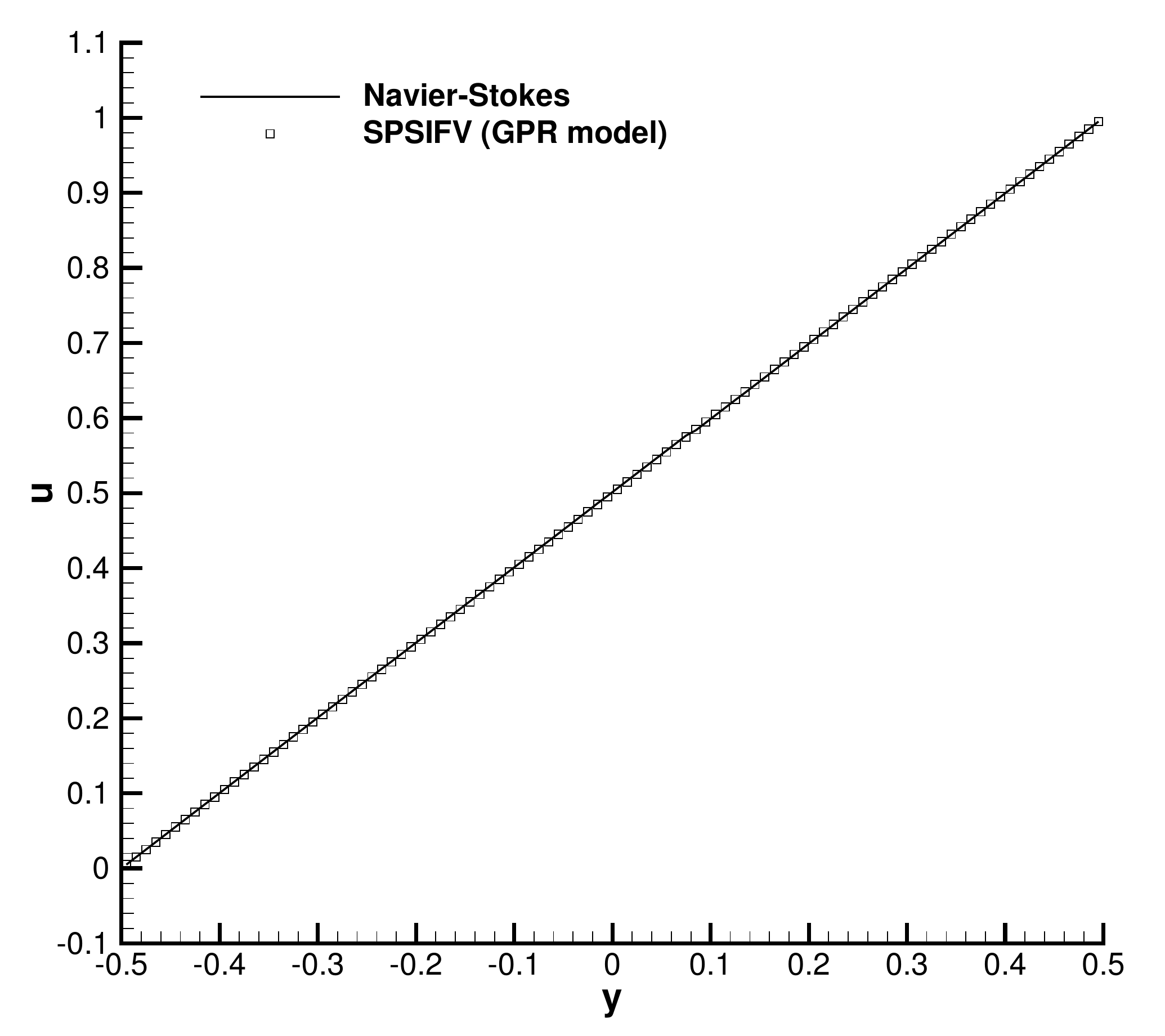}    & 
			\includegraphics[width=0.4\textwidth]{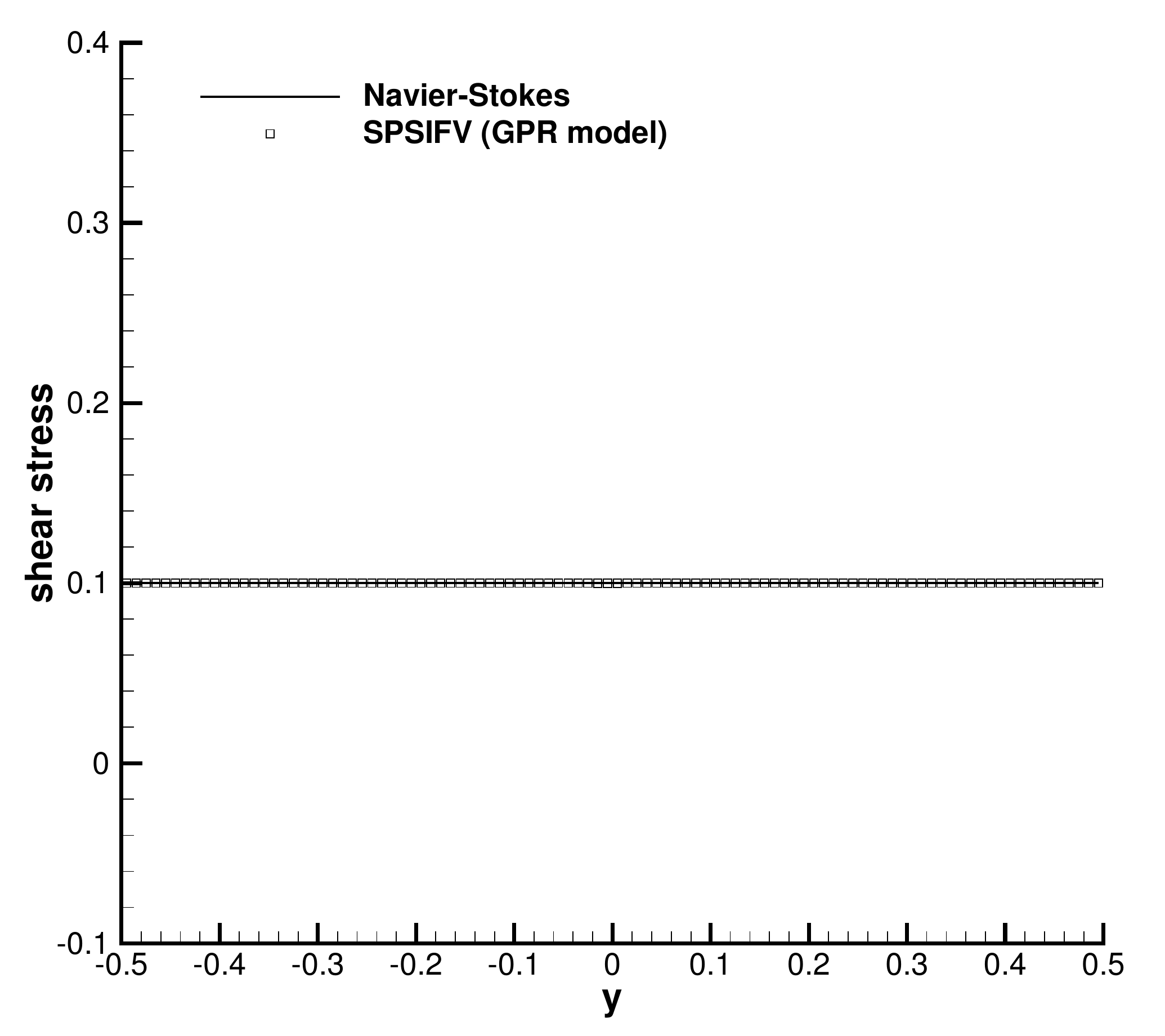}   
		\end{tabular} 
		\caption{Couette flow at Mach number $M=10^{-2}$ and Reynolds number $Re=10$. Results obtained with the new SPSIFV scheme applied to the GPR model and comparison with the Navier-Stokes reference solution at time $t=0.5$ (top row) and $t=100$ (bottom row). 
			Velocity component $u$ (left) and shear stress tensor component $\sigma_{12}$ (right).   } 
		\label{fig.couetteRe10}
	\end{center}
\end{figure}

\begin{figure}[!htbp]
	\begin{center}
		\begin{tabular}{cc} 
			\includegraphics[width=0.4\textwidth]{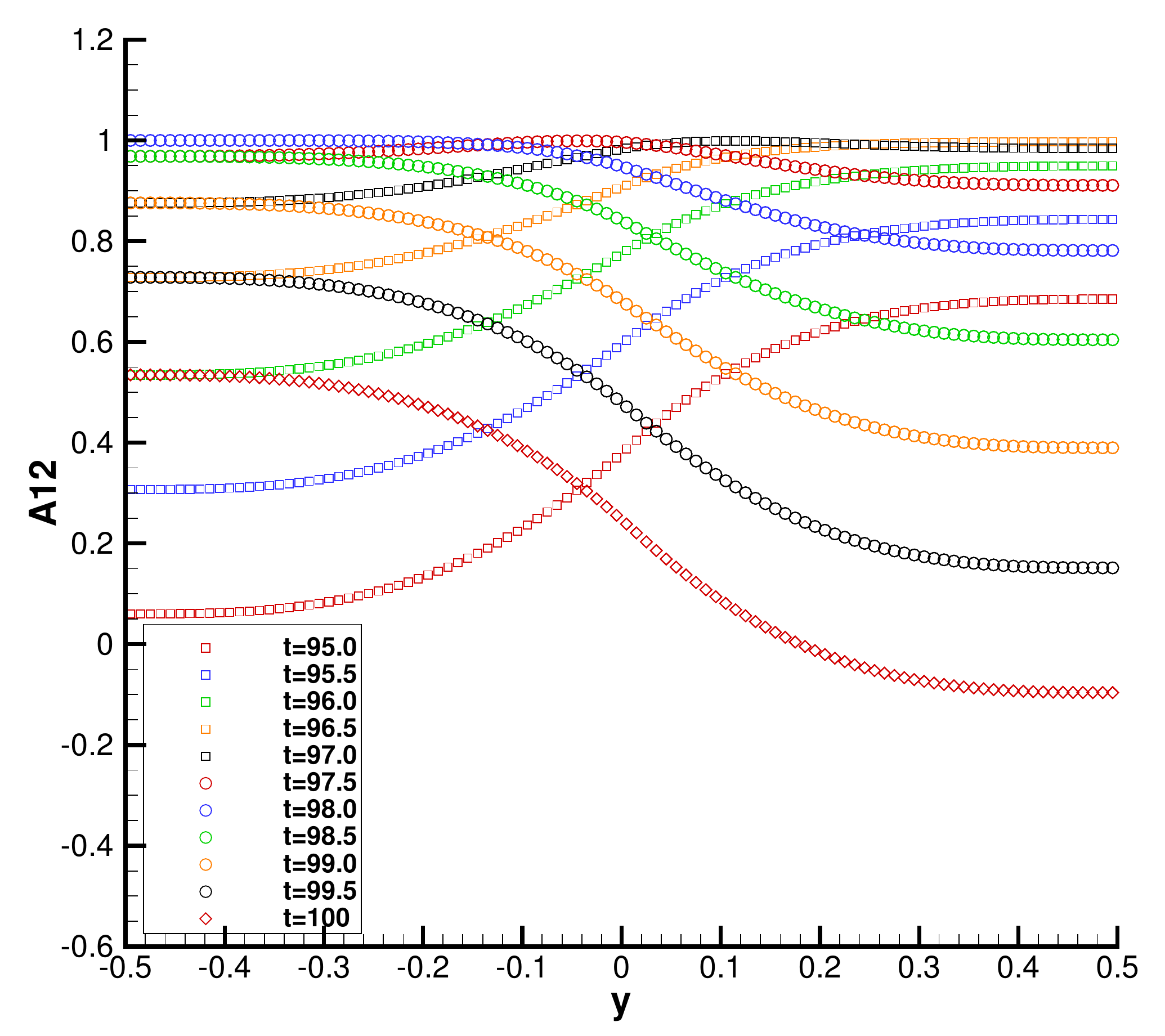}    & 
			\includegraphics[width=0.4\textwidth]{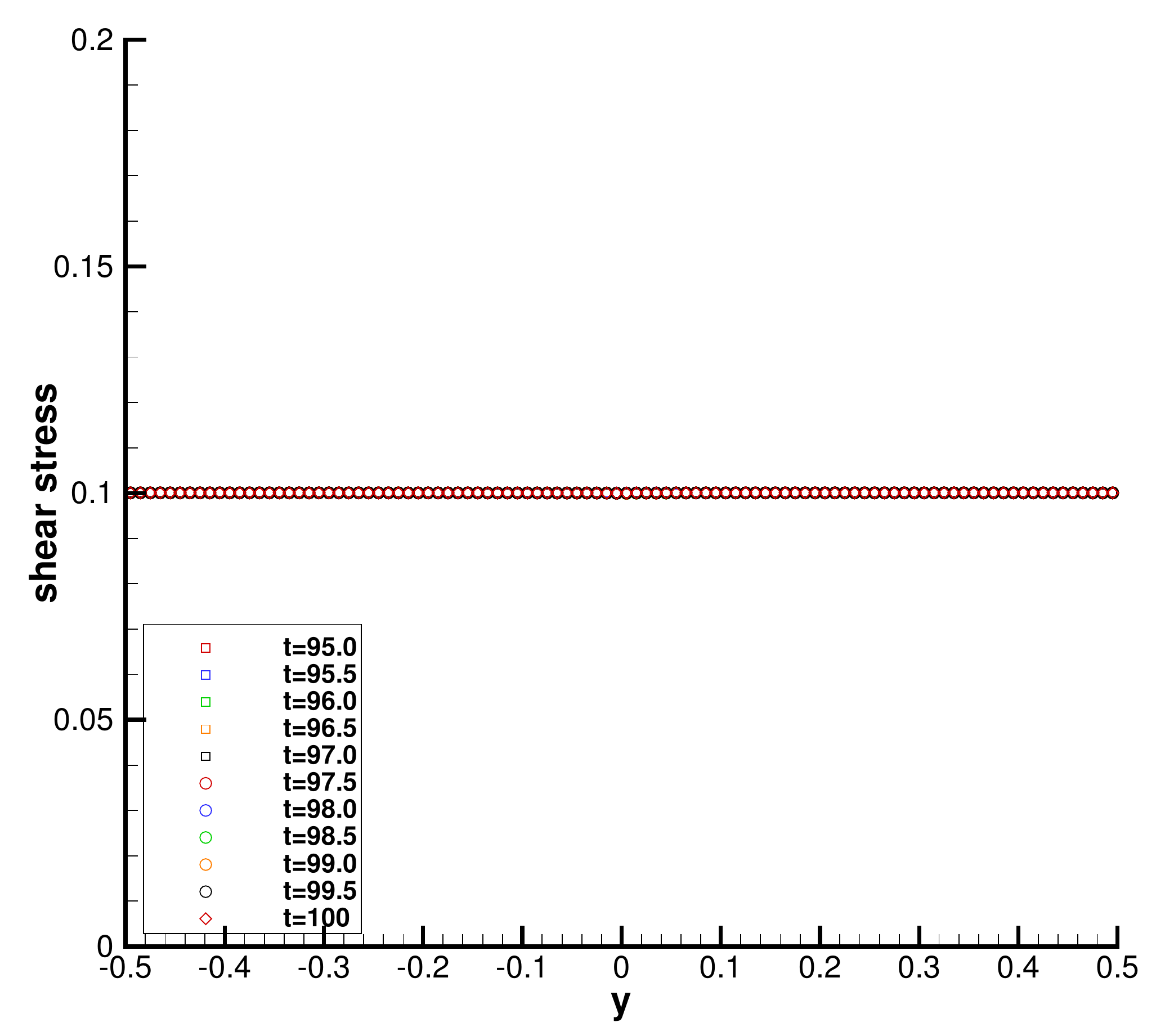}  
		\end{tabular} 
		\caption{Couette flow at Mach number $M=10^{-2}$ and Reynolds number $Re=10$. Temporal 
		evolution of the distortion field component  $A_{12}$ from $t=95$ to $t=100$ (left) and 
		corresponding temporal evolution of the shear stress component $\sigma_{12}$ (right). It 
		can clearly be noticed how the unsteady distortion field $A_{ik}$ corresponds to a 
		perfectly stationary shear stress $\sigma_{ik}$. } 
		\label{fig.couetteRe10AS}
	\end{center}
\end{figure}

In this section we discuss the Couette flow \cite{boundaryLayer}, which is a very elementary but important flow that allows to measure the viscosity of a fluid in rotational viscometers.
The setup of the problem used in this paper is as follows: the computational domain is $\Omega = [-0.5,+0.5]^2$, covered by $4 \times 100$ control volumes. The boundary conditions in 
$x$-direction are periodic. The initial condition for the velocity is $\mathbf{v}=0$, while density is set to $\rho=1$ and pressure to $p=10^4 / \gamma$ everywhere in $\Omega$. The initial
condition for the distortion field is $\AAA = \mathbf{I}$ and the thermal impulse is set to $\J = 0$. The remaining parameters of the GPR model are chosen as $\gamma = 1.4$, 
$\rho_0 = 1$, $c_s = 8$ and $\alpha = 0$. The fluid is set in motion via the upper wall that is 
moving with velocity $\mathbf{v}=(1,0,0)$, while the lower wall in $y=-0.5$ is a non-moving
no-slip wall. We set the relaxation time $\tau_1$ so that the viscosity coefficient results as $\mu = 0.1$, hence the Reynolds number based on the wall velocity and the characteristic
length of the problem is $Re=10$. The obtained computational results are shown for various times in Fig. \ref{fig.couetteRe10}, where we also compare the results of the new SPSIFV scheme 
applied to the GPR model with the semi-implicit finite volume scheme proposed in \cite{DumbserCasulli2016} for the compressible Navier-Stokes equations. We report data for the velocity profile
$u(y)$ and of the shear stress tensor component $\sigma_{12}(y)$. One can observe that the match between the numerical solution of the GPR model and the Navier-Stokes reference solution is excellent. 
For long times ($t=100$), the well-known linear velocity profile of the Couette flow is reached in both cases and the shear stress becomes constant along the $y$ axis. 
At this point, it is very interesting to note that the shear stress becomes a constant in both space and time, but the distortion field $\AAA$ is \textit{not} constant, neither in space 
nor in time, see Fig. \ref{fig.couetteRe10AS}, where the time evolution of the shear stress component $\sigma_{12}$ is plotted together with the time evolution of $A_{12}$ at late times ($95 \leq t \leq 100$). 
This behaviour was already discussed in \cite{PeshRom2014} and explained by the rotations of fluid elements, which is a geometric information that is contained in the GPR model, but not in the Navier-Stokes 
equations. Therefore, the Couette flow is \textit{not} a stationary solution of the GPR model, while it is a stationary solution of the Navier-Stokes equations.

\subsection{Poiseuille flow at low Mach number} 

\begin{figure}[!htbp]
	\begin{center}
		\begin{tabular}{ccc} 
			\includegraphics[trim= 0 10 10 0,clip,width=0.3\textwidth]{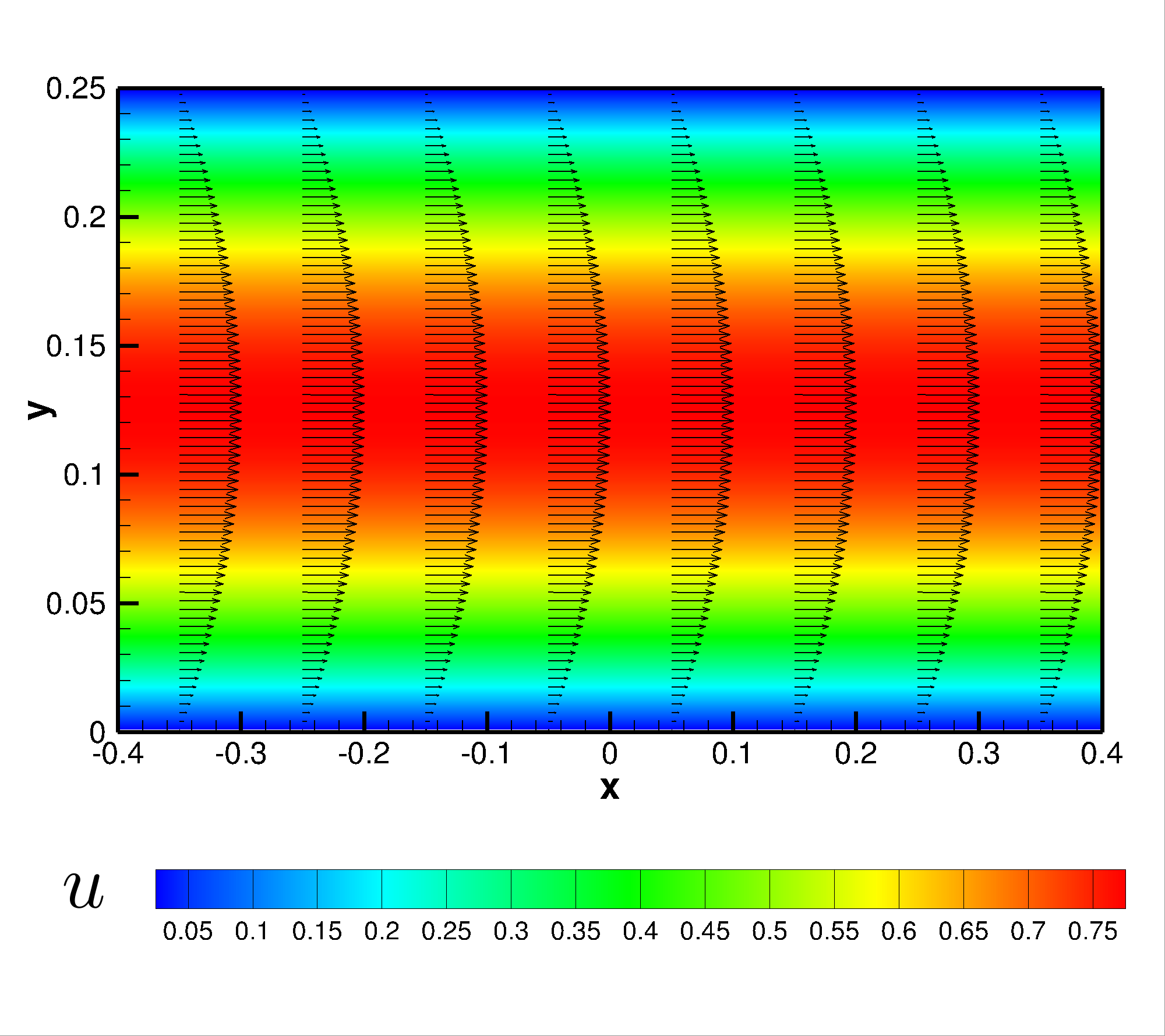}  & 
			\includegraphics[trim= 0 10 10 0,clip,width=0.3\textwidth]{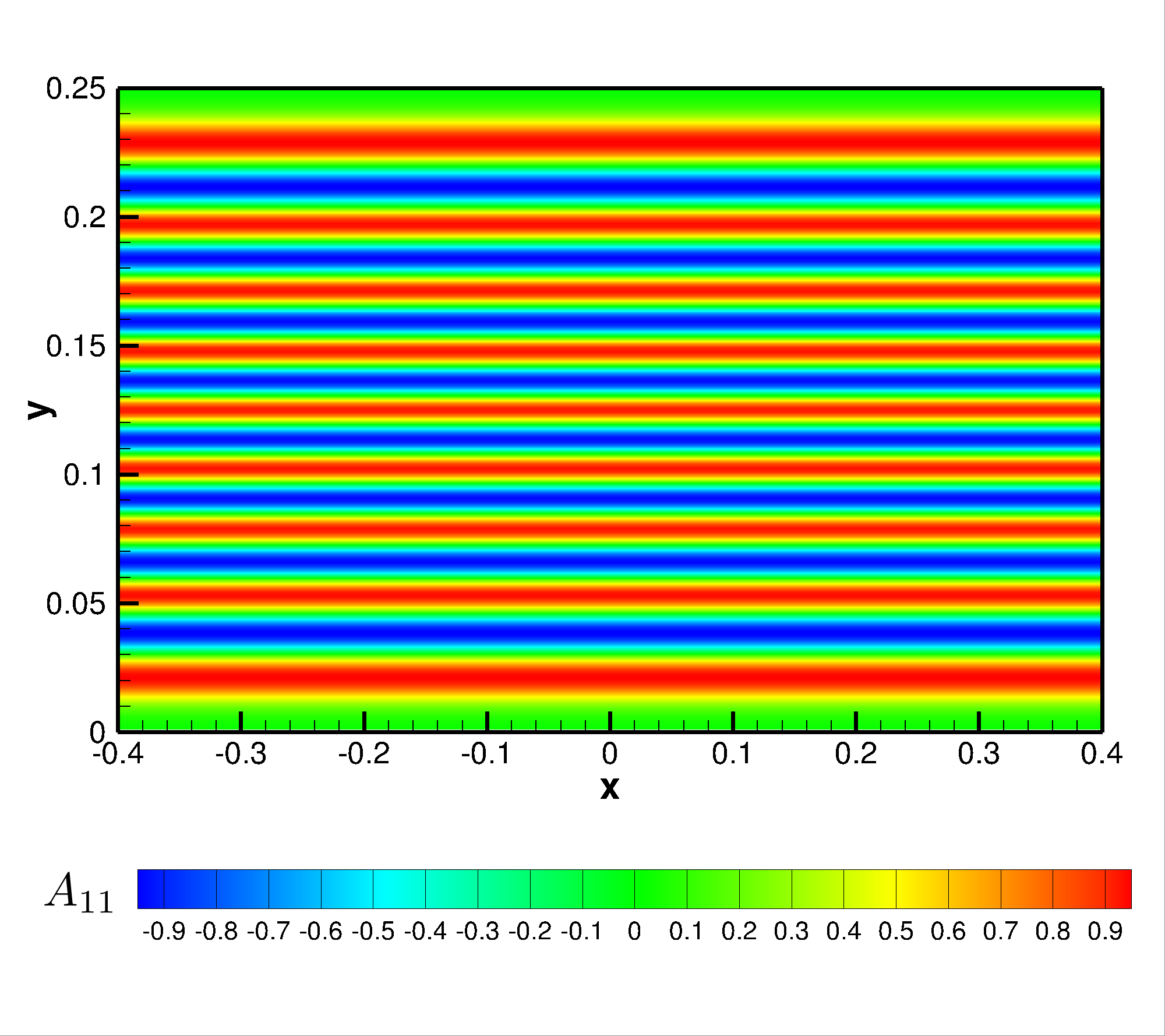}  & 
			\includegraphics[trim= 0 10 10 0,clip,width=0.3\textwidth]{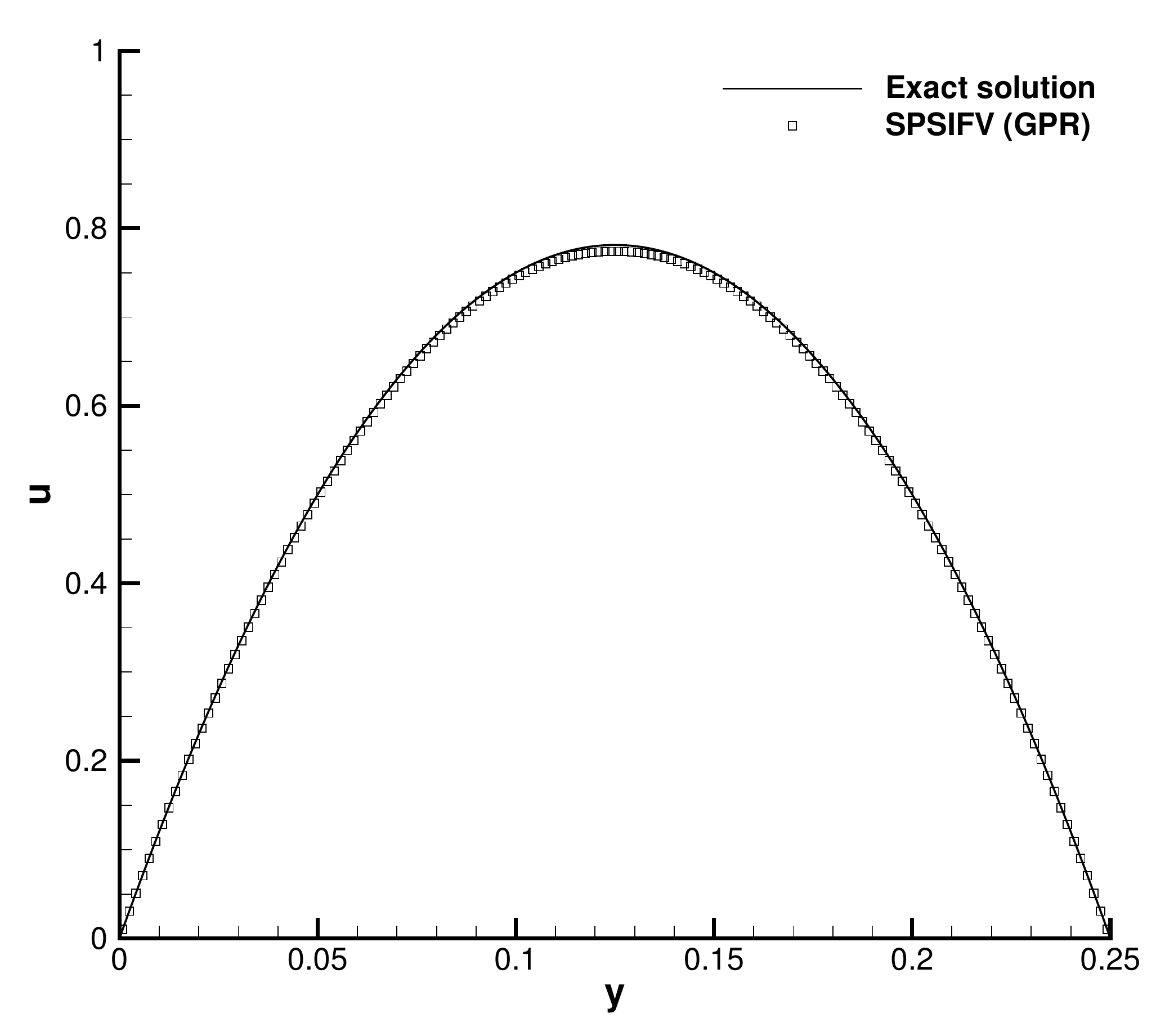}   
		\end{tabular} 
		\caption{Poiseuille flow at Mach number $M=6.6 \cdot 10^{-3}$ and Reynolds number 
		$Re_d=19.5$. Results obtained at time $t=10$ 	with the new SPSIFV scheme applied to the 
		GPR 
		model. Velocity vectors and color contours of the velocity component $u$ (left), $ A_{11} $ 
		component fo the distortion field (middle), and comparison of the velocity component $u$ on 
		a 1D cut along the $y$ axis with the exact solution of the Hagen-Poiseuille flow (right).  
		} 
		\label{fig.poiseuille}
	\end{center}
\end{figure}

The test case considered here is the steady flow of a viscous Newtonian fluid in a rectangular duct of length $L$ and diameter 
$d$ in the presence of a constant source term $\mathbf{S} = \left( 0, \mathbf{g}, 0, 0, \mathbf{g} \cdot \mathbf{v} \right)^T$ that is added to the right hand side of the PDE system \eqref{eqn.GPR} in order to drive the flow and which can be discretized explicitly together with the convective terms without changing anything else in the numerical scheme. We choose $\mathbf{g}=(1,0,0)^T$ and a computational domain $\Omega = [-L/2,L/2] \times [0,d]$ with $L=1$ and $d=0.25$ and 
periodic boundary conditions in $x$-direction. The source term replaces the pressure gradient $ \partial p / \partial x$ that  usually drives the \emph{Hagen-Poiseuille flow}, which is a well-known solution of the Navier-Stokes equations and which leads to 
the parabolic velocity profile  
\begin{equation} 
\label{eqn.hp}
u(y) = - \halb \frac{1}{L} \frac{\rho}{\mu} y(y-d). 
\end{equation} 
If we run the problem in the low Mach number regime, we can expect Eqn. (\ref{eqn.hp}), which is valid for an incompressible fluid, to hold also for the weakly compressible case. We therefore choose the following initial data and set of parameters: 
 $\gamma=1.4$, $\rho_0=1$, 
$c_v=1$, $c_s=8$, $p=10^4$, $u=v=0$, $\AAA=\mathbf{I}$, $\mathbf{J}=0$, $\alpha=\kappa=0$ and 
$\mu=10^{-2}$. The maximum flow velocity
is $u_{\max} = 0.78125$, which means the Mach number is $M=6.6 \cdot 10^{-3}$ and the Reynolds number based on the diameter is $Re_d = u_{\max} d / \nu = 19.5$. 
The computational domain was discretized with the new structure-preserving semi-implicit finite volume scheme using 
$10 \times 150$ cells and running the problem until a final time of $t=10$ so that the solution becomes stationary. 
The obtained computational results are shown in Figure \ref{fig.poiseuille}, where the velocity contours, the velocity vectors
and a comparison with the exact solution of the Hagen-Poiseuille profile \eqref{eqn.hp} are shown.  
One can observe that the obtained numerical results are in good agreement with the reference 
solution. Similar to the previous examples of fluid flows, we note a highly heterogeneous profile 
of the distortion field in Figure 17 despite homogeneous and stationary profile of the velocity and 
thus, of the viscous stress.

\subsection{Lid-driven cavity at low Mach number} 

A classical benchmark problem for the numerical solution of the incompressible Navier-Stokes equations is the well-known lid-driven  cavity problem, see e.g. \cite{Ghia1982,TavelliNS}. It can also be used to validate compressible flow solvers in the low Mach number  regime, see e.g. \cite{GPRmodel,TavelliDumbser2017,DumbserCasulli2016}. 
For the computational setup in this paper the computational domain is given by $\Omega = [-0.5,0.5] \times [-0.5,0.5]$. The initial
condition is simply given by $\rho=1$, $\mathbf{v}=0$, $p=10^4 / \gamma$,  $\AAA=\mathbf{I}$ and $\mathbf{J}=0$. The parameters of 
the GPR model are set to $\gamma=1.4$, $c_v = 1$, $c_s = 8$, $\rho_0=1$ and $\alpha=0$, i.e. heat 
conduction is neglected.  
The dynamic viscosity is chosen as $\mu=10^{-2}$ so that the Reynolds number of the test problem is $Re=100$. 
The fluid flow inside the cavity is induced by the moving upper boundary, whose velocity is set to $\mathbf{v}=(1,0,0)$. On all  
other boundaries, a no-slip wall boundary condition with $\mathbf{v}=0$ is imposed. 
With the chosen initial and boundary conditions, the Mach number of this test problem is $M=10^{-2}$ with respect to the lid 
velocity. 

\begin{figure}[!htbp]
	\begin{center}
		\begin{tabular}{cc} 
			\includegraphics[width=0.47\textwidth]{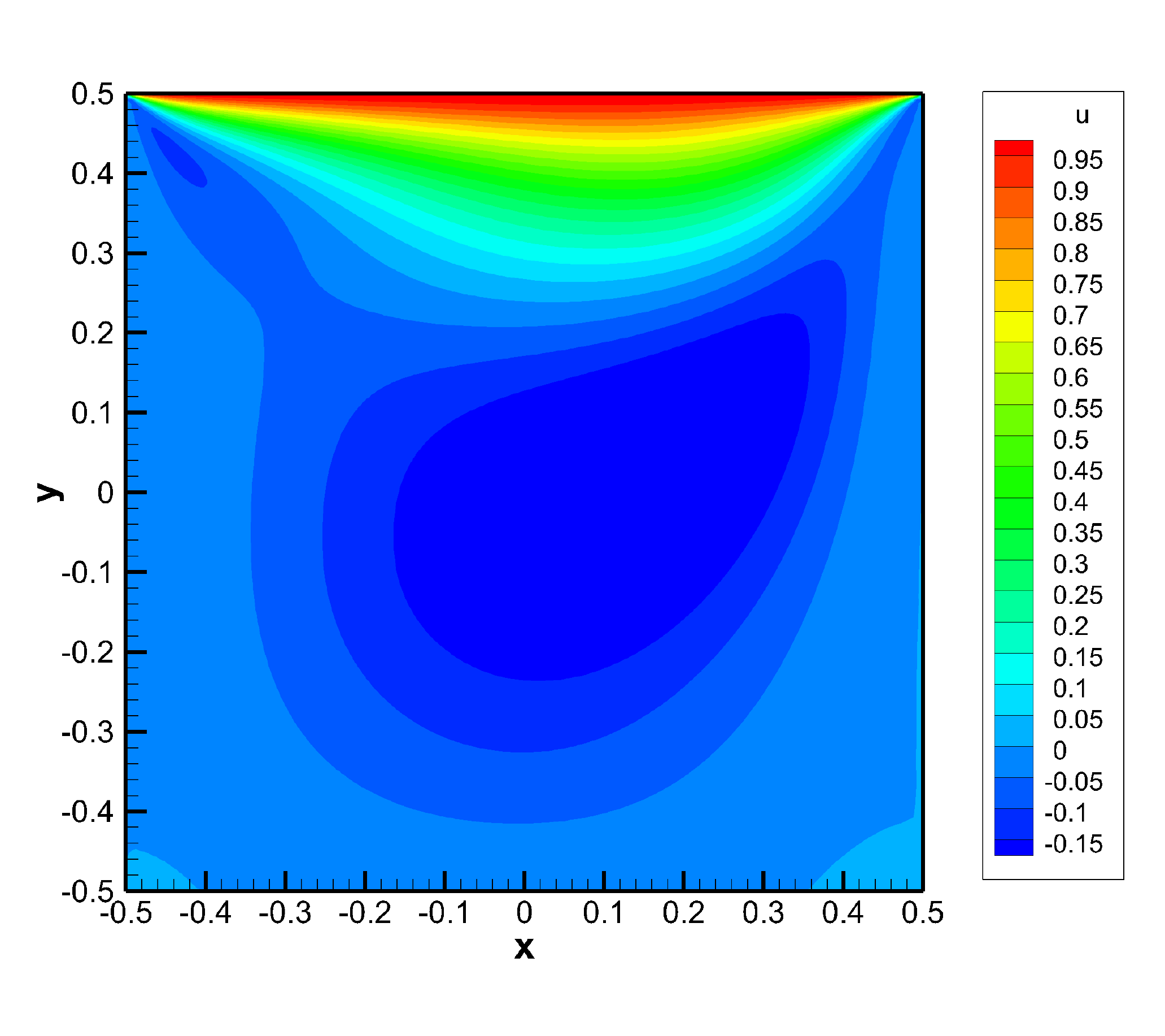}  & 
			\includegraphics[width=0.47\textwidth]{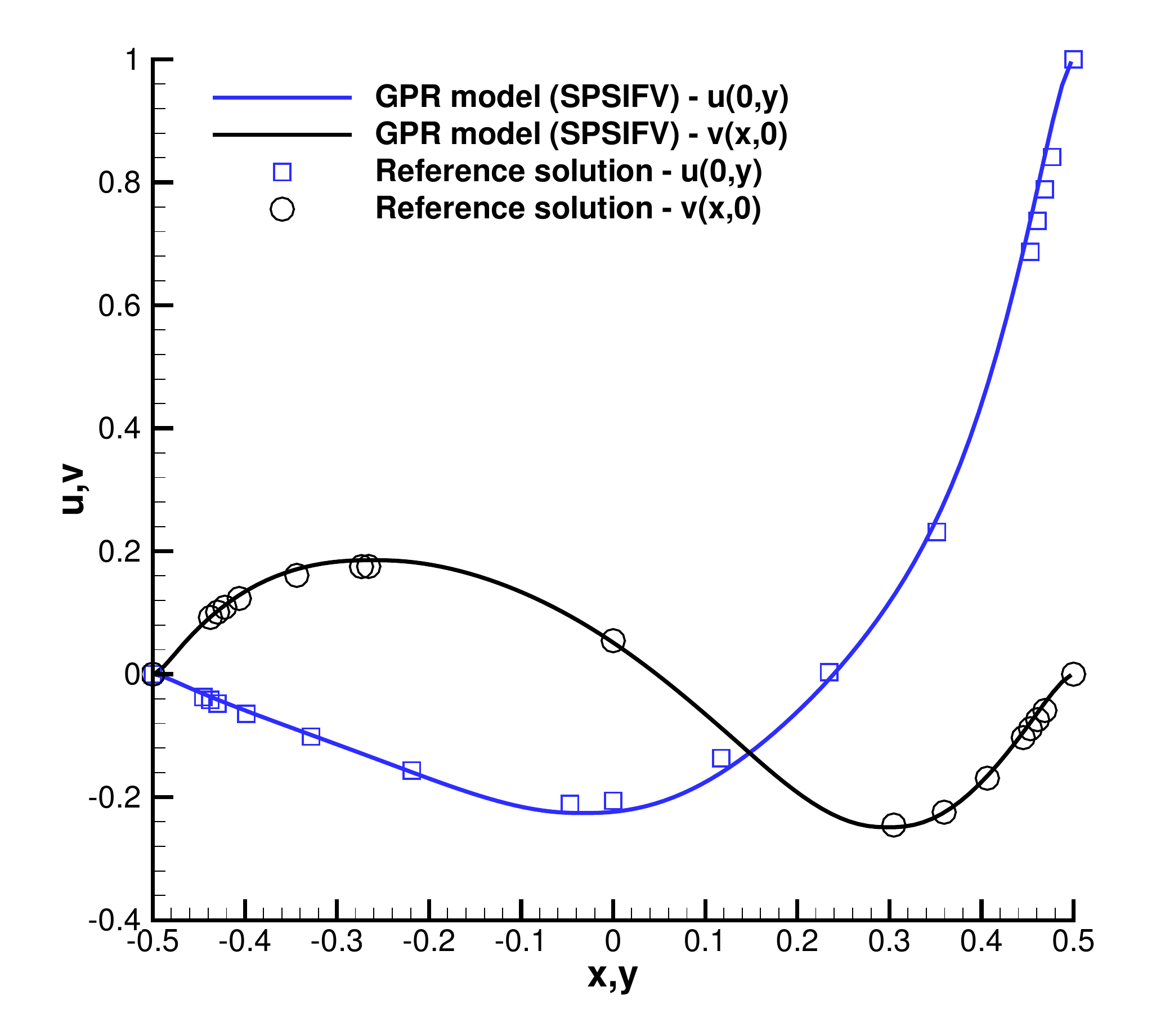}   
		\end{tabular} 
		\caption{Lid-driven cavity problem at Mach number $M=10^{-2}$ and Reynolds number $Re=100$. Results obtained at time $t=10$ with the new SPSIFV scheme applied to the GPR model. Color contours of the velocity component $u$ (left) and  
				 comparison of the velocity components $u$ and $v$ on 1D cuts along the $x$ and $y$ axis with the reference solution of Ghia \textit{et al.} \cite{Ghia1982} (right).  } 
		\label{fig.cavity}
	\end{center}
\end{figure}

The new structure-preserving semi-implicit finite volume scheme is run until a final time of $t=10$ using a computational grid 
composed of $200 \times 200$ elements. The numerical results are shown in Fig. \ref{fig.cavity}, where also a comparison with 
the Navier-Stokes reference solution of Ghia \textit{et al.} \cite{Ghia1982} is provided. We note a very good agreement between the  
numerical solution of the GPR model and the incompressible Navier-Stokes reference solution. At this point we would like to 
stress that the new SPSIFV scheme is able to solve this test problem efficiently also at low Mach number, while an explicit
method as the one used in \cite{GPRmodel} would require a very large number of time steps due to the CFL condition based on 
the sound speed. 

At this point, we also provide a quantitative performance comparison of the new SPSIFV scheme with a classical second order
MUSCL-Hancock TVD finite volume method, see \cite{toro-book}. All runs are performed on one single core of an Intel i9-7900X 
CPU with 3.3 GHz nominal clock speed and 32 GB of RAM. The total wall clock time needed by the SPSIFV scheme to reach the final 
simulation time of $t=10$ was 2031 s, while the explicit second order MUSCL-Hancock scheme needed 5641 s to complete the simulation. 
The time needed to update one single control volume was $4.2 \mu s$ for the SPSIFV scheme and $1.9 \mu s$ for the explicit 
second order TVD method. 

The computational efficiency of the semi-implicit scheme can be highlighted even further when increasing the initial pressure to $p=10^5$, i.e. by further reducing the
Mach number of the test problem. In this case, the total wall clock time needed by the new SPSIFV scheme to reach $t=10$ was 
2666 s,  while the explicit MUSCL-Hancock scheme needed 20392 s to complete the simulation. Here, the CPU time for one single element 
update was $5.5 \mu s$ for the semi-implicit scheme and $1.9 \mu s$ for the explicit TVD scheme. 

These results clearly show that the semi-implicit scheme is not only faster for this type of low Mach 
number problem, as expected, but that also the absolute computational cost per element and time step is quite competitive with the 
one of the explicit scheme, despite the need to solve a global pressure system in each time step.

\subsection{2D explosion problems} 
In this subsection we consider two circular explosion problems (EP1 and EP2), one in the fluid 
limit of the model, one in the solid limit. Given the computational domain $\Omega=[-1;1]^2$ the 
initial condition reads 
\begin{equation}
\mathbf{Q}(x,y,0)=\begin{cases}
\mathbf{Q}_{in} \quad \text{if} \quad r\leq R \\
\mathbf{Q}_{out} \quad \text{if} \quad r>R. \\
\end{cases}
\end{equation}
where $\mathbf{Q}_{in}$ and $\mathbf{Q}_{out}$ are the internal and external states, respectively, and $r=\sqrt{x^2+y^2}$ is the radial coordinate, while $R=0.5$. 

\begin{figure}[!htbp]
	\begin{center}
		\begin{tabular}{cc} 
			\includegraphics[width=0.47\textwidth]{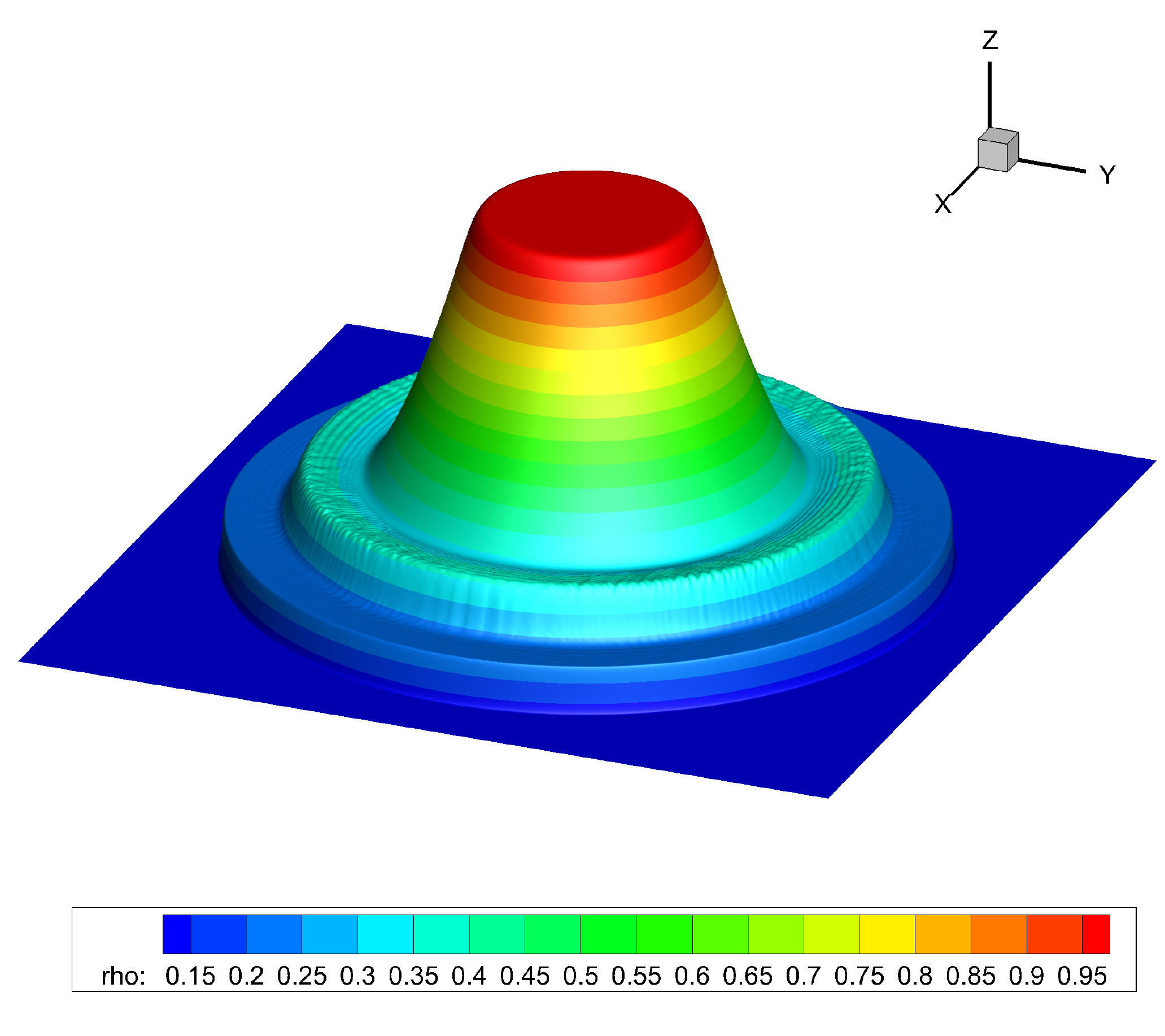}  &
			\includegraphics[width=0.47\textwidth]{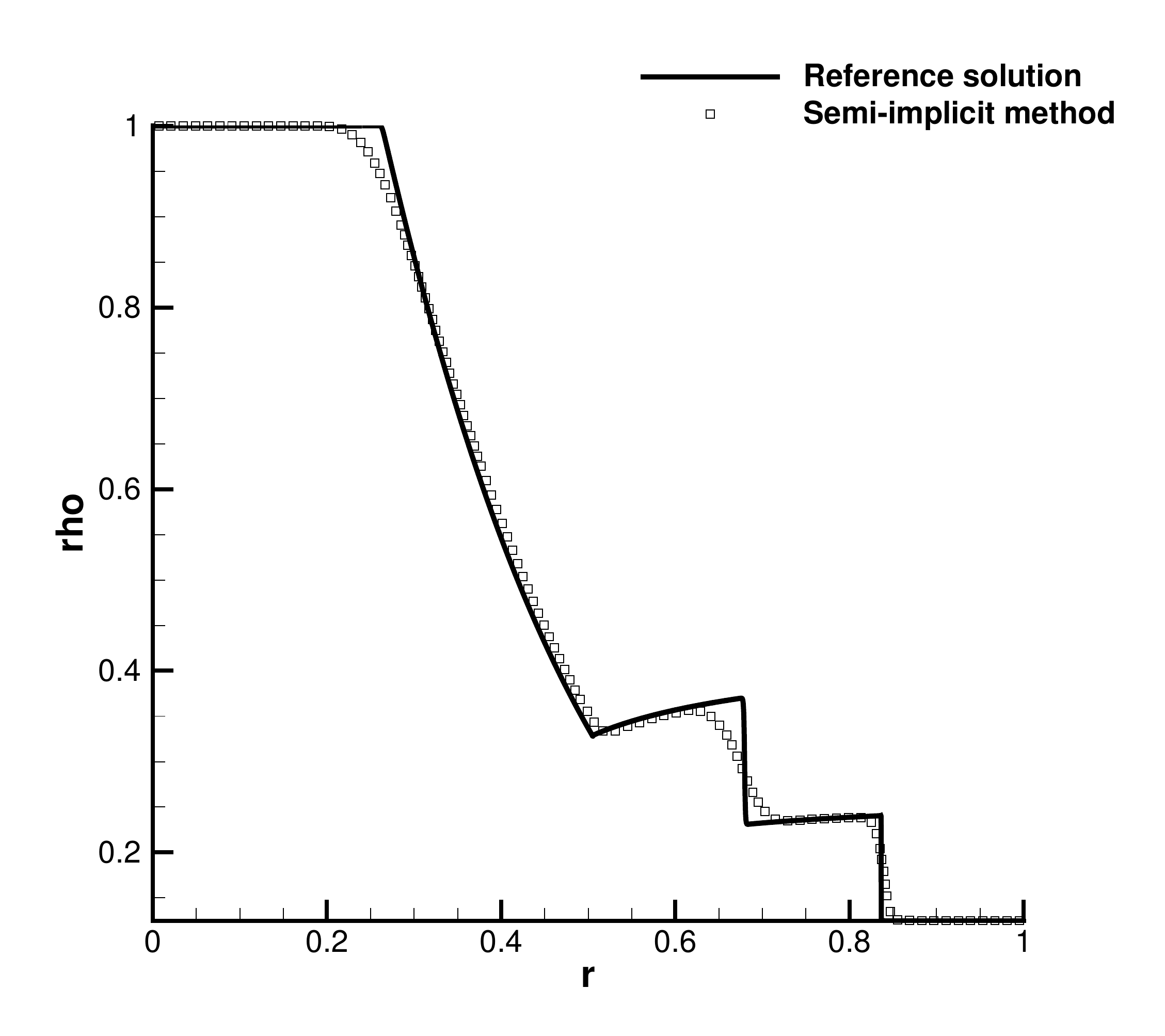}  \\ 
			\includegraphics[width=0.47\textwidth]{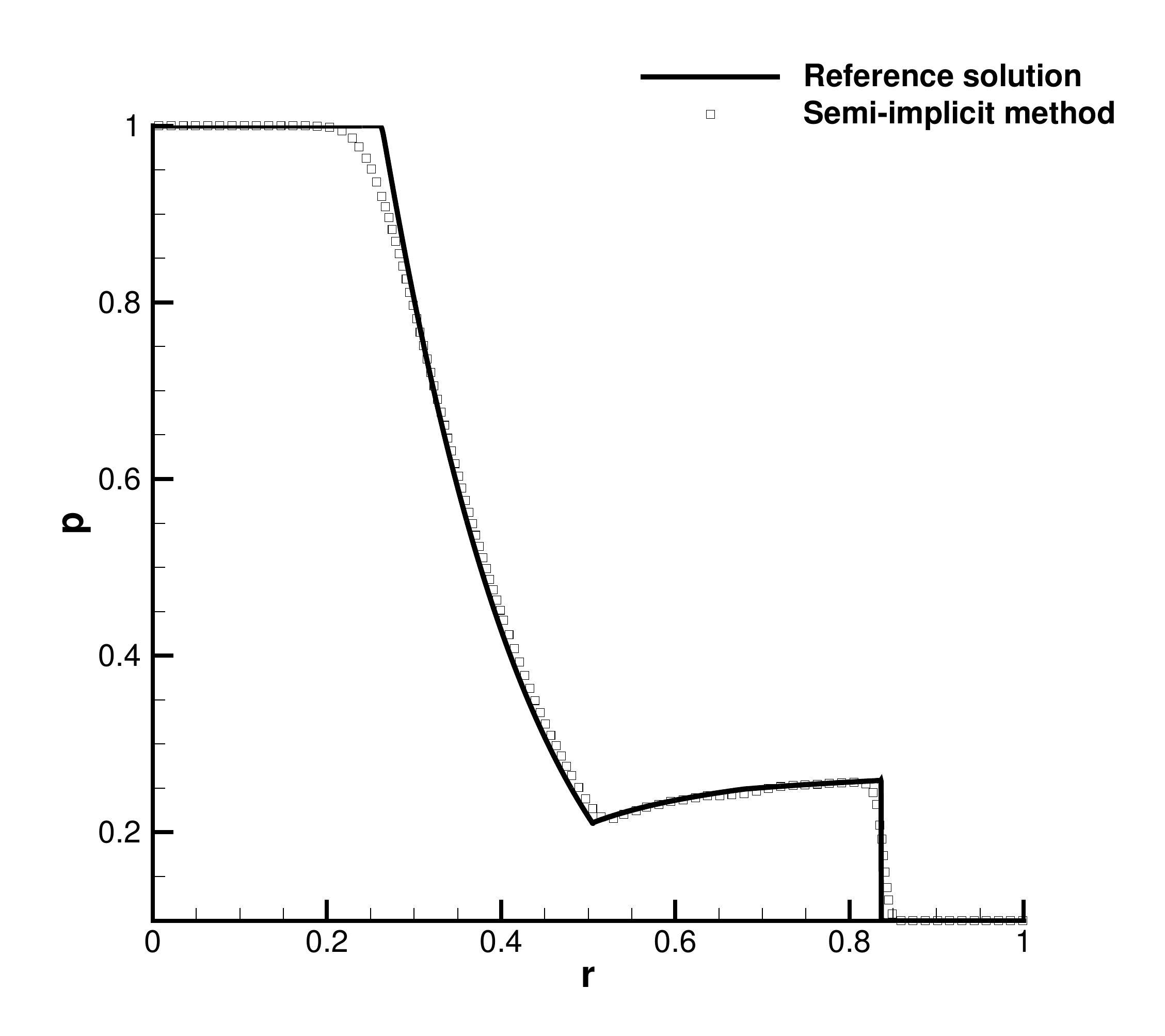}  & 
			\includegraphics[width=0.47\textwidth]{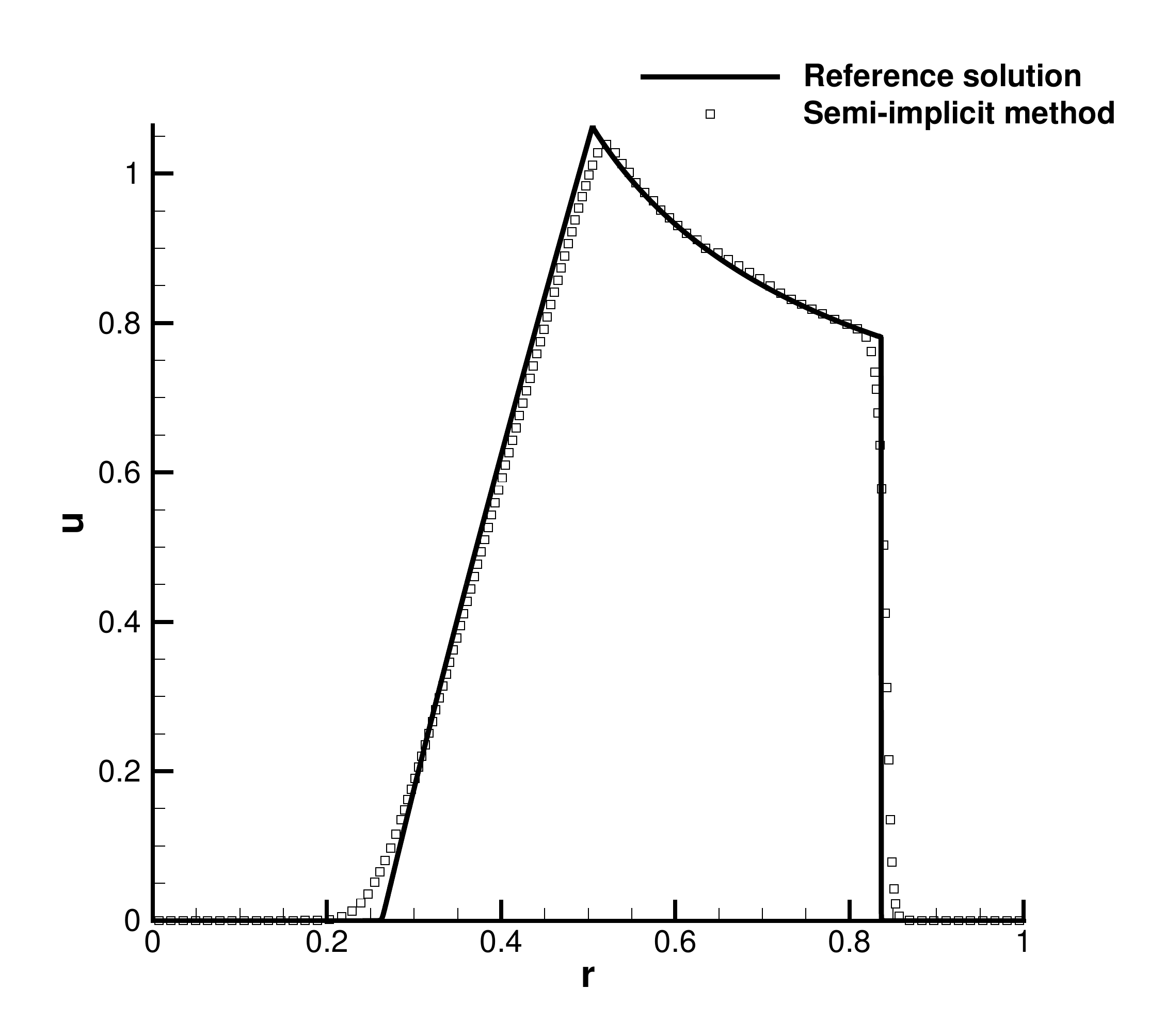}  \\ 
		\end{tabular} 
		\caption{2D explosion problem EP1 at time $t=0.2$. 3D density contour color plot (top left) and 1D cut along the $x$ axis with a  comparison of the Euler reference solution (solid line) against the numerical solution of the GPR model obtained 
		with the new SPSIFV scheme (square symbols) in the stiff relaxation limit ($\tau_1=\tau_2=10^{-5}$).   } 
		\label{fig.ep2dfluid}
	\end{center}
\end{figure}

\paragraph{EP1: Fluid limit of the model} We first solve the governing PDE system in the fluid 
limit, 
i.e. $\tau_1 \ll 1$ and $\tau_2 \ll 1$. In the inner state, the density and the pressure are 
$\rho_{in}=1$ and $p_{in}=1$, while in the outer state we impose $\rho_{out}=0.125$ and 
$p_{out}=0.1$. In addition, in the entire domain the initial velocity is  set to $\mathbf{v}=0$, 
the initial thermal impulse vector is equal to $\mathbf{J}=0$ and the initial distortion field is 
imposed as  $\mathbf{A}=\sqrt[3]{\rho} \, \mathbf{I}$. The other parameters of the GPR model are 
set to $\gamma=1.4$, $c_v=2.5$, $\rho_0=1$, $\tau_1 = \tau_2 = 10^{-5}$. The  computational mesh is 
composed by $1000 \times 1000$ control volumes and the final time of the simulation is $t=0.2$. 
Moreover, we  compute a reference solution by solving the 1D Euler equations of gasdynamics, where 
the cylindrical symmetry was properly accounted for via an algebraic source term in the PDE system, 
see \cite{toro-book} for details. In this case the numerical solution has been obtained  by using a 
robust second order TVD finite volume scheme (see \cite{leer5,toro-book}) on a very fine grid. In 
Figure \ref{fig.ep2dfluid} a density contour plot at the final time is depicted and the numerical 
results obtained with the SPSIFV scheme are compared against 
the 1D reference solution. From the obtained results we can observe that the agreement is very good. 

\begin{figure}[!htbp]
	\begin{center}
		\begin{tabular}{cc} 
			\includegraphics[width=0.47\textwidth]{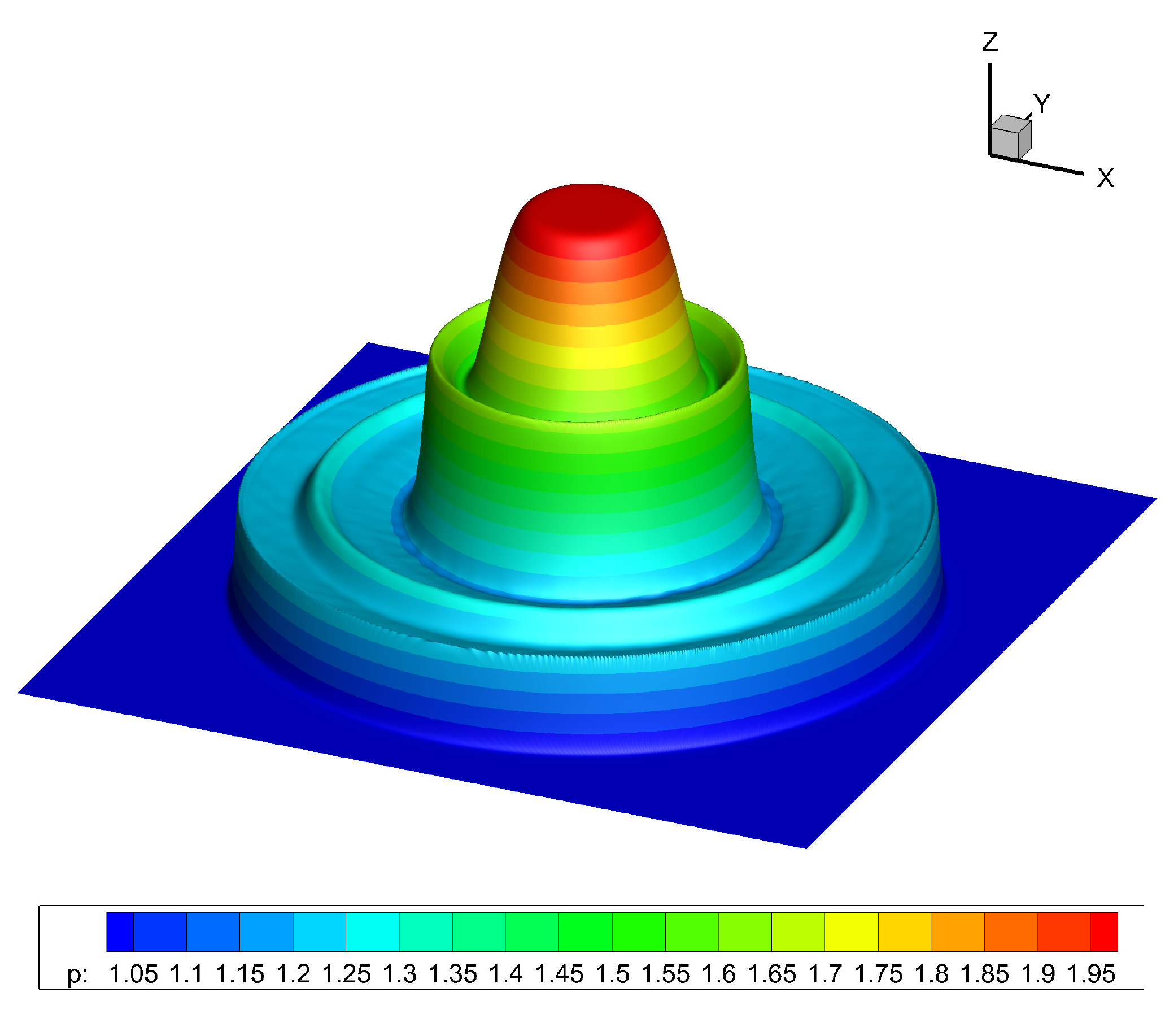}  &
			\includegraphics[width=0.47\textwidth]{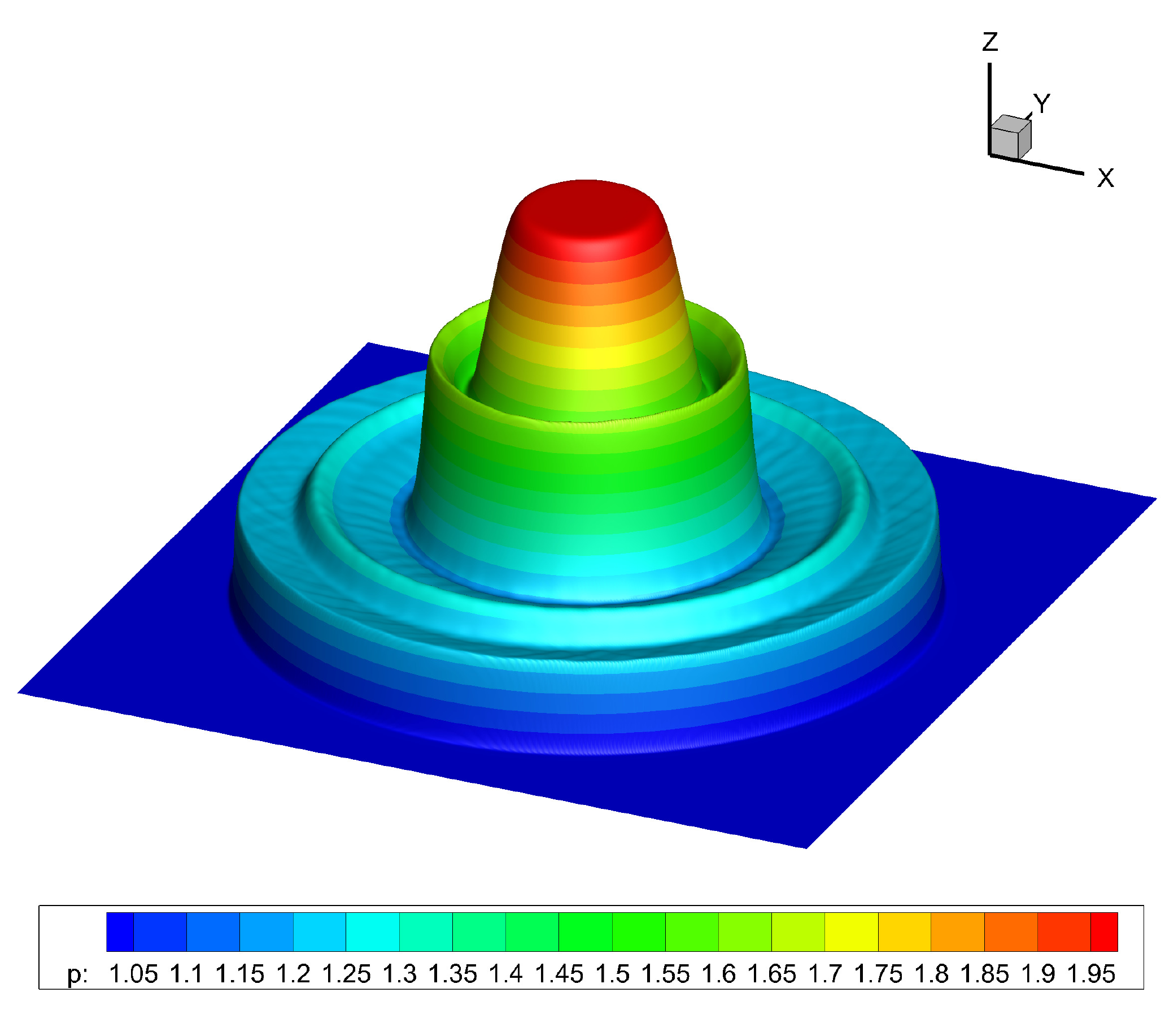}   
		\end{tabular} 
		\caption{2D explosion problem EP2 for the homogeneous GPR model with $\tau_1 = \tau_2 = 10^{20}$ at time $t=0.15$. 3D pressure contour color plot for the new SPSIFV scheme (left) and a standard explicit MUSCL TVD finite volume method (right).   } 
		\label{fig.ep2dsolid1}
	\end{center}
\end{figure}

\begin{figure}[!htbp]
	\begin{center}
		\begin{tabular}{cc} 
			\includegraphics[width=0.47\textwidth]{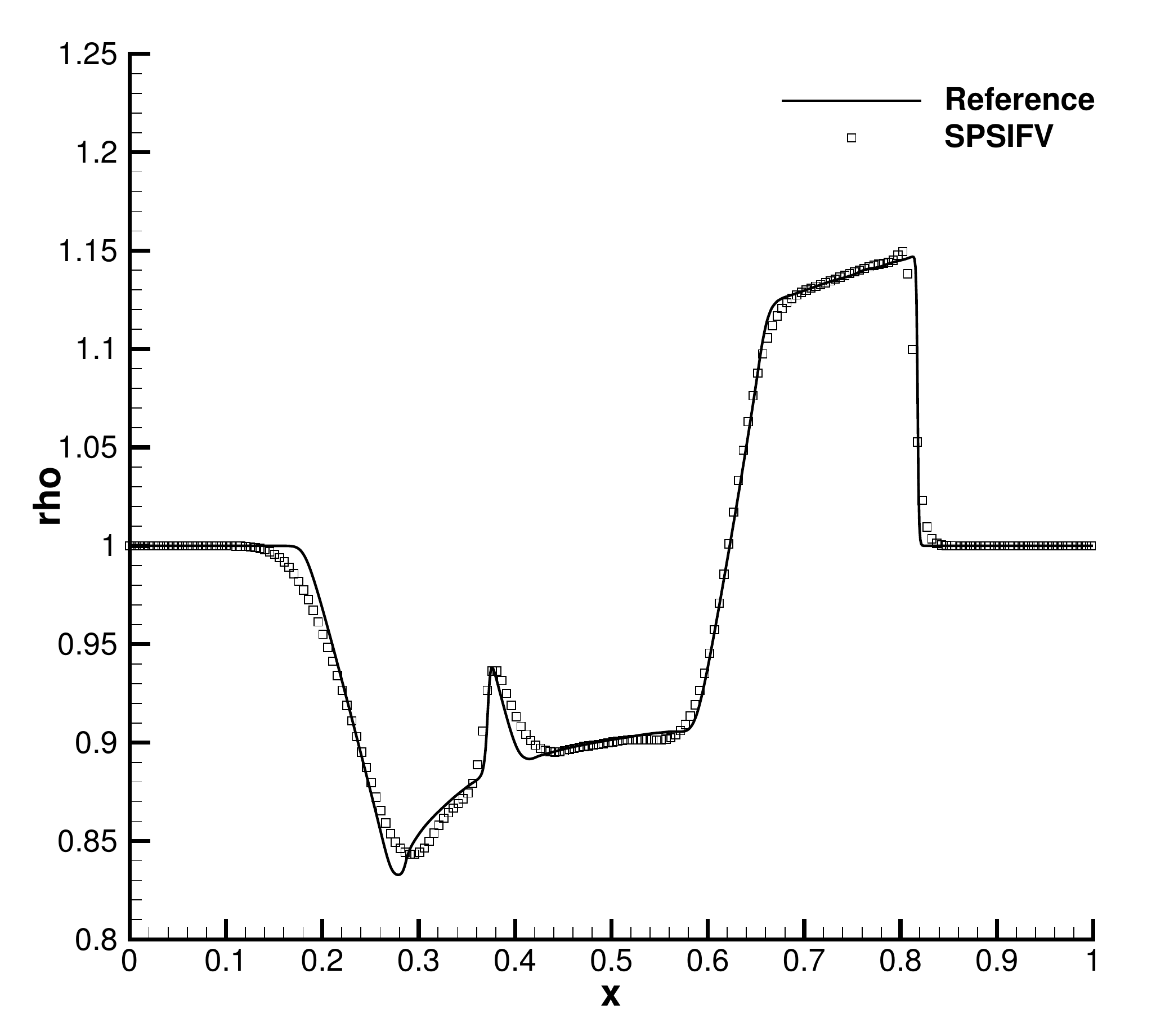}  &
			\includegraphics[width=0.47\textwidth]{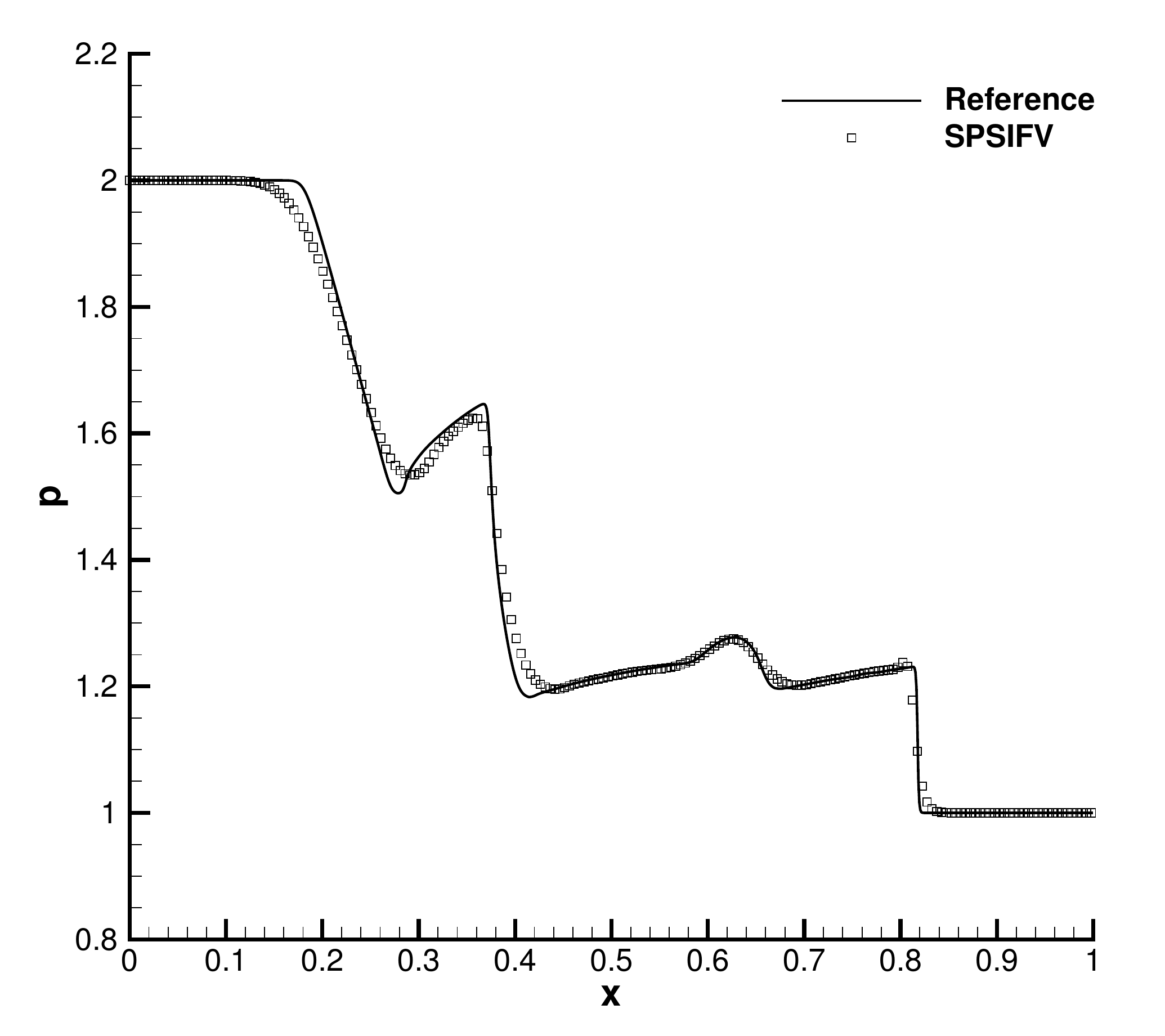}   \\ 
			\includegraphics[width=0.47\textwidth]{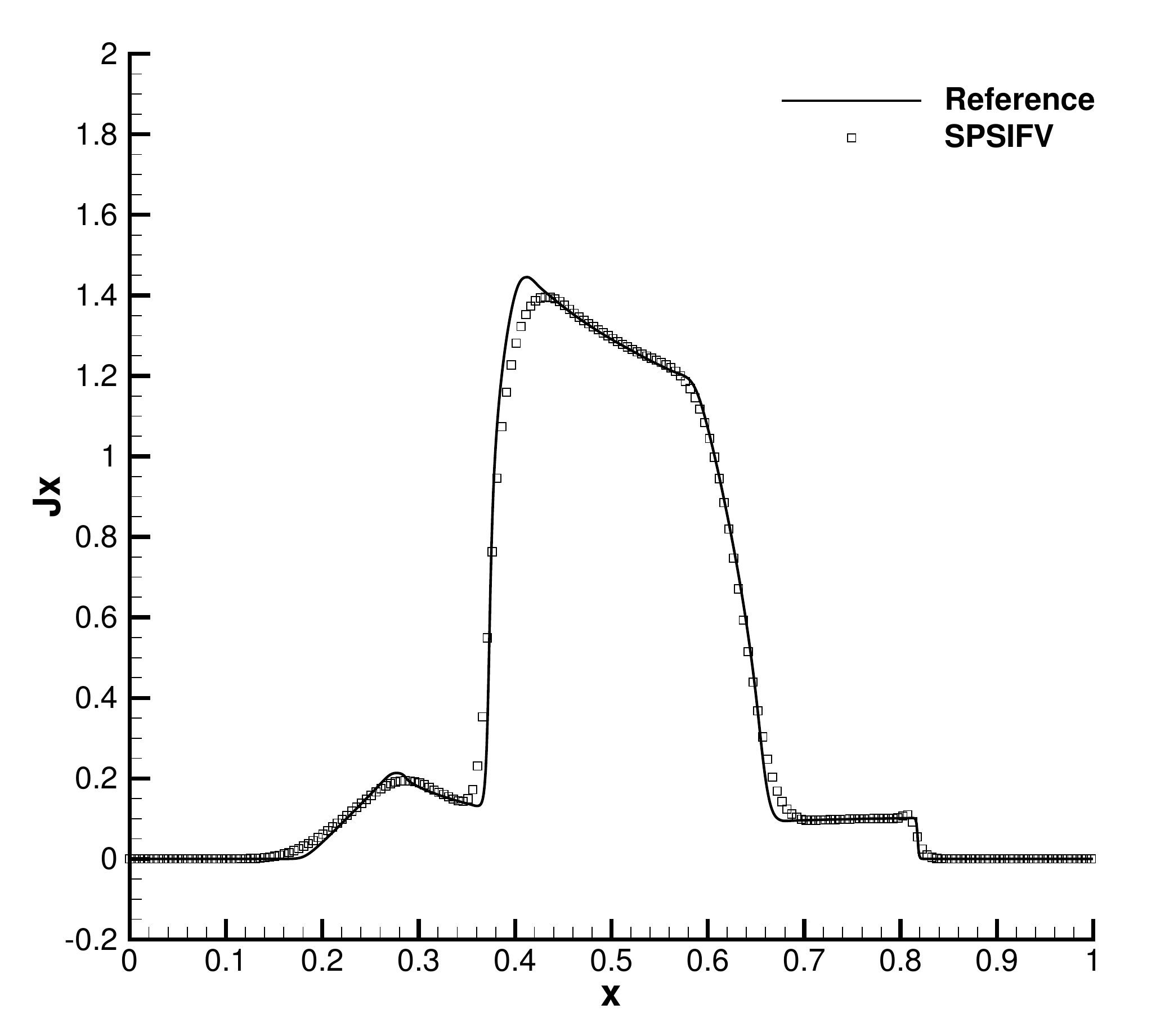}  & 
			\includegraphics[width=0.47\textwidth]{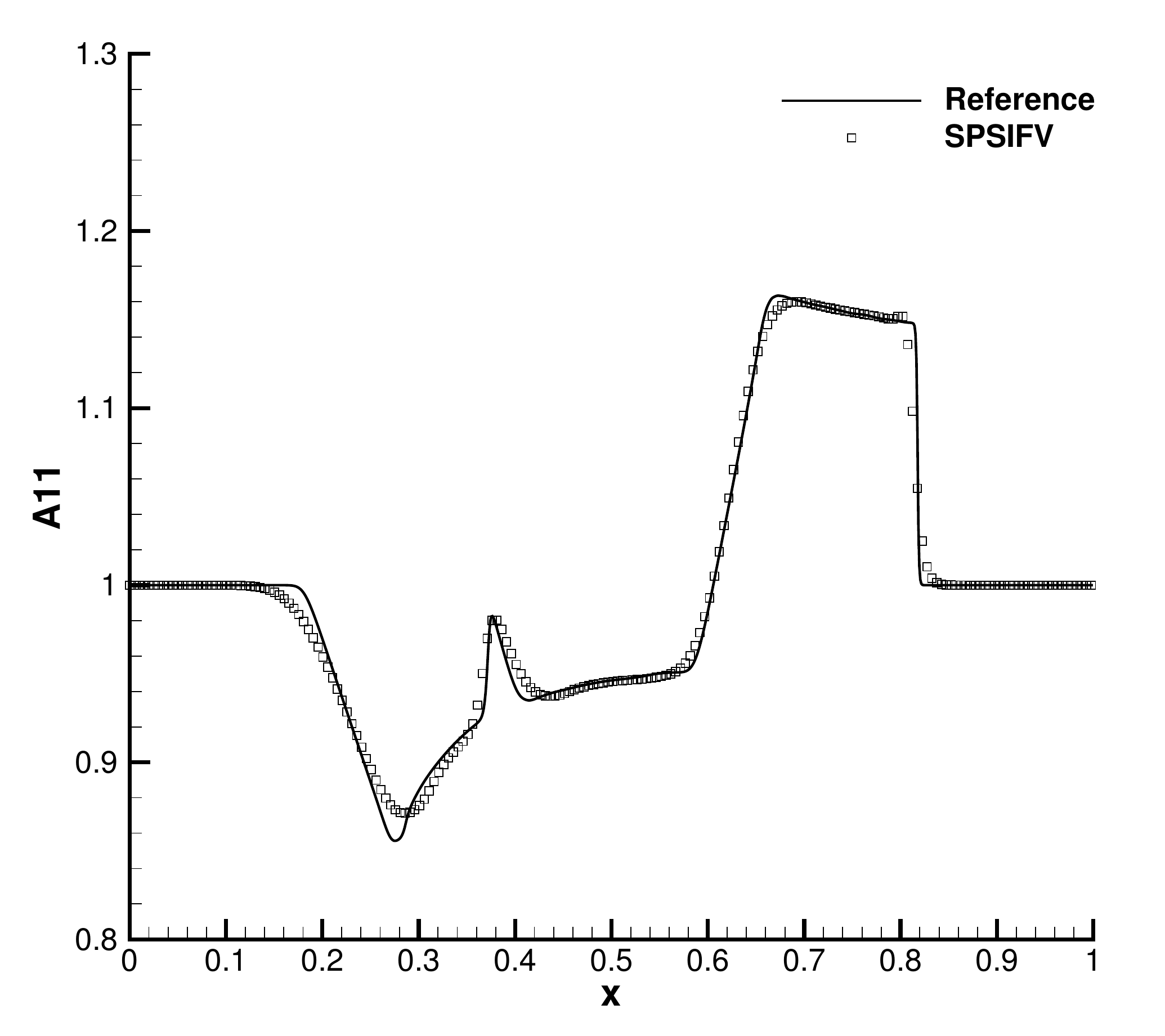}  \\ 
		\end{tabular} 
		\caption{Cut along the $x$ axis for the 2D explosion problem EP2 at time $t=0.15$. Comparison of a fine grid reference solution (solid line) against the numerical solution of the homogeneous GPR model ($\tau_1=\tau_2=10^{20}$)  obtained with the new SPSIFV scheme (square symbols). Density $\rho$ (top left), pressure $p$ (top right), thermal impulse component $J_1$ (bottom left) and distortion field component $A_{11}$ (bottom right).   } 
		\label{fig.ep2dsolid2}
	\end{center}
\end{figure}

\paragraph{EP2: Solid limit of the model} We now solve the governing PDE system in the solid limit, 
i.e. 
$\tau_1 \to \infty$ and $\tau_2 \to \infty$. Here, we choose the following initial condition. We 
set $\rho = 1$, $\mathbf{v}=0$, $\AAA = \mathbf{I}$ and $\mathbf{J}=0$
everywhere in the computational domain, while we choose $p_{in} = 2$ and $p_{out}=1$ for the pressure. The other parameters of the GPR model are set to $\gamma=1.4$, $c_v=2.5$, $\rho_0=1$, $\tau_1 = \tau_2 = 10^{20}$. The computational mesh is composed of $500 \times 500$ and the final simulation time is set to $t=0.15$. We solve the problem twice, once with the new SPSIFV scheme and another time
with a standard second order accurate MUSCL-Hancock finite volume scheme, see \cite{toro-book}. The computational results obtained
with both schemes are shown via contour plots and 1D cuts along the $x$-axis in Figs. \ref{fig.ep2dsolid1} and \ref{fig.ep2dsolid2}, respectively. Overall, we can observe a good agreement between the two solutions. Since we run the present test case with 
$\tau_1 = \tau_2 = 10^{20}$, i.e. the governing PDE system becomes homogeneous, we can again compare the $L_1$ error norms of the 
curl of $\AAA$ and $\J$ obtained with the two different schemes. As already shown in the first test problem, the new SPSIFV method is 
able to maintain the error close to machine zero, while the standard TVD finite volume scheme produces errors in the curl of $\AAA$ and $\J$ that are about ten orders of magnitude larger. 

Finally, we also provide detailed CPU times for the two schemes so that the reader can assess the computational efficiency of the 
proposed SPSIFV scheme. The present test problem is \textit{not} a low Mach number problem, i.e. we expect the semi-implicit scheme
to be less efficient than the explicit method. The total CPU time needed by the SPSIFV method on one single core of an Intel i9-7900X 
CPU with 3.3 GHz of clock frequency and 32 GB of RAM was 298.3 s, while the explicit second order TVD scheme needed 207.2 s, which is
only about 31 \% less than the semi-implicit scheme. Note that the new semi-implicit scheme needs to solve an implicit pressure 
system in each time step. The total wall clock time can also be normalized by the number of time steps and by the number of elements,  leading to the time that is needed to update one element. The time needed by the SPSIFV scheme for one single element update was $4.4 \mu s$,  while it was $2.39 \mu s$ for the explicit second order TVD scheme. 
From these results we can conclude that the computational efficiency of the proposed semi-implicit  finite volume scheme is still  competitive even for non low Mach number flows, while it is obviously much faster than an explicit scheme 
for low Mach number flow problems, see the results obtained for the Taylor-Green vortex and for the lid-driven cavity problem.

%
%
%
%
%
%

\begin{figure}[!htbp]
	\begin{center}
		\includegraphics[width=0.7\textwidth]{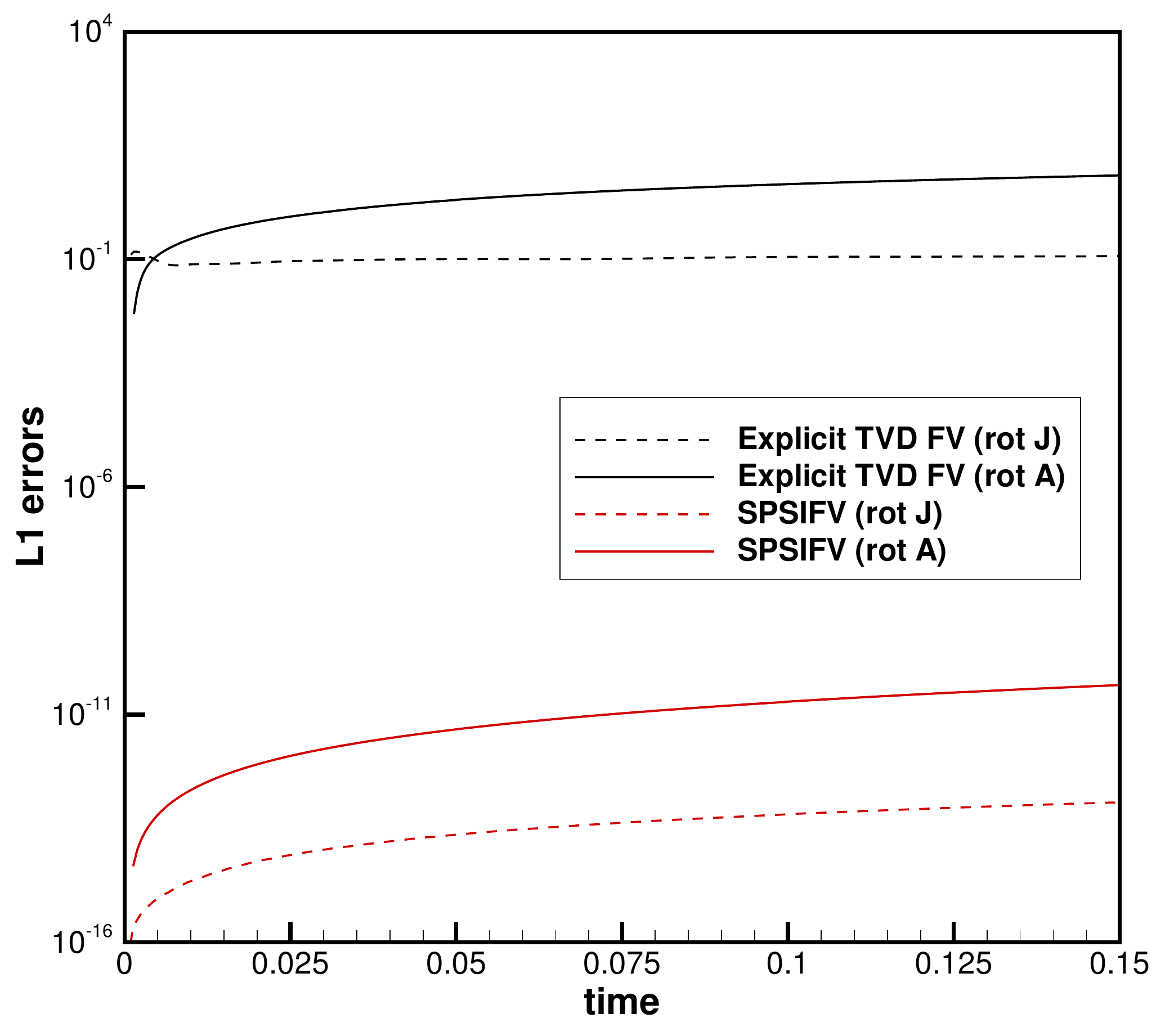}  
		\caption{Time series of the $L_1$ error norms of the curl of $\AAA$ and $\mathbf{J}$ for the 2D explosion problem EP2 until time $t=0.15$ using a standard second order MUSCL-Hancock-type TVD finite volume scheme (black) and the new structure-preserving semi-implicit finite volume scheme (red). The new structure-preserving method is able to preserve all curl-free conditions of the GPR model essentially up to machine precision.   } 
		\label{fig.ep2d.curlerrors}
	\end{center}
\end{figure}

\section{Conclusions}
\label{sec.concl} 
We have presented a new structure-preserving staggered semi-implicit finite volume method for the unified GPR model of  continuum mechanics. The scheme is consistent with the low Mach number limit of the equations,  it is exactly curl-free 
for the homogeneous part of the PDE system in the absence of source terms and is consistent with 
the Navier-Stokes-Fourier limit
of the model in the stiff relaxation limit when $\tau_1 \to 0$ and $\tau_2 \to 0$. To the best knowledge of the authors,
this is the first time that an exactly curl-free scheme has been proposed for the equations of nonlinear large-strain  
hyperelasticity in Eulerian coordinates.

Future work will consist in an extension of the present approach to general unstructured meshes in multiple space dimensions and to higher order of accuracy at the aid of staggered 
semi-implicit discontinuous Galerkin (DG) finite element schemes, following the ideas outlined in \cite{DumbserCasulli,TavelliDumbser2016,TavelliDumbser2017,FambriDumbser,AMRDGSI}. 
In the near future we also plan an extension of this new family of efficient semi-implicit finite volume schemes to Baer-Nunziato-type models of compressible multi-phase flows \cite{BaerNunziato1986,SaurelAbgrall,SaurelAbgrall2,SaurelSurfTension} and to the conservative two-phase flow model \cite{RomenskiTwoPhase2007,RomenskiTwoPhase2010}, 
where low Mach number problems are particularly important due to the simultaneous presence of two 
different phases with substantially different sound speeds. Further extensions will also concern 
compressible  multi-phase flow models with surface tension 
\cite{HypSurfTension,SHTCSurfaceTension}, as well as a recent 
hyperbolic reformulation of the Schr\"odinger equation \cite{Dhaouadi2018}, which are also endowed with
curl constraints.
First preliminary results of the authors indicate that the use of exactly curl-free schemes for hyperbolic 
models with curl involutions, like the ones proposed in \cite{HypSurfTension,Dhaouadi2018} might be as important as the use of exactly divergence-free schemes in the context of the magnetohydrodynamics (MHD)  equations.

\section*{Acknowledgements}

The research presented in this paper was partially funded by the European Union's Horizon 2020 Research and Innovation  Programme under the project \textit{ExaHyPE}, grant no. 671698 (call FETHPC-1-2014). 
The authors also acknowledge funding from the Istituto Nazionale di Alta Matematica (INdAM) through the GNCS group and the program \textit{Young Researchers Funding 2018} via the research project \textit{Semi-implicit structure-preserving schemes for continuum mechanics}. 
Results by E.R. obtained in Sec. 2 were carried out within the framework of the state contract of the Sobolev Institute of Mathematics (project no. 0314-2019-0012).  
M.D. and I.P. acknowledges the financial support received from the Italian Ministry of Education, 
University and Research (MIUR) in the 
frame of the Departments of Excellence Initiative 2018--2022 attributed to DICAM of the University of Trento (grant L. 232/2016). W.B., M.D. and I.P. also received financial support in the frame of the PRIN Project 2017 (No. 2017KKJP4X entitled \textit{Innovative numerical methods for evolutionary partial differential equations and applications}). MD has also received funding from the University of Trento via the  
\textit{Strategic Initiative Modeling and Simulation}. I.P. has further received funding 
from  
the University of Trento via the \textit{UniTN Starting Grant initiative}. 


\appendix

\section{Sound speeds at the equilibrium}
\label{sec.appendix.sound.speed}

It is likely impossible to get analytical expression for the characteristic speeds of the GPR model 
\eqref{eqn.GPR} in the general case. However, for understanding the type of the waves the model can 
have, it might be 
useful to have the formulas for characteristic speeds at the equilibrium, i.e. $ \rho = \rho_0 $, $ 
\AAA = \mathbf{I} $, $ \J = 0 $. Thus, in the equilibrium, the characteristic polynomial reads
\begin{equation}
\tilde{\lambda}^9 (c_s^2 - \tilde{\lambda}^2)^2 (a_0 + a_2 \tilde{\lambda}^2 - a_4 \tilde{\lambda}^4) = 0,
\end{equation}
with $\tilde{\lambda}=\lambda - u$ and where 
\begin{subequations}
	\begin{align}
		a_0 &= \alpha^2 T (4 c_s^2 + 3 c_v (\gamma - 1) T), \\
		a_2 &= 4 c_s^2 c_v + 3 (\alpha^2 + c_v^2 (\gamma-1) \gamma) T,\\
		 a_4 &= 3 c_v,
	\end{align}
\end{subequations}
and $ T = \pd E_1(\rho,s)/\pd S$ is the equilibrium temperature.
Thus, there are 8 types of wave: pure advective waves corresponding to $ \lambda^9 $, 2 shear 
waves with characteristic speed $\lambda_s = u \pm c_s$, which may be different out of equilibrium, 
corresponding to $ (c_s^2 - \lambda^2)^2 $, 
4 thermo-acoustic waves 
\begin{equation}
\lambda_{ta} = u \pm \sqrt{\frac{a_2 \mp \sqrt{a_2^2 + 4 a_0 a_4}}{2 a_4}}
\end{equation}
corresponding to $ a_0 + a_2 \lambda^2 + a_4 \lambda^4 = 0$ which couple the 
the longitudinal waves and two (fast and slow) thermal waves. Note that if we put $ \alpha = 0$, 
then 
the speeds of the thermo-acoustic waves reduce to $ c_l^2 = c_0^2 + \frac43 c_s^2$, i.e. to the 
standard expression for the equilibrium speeds of longitudinal waves in solids. 
The remaining eigenvalues are $\lambda_a = u$ associated with pure advection.

\bibliography{SIGPR}

\end{document}